 \documentclass[final,5p,times,twocolumn,authoryear]{elsarticle}

\usepackage{amssymb}
\usepackage{amsmath}
\usepackage{pifont}
\usepackage{xcolor}

\newcommand{\draft}[1]{#1}
\newcommand{\draftt}[1]{#1}

\newcommand{\cmark}{\ding{51}}%
\newcommand{\xmark}{\ding{55}}%

\renewcommand\vec{\boldsymbol}

\journal{New Astronomy}

\def\colorB{black}
\def\colorDO{black}
\def\colorDT{black}
\def\colorDTh{black}

\begin{document}

\begin{frontmatter}



\title{OSIRIS-GR: General relativistic activation of the polar cap of a compact neutron star}


\author[first]{R. Torres}
\author[first]{T. Grismayer}
\author[first]{F. Cruz}
\author[first,second]{R.A. Fonseca}
\author[first]{L.O. Silva}

\affiliation[first]{organization={GoLP/Instituto de Plasmas e Fusão Nuclear},
            addressline={Instituto Superior Técnico, Universidade de Lisboa}, 
            postcode={1049-001}, 
            state={Lisboa},
            country={Portugal}}

\affiliation[second]{organization={DCTI/ISCTE},
            addressline={Instituto Universitário de Lisboa}, 
            postcode={1649-026}, 
            state={Lisboa},
            country={Portugal}}

\begin{abstract}
\color{\colorDO}We \color{\colorDT}present ab \color{\colorDO}initio global general-relativistic Particle-in-cell (GR-PIC) simulations of compact millisecond neutron star magnetospheres in the axisymmetric \draft{aligned} rotator configuration. We investigate the role of GR and plasma supply on the polar cap particle acceleration efficiency - the precursor of coherent radio emission - employing a new module for the PIC code OSIRIS, designed to model plasma dynamics around compact objects with \draft{fully} self-consistent GR effects. We provide a detailed description of the main sub-algorithms of the novel PIC algorithm, including a charge-conserving current deposit scheme for curvilinear coordinates. We \color{\colorDT}demonstrate efficient \color{\colorDO}particle acceleration in the polar caps of compact neutron stars with denser magnetospheres, numerically validating the spacelike current extension provided by force-free models. We show that GR relaxes the minimum required poloidal magnetospheric current for the transition of the polar cap to the accelerator regime, thus justifying the observation of weak pulsars beyond the expected death line. We denote that spin-down luminosity intermittency and radio pulse nullings for older pulsars might arise from the interplay between the polar and outer gaps. Also, narrower radio beams are expected for weaker \draft{low-obliquity} pulsars.\color{black}
\end{abstract}





\end{frontmatter}




\section{Introduction}
\label{Sec: Intro}

Neutron stars are exquisite physics laboratories coupling extreme magnetic fields with strong gravitational fields. These compact astrophysical objects were discovered in 1967 as radio pulsars \citep{Hewish_1968}. Since then \color{\colorDT}pulsars \color{black}have been observed as bright sources of electromagnetic radiation, from radio to gamma rays. Despite recent advances in understanding the intricate dynamics of their magnetospheres, propelled in part by the available computational power, we still lack a complete picture that explains the observed radiation features.

\citet{Goldreich_1969} noted that an isolated neutron star modelled by a rotating spheric perfect conductor that is threaded by a dipolar magnetic field would induce an electric field strong enough to extract charged particles from its surface, populating its magnetosphere with a dense charge-separated plasma. In an ultra-strong magnetic field, the electromagnetic force dominates, leading to the force-free condition, where plasma shorts out the electric field parallel to the magnetic field. Consequently, particles move along magnetic field lines, which corotate rigidly with the neutron star. A crucial element in the magnetosphere is the light cylinder radius, $R_{\mathrm{LC}}=c/\Omega_*$, the distance from the rotational axis after which corotation is not possible due to its tangential speed exceeding the light speed. This boundary separates the closed corotating field lines from the open bundle originating from the polar cap of the neutron star. The outflowing particles generate poloidal currents that sweep back the open field lines, inducing a toroidal component. Beyond the light cylinder radius, an equatorial current sheet forms to support the discontinuity of the toroidal magnetic field component. In this region, the poloidal electric field and the toroidal magnetic field result in an outflowing Poynting flux that we observe as a spin-down luminosity. \citet{Contopoulos_1999} confirmed this model by numerically solving the ``pulsar equation" \citep{Scharlemann_1973,Michel_1973}, the time-independent version of force-free electrodynamics. This work established the standard magnetospheric picture, providing the complete magnetospheric current closure with a y-shaped current layer flowing between the open and closed field lines.

Observations of pulsar wind nebulae (PWNe), i.e. rotation-powered structures formed from the ejection of plasma advected from the inner pulsar magnetosphere \citep{Kirk_2009, Mitchell_2022}, support the force-free solution given that there is a dense plasma to screen the non-ideal electric field everywhere \citep{Gruzinov_2005,Spitkovsky_2006,Kalapotharakos_2009,Timokhin_2007,Bai_2010,Petri_2012,Philippov_2014a}. However, these models do not constrain the mechanism that provides the plasma to drive the required magnetospheric currents. Without external plasma sources, there must exist dissipative regions - gaps - within the magnetosphere where particles accelerate and produce pairs that cancel the accelerating electric field. These gaps would operate at much shorter time scales (microseconds) than the magnetospheric time scales (milliseconds to seconds, thus quasi-stationary), consistent with the force-free solution. Several studies explored the observational implications of the polar cap gap \citep{Daugherty_1982,Usov_1995,Daugherty_1996,Ruderman_1975}, slot gap \citep{Arons_1979,Arons_1983,Muslimov_2003,Muslimov_2004}, and outer gap models \citep{Cheng_1986,Romani_1996,Takata_2004,Hirotani_2008}\color{black}, highlighting the importance of these gaps not only to provide the necessary plasma to the magnetosphere but also to explain the observed electromagnetic spectrum \color{\colorDT}(e.g., \citet{Tao_2021})\color{black}.

The radio beam in radio pulsars originates from a region above the pair formation front of the polar cap gap due to the nonlinear screening of the accelerating electric field. The misalignment of the magnetic and rotation axis leads to a pulsed behaviour as the beam crosses our line of sight periodically, similar to a lighthouse. These radio pulses contain microstructures, some possessing single \citep{Drake_1968,Deshpande_1999,Vivekanand_1999} or multiple drifting components \citep{Qiao_2004,Basu_2018}. Models to explain this behaviour assume a limitation in the extracted charged particles from the stellar surface such that the spark filaments no longer corotate with it. Hence, the sub-pulses undergo an $E\times B$ drift different from the field lines due to the parallel electric field at the base of those field lines. \citet{Ruderman_1975} (hereafter RS75) assumed the case where no particles could be lifted from the stellar surface, thus forming a vacuum gap above it. The intermediate model of \citet{Gil_2003} allows a thermionic outflow from the surface forming a partially screened gap (PSG). This model accommodates a necessary reduction of the drifting speed and provides a better estimate of the X-ray luminosity from the backflow bombardment of particles. The space charge limited flow model (SCLF) by \citet{Arons_1979}, where particles can freely escape from the stellar surface, does not provide a natural explanation for this phenomenon. The binding energies of either ions or electrons should not be too high, thus motivating the SCLF model \citep{Medin_2010}. Alternatively, studies have highlighted the possibility of strongly bound particles in bare strange quark stars \citep{Xu_1998,Xu_1999,Yu_2011}, favouring the RS75 or PSG models. 

First principle global magnetospheric models that capture the kinetic plasma scales are the best way to introduce the effects of these dissipative regions. Several works have proposed tools to tackle this very computationally demanding task in one-dimension \citep{Timokhin_2010,Timokhin_2012}, two-dimensions \draft{\citep{Chen_2014,Belyaev_2015a,Belyaev_2015b,Cerutti_2013,Cerutti_2015,Philippov_2015a,Cerutti_2017,Cruz_2021,Hu_2022, Bransgrove_22,Cruz_2023}} or three-dimensions \draft{\citep{Philippov_2015b,Cerutti_2016b,Philippov_2018,Kalapotharakos_2018,Brambilla_2018,Cerutti_2020,Hakobyan_2023}}. The work by \draft{\citet{Philippov_2015b}} concluded that the polar cap of low obliquity rotators has weak particle acceleration, suppressing the pair production and shutting down the radio pulsar mechanism. Several authors later introduced the general-relativistic frame-dragging effect \citep{Beskin_1990,Muslimov_1992} into \color{\colorDO}Faraday's induction equation\color{\colorB}, demonstrating that this effect could potentiate the ignition of the polar cap even for the aligned rotator \color{\colorDO}\draft{\citep{Philippov_2015a,Chen_2020,Philippov_2020,Bransgrove_22}} as predicted by general-relativistic force-free models \citep{Belyaev_2016,Gralla_2016,Huang_2018}. Although such kinetic models are able to capture the polar cap ignition, qualitative corrections from neglecting the lapse function and general-relativistic magnetic field topology \citep{Torres2023a} are not resolved. One of such parameters of utmost importance is the spacelike angle $\theta_{\mathrm{SL}}$, expected to limit the poloidal extension of the observed pulsar radio emission generation volume. This angle was obtained analytically for force-free models \citep{Belyaev_2016,Gralla_2016}, but was not yet verified using general-relativistic particle-in-cell simulations. Thus, ab \color{\colorB}initio simulations of neutron star magnetospheres with \draft{fully} self-consistent general-relativistic effects still lack in the community.

The purpose of this paper is to present \color{\colorDO}the first \color{\colorB}simulations of compact \color{\colorDO}millisecond \color{\colorB}neutron stars including self-consistent general-relativistic effects\color{\colorDT}, using a charge-conserving GR-PIC code\color{\colorB}. Also, we \color{\colorDT}summarize \color{\colorB} all the necessary techniques required for the implementation of a GR-PIC code. {As far as we know, a systematic description of the GR-PIC algorithm does not exist in the literature}. We detail the methodology and algorithms used to extend the particle-in-cell code OSIRIS to include GR effects in Sect.~\ref{Sec: GR-PIC}. In Sects.~\ref{Sec: Results sec3} and~\ref{Sec: Results sec4}, we perform numerical experiments of such magnetospheres by varying the plasma injection methods to explore the full range of magnetospheric solutions, from quasi-force-free to full-charge-separated solutions, expanding on the consequent observational implications. {\color{\colorDTh}We show that \draft{the} correct \draft{dimensions} of the polar cap and spacelike current region \draft{can} only \draft{be obtained} when considering the lapse function effects. Also, we demonstrate that charge-separated magnetospheric solutions possess smaller values of the spacelike angle with respect to force-free (FF) estimates. Therefore,} \color{\colorDT}we conjecture the existence of narrower radio beams from weak pulsars, some still observable beyond the expected death line. \color{\colorB}Conclusions and future prospects are presented in Sect.~\ref{Sec: Conclusions}.

\section{GR-PIC method}
\label{Sec: GR-PIC}
In this section, we detail the extension module of general relativity to the OSIRIS particle-in-cell framework \citep{Osiris}. We describe the main equations and integration algorithms implemented in OSIRIS including the field solver, equations of motion, charge conservation, and boundary conditions. A general description of the techniques and methodologies (independent of the internal specifics/implementation of the PIC code) is provided to facilitate their implementation in other PIC codes.

Throughout the paper, we use units in which $c=G=1$, the $(-,+,+,+)$ metric signature, and stellar properties are characterized by an asterisk in subscript (e.g., $R_*$ for the stellar radius). In addition, we adopt the 3+1 formalism of general relativity \citep{Thorne1982}.

\subsection{Metric}
\label{Sec: Metric}
The astrophysical objects we are interested in constitute the class of compact objects where the background metric is no longer a flat (Minkowski) metric. We focus our study on the dynamics of the magnetospheres of black holes and neutron stars. The electromagnetic energy density in such environments is negligibly small compared to their corresponding total mass density. Consequently, we can assume a fixed background metric for these systems. Black holes and neutron stars are modeled by the Kerr and the Hartle-Thorne metric \citep{Hartle1968}, respectively. However, as neutron stars rotate at significantly smaller angular velocities, one can invoke the slow-rotation approximation and simplify the corresponding metric ($R_*\Omega_*\ll1$) for neutron stars. In this approximation, and keeping only the linear terms of the angular velocity, the exterior metric takes the form \citep{Hartle1967}:
\begin{align}
    \mathrm{d}s^2 =& -\alpha^2\mathrm{d}t^2+{\gamma}_{ij}\left(\mathrm{d}x^i+\beta ^i\mathrm{d}t\right)\left(\mathrm{d}x^j+\beta ^j\mathrm{d}t\right)\nonumber\\
    =&-\alpha^2\mathrm{d}t^2+\alpha^{-2}\mathrm{d}r^2 + r^2\mathrm{d}\Omega^2-2\omega(r)r^2\sin^2{\theta}\mathrm{d}\phi\mathrm{d}t, \label{eq1}
\end{align}
where $\alpha=\sqrt{1-R_s/r}$ and $\beta^i=\left(0,0,-\omega\right)$ are the lapse function and shift vector, respectively, ${\gamma}_{ij}$ is the spatial part of the metric, $R_s$ is the Schwarzschild radius, and $\mathrm{d}\Omega^2 = \mathrm{d}\theta^2 + \sin^2{\theta}\mathrm{d}\phi^2$. Interestingly, this metric is equivalent to the slow-rotation approximation of the Kerr metric, which leads us to adopt the Boyer-Lindquist coordinate system ($t$,$r$,$\theta$,$\phi$). Currently, OSIRIS has the Minkowski, the Schwarzschild, and the slow-rotation approximation of the Kerr metric implemented. The Minkowski metric is recovered in the zero compactness limit ($R_s\rightarrow0$), and the Schwarzschild metric by neglecting the frame-dragging effect ($\omega(r)\rightarrow0$). The latter effect is a correction that captures the differential rotation $\omega(r)$ associated with a free-falling inertial frame \citep{Ravenhall1994}:
\begin{equation}
    \omega(r) \equiv \frac{\mathrm{d}\phi}{\mathrm{d}t} \approx 0.21\Omega_*\frac{R_s}{R_*-R_s}\left(\frac{R_*}{r}\right)^3=\frac{\omega_0}{r^3}.
\end{equation}

We define fiducial observers (FIDOs) with world lines normal to spatial hypersurfaces of constant ``universal" time. The lapse function relates the rate at which the local fiducial observer clock ticks in relation to the universal time. Hence, the lapse function translates local time-measured quantities to the universal time. The shift vector measures the angular velocity at which the coordinate frame shears in relation to the FIDOs. We select FIDOs corotating with the absolute space, thus considering zero angular momentum observers (ZAMOs). In these hypersurfaces, we use the orthonormal basis vectors and vectorial components, i.e. $e_{\hat i}\equiv e_i / \sqrt{\gamma_{ii}}$ and $A^{\hat i}\equiv \sqrt{\gamma_{ii}}A^i$ such that $\vec{A}=A^ie_i=A^{\hat i}e_{\hat i}$. This basis allows for a trivial extension of three-dimensional vectorial operations to curved geometries. Also, the definition of line, area, and volume elements is modified according to eq.~\eqref{eq1} - see~\ref{AP:A}.

\subsection{Grid and coordinates}
So far, we have addressed the metric that defines the system dynamics. In this section, we exploit the usage of different coordinate systems for the same metric (e.g., cartesian or spherical for the Minkowski metric). 

Compact objects are naturally described in spherical coordinates. Using body-fitted coordinate grids simplifies the inclusion of system symmetries and boundary conditions. Another degree of freedom is the usage of linear or non-uniform grid spacings. A common feature in similar codes is to allow uniform grid spacings in $\log r $ (logarithmic coordinate) and $-\cos\theta$ (equal area coordinate). The first allows for higher resolution close to the central object, while the second enforces constant charge density of each macro-particle in the meridional direction \citep{Belyaev_2015b}. These two properties have shown a reduction of numerical noise on the axis and increased code stability. Therefore, the {OSIRIS-GR} module has these two options available. Linear grid spacing options are also available, although these are not explored in the current paper. Another possibility available is to adopt logical coordinates, i.e. ($\tilde r$,$\tilde \theta$)=($\log r$,$-\cos\theta$), which possesses the two properties described above and retains uniform grid spacings.The background metric in the new coordinate system takes the form:
\begin{align}
    \mathrm{d}s^2 = -\alpha^2\mathrm{d}t^2 &+ \exp{\left(2\tilde r\right)}\left(\frac{\mathrm{d}\tilde r^2}{\alpha^{2}}+\frac{\mathrm{d}\tilde \theta^2}{1-\tilde \theta^2}\right)+\nonumber\\
    &+\exp{\left(2\tilde r\right)}\left(\left(1-\tilde \theta^2\right)\left(\mathrm{d} \phi^2-2\omega(\tilde r)\mathrm{d} \phi\mathrm{d}t\right)\right).\label{eqlineele}
\end{align}
It is important to highlight that vectors and tensors are identical if written on the orthonormal basis, independently of the chosen coordinates:
\begin{equation}
    \tilde A^{\hat r} = \sqrt{{\gamma}_{\tilde r\tilde r}}\tilde A^{r}=\frac{\exp{\left(\tilde r\right)}}{\alpha}\exp{\left(-\tilde r\right)}A^{r} = \frac{1}{\alpha}A^{r} = \frac{1}{r}\frac{A^{\hat r }}{\alpha^{-1}}=A^{\hat r}.
\end{equation}
We should also mention that this is only possible for orthogonal metric systems, i.e. diagonal spatial metrics.

\color{\colorDO}Also, the OSIRIS-GR module introduces a new possibility towards modelling compact object magnetospheres \color{\colorDT}via \color{\colorDO}the half-domain setup. This simulation mode differs from the standard pole-to-pole meridional extension ($0\leq\theta\leq\pi$) by reducing the poloidal plane in half, thus possessing a pole-to-equator meridional extension ($0\leq\theta\leq\pi/2$). This novel numerical setup captures the global magnetospheric current closure while reducing the computational cost of emulating the complete poloidal plane. In this way, a higher number of macro-particles can be employed, increasing the statistical significance and reducing the associated numerical noise. This simulation mode is particularly interesting to study the pulsar polar cap dynamics, where the asymmetric mode development of the equatorial current sheet may be neglected to leading order.\color{\colorB}


\subsection{Maxwell's equations}
\begin{figure}
    \centering
    \includegraphics[width=.7\columnwidth]{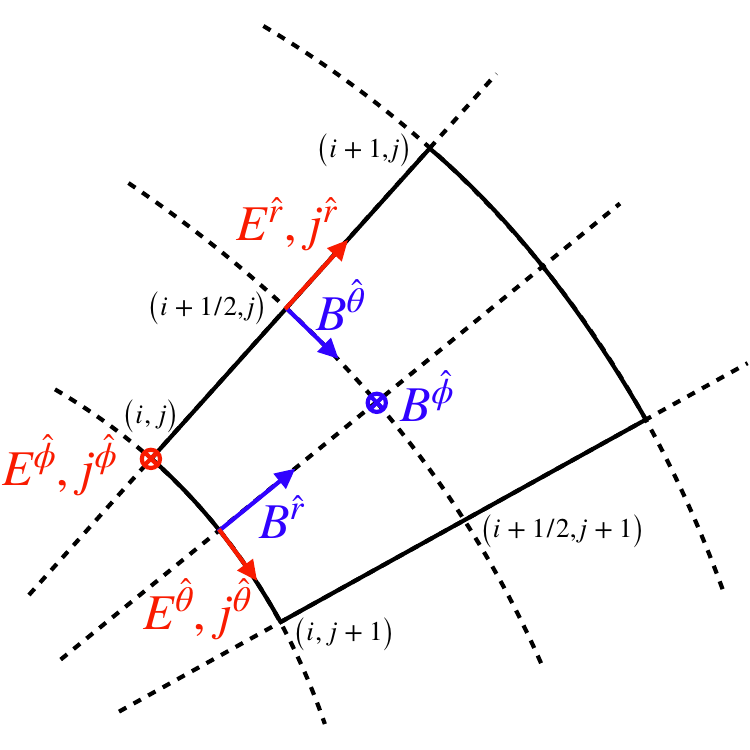}
    \caption{Spherical Yee cell adopted in {OSIRIS-GR} with the location in the numerical cell where each electromagnetic component is defined.}
    \label{fig:Yee}
\end{figure}

In the 3+1 formalism, Maxwell's equations take the form \citep{Komissarov2011}:
\begin{align}
    &\vec{\nabla}\cdot\vec{E} = 4\pi\rho,\label{eq5}\\
    &\vec{\nabla}\cdot\vec{B} = 0,\\
    &\vec{\nabla}\times\left(\alpha\vec{E}+\vec{\beta}\times\vec{B}\right) = - \frac{\partial\vec{B}}{\partial t},\\
    &\vec{\nabla}\times\left(\alpha\vec{B}-\vec{\beta}\times\vec{E}\right) = \frac{\partial\vec{E}}{\partial t} + 4\pi\left(\alpha\vec{j} - \rho\vec{\beta}\right),
\end{align}
where $\vec{E}$, $\vec{B}$, $\vec{j}$ and $\rho$ are physical quantities measured by fiducial observers in co-rotation with the absolute space (ZAMOs). In spherical coordinates, Maxwell's equations are better described in the integral form using the Kelvin–Stokes theorem on a Yee-lattice \citep{Belyaev_2015a, Cerutti_2015}. The {OSIRIS-GR} module adopts the spherical cell shown in Fig.~\ref{fig:Yee}.
In this way, the curl terms can be written as
\begin{align}
    &\left(\vec{\nabla}\times\vec{\tilde E}\right)^{\hat r}_{i,j+1/2} = \frac{\alpha_i}{A^{\hat r}_{i,j+1/2}}\left(l^{\hat\phi}_{i,j+1}E^{\hat\phi}_{i,j+1}-l^{\hat\phi}_{i,j}E^{\hat\phi}_{i,j}\right),\label{eq9}\\
    &\left(\vec{\nabla}\times\vec{\tilde E}\right)^{\hat \theta}_{i+1/2,j} = \frac{1}{A^{\hat \theta}_{i+1/2,j}}\left(l^{\hat\phi}_{i,j}\alpha_iE^{\hat\phi}_{i,j}-l^{\hat\phi}_{i+1,j}\alpha_{i+1}E^{\hat\phi}_{i+1,j}\right),\\
    &\left(\vec{\nabla}\times\vec{\tilde E}\right)^{\hat \phi}_{i+1/2,j+1/2} = \frac{l^{\hat r}_{i+1/2}}{A^{\hat \phi}_{i+1/2,j+1/2}}\left(\alpha_{i+1/2}\left(E^{\hat r}_{i+1/2,j}-E^{\hat r}_{i+1/2,j+1}\right)+\right.\nonumber\\
    &+\left.\beta^{\hat \phi}_{i+1/2,j+1}B^{\hat \theta}_{i+1/2,j+1}-\beta^{\hat \phi}_{i+1/2,j}B^{\hat \theta}_{i+1/2,j}\right)+\nonumber\\
    &+\frac{1}{A^{\hat \phi}_{i+1/2,j+1/2}}\left(l^{\hat\theta}_{i+1,j+1/2}\alpha_{i+1}E^{\hat\theta}_{i+1,j+1/2}-l^{\hat\theta}_{i,j+1/2}\alpha_{i}E^{\hat\theta}_{i,j+1/2}+\right.\nonumber\\
    &+\left.l^{\hat\theta}_{i+1,j+1/2}\beta^{\hat \phi}_{i+1,j+1/2}B^{\hat r}_{i+1,j+1/2}-l^{\hat\theta}_{i,j+1/2}\beta^{\hat \phi}_{i,j+1/2}B^{\hat r}_{i,j+1/2}\right),\label{eq11}\\
    &\left(\vec{\nabla}\times\vec{\tilde B}\right)^{\hat r}_{i+1/2,j} = \frac{\alpha_{i+1/2}}{A^{\hat r}_{i+1/2,j}}\left(l^{\hat\phi}_{i+1/2,j+1/2}B^{\hat\phi}_{i+1/2,j+1/2}-\right.\nonumber\\
    &-\left.l^{\hat\phi}_{i+1/2,j-1/2}B^{\hat\phi}_{i+1/2,j-1/2}\right),\\
    &\left(\vec{\nabla}\times\vec{\tilde B}\right)^{\hat \theta}_{i,j+1/2} = \frac{1}{A^{\hat \theta}_{i,j+1/2}}\left(l^{\hat\phi}_{i-1/2,j+1/2}\alpha_{i-1/2}B^{\hat\phi}_{i-1/2,j+1/2}-\right.\nonumber\\
    &-\left.l^{\hat\phi}_{i+1/2,j+1/2}\alpha_{i+1/2}B^{\hat\phi}_{i+1/2,j+1/2}\right),
\end{align}
\begin{align}
    &\left(\vec{\nabla}\times\vec{\tilde B}\right)^{\hat \phi}_{i,j} = \frac{l^{\hat r}_{i}}{A^{\hat \phi}_{i,j}}\left(\alpha_{i}\left(B^{\hat r}_{i,j-1/2}-B^{\hat r}_{i,j+1/2}\right)+\right.\nonumber\\
    &+\left.\beta^{\hat \phi}_{i,j-1/2}E^{\hat \theta}_{i,j-1/2}-\beta^{\hat \phi}_{i,j+1/2}E^{\hat \theta}_{i,j+1/2}\right)+\nonumber\\
    &+\frac{1}{A^{\hat \phi}_{i,j}}\left(l^{\hat\theta}_{i+1/2,j}\alpha_{i+1/2}B^{\hat\theta}_{i+1/2,j}-l^{\hat\theta}_{i-1/2,j}\alpha_{i-1/2}B^{\hat\theta}_{i-1/2,j}+\right.\nonumber\\
    &+\left.l^{\hat\theta}_{i-1/2,j}\beta^{\hat \phi}_{i-1/2,j}E^{\hat r}_{i-1/2,j}-l^{\hat\theta}_{i+1/2,j}\beta^{\hat \phi}_{i+1/2,j}E^{\hat r}_{i+1/2,j}\right),\label{eq14}
\end{align}
with $\vec{\tilde E} \equiv\alpha\vec{E}+\vec{\beta}\times\vec{B}$ and $\vec{\tilde B}\equiv\alpha\vec{B}-\vec{\beta}\times\vec{E}$, respectively. Expressions for the line and area elements are detailed in~\ref{AP:A}. Up to this point, we have assumed $\beta^{\hat \phi}$ as the only non-zero component, with a purely radial dependence in $\alpha$, and neglected all azimuthal derivatives (axisymmetry). The evolution equations are then expressed as in the typical \citet{Yee} scheme
\begin{align}
    &\vec{B}^{n+1/2}=\vec{B}^{n}-\frac{\Delta t}{2}\left(\vec{\nabla}\times\vec{\tilde E}\right)^{n},\\
    &\vec{E}^{n+1}=\vec{E}^{n}+\Delta t\left(\vec{\nabla}\times\vec{\tilde B}\right)^{n+1/2}-4\pi\Delta t J^{n+1/2},\label{eq16}\\
    &\vec{B}^{n+1}=\vec{B}^{n+1/2}-\frac{\Delta t}{2}\left(\vec{\nabla}\times\vec{\tilde E}\right)^{n+1},\label{eq17}
\end{align}
with $J\equiv\alpha\vec{j} - \rho\vec{\beta}$ and $n$ being the temporal index. Equation \eqref{eq16} also shows that to correctly push in time the electric field we need to gather the particle's half-step charge density and current (see Sect.~\ref{Sec: Charge conservation} and~\ref{AP:B}). Due to the complex coupling introduced by equations \eqref{eq11} and \eqref{eq14}, the poloidal components are pushed in time before the azimuthal ones in equations \eqref{eq16} and \eqref{eq17}. By applying the Newtonian limit, we retrieve the Minkowski version of the evolution equations in \citet{Belyaev_2015a, Cerutti_2015}. Notice that \eqref{eq9}-\eqref{eq14} can be applied to any axisymmetric system given a diagonal spatial metric (e.g. spherical, cylindrical, toroidal).

\subsection{Particle pusher}
\label{Sec: Particle Pusher}
In the 3+1 formalism, the equations of motion for a charged particle \citep{Thorne1982,Philippov_2015a,Parfrey_2019} are given by 
\begin{align}
&\frac{\mathrm{d}{r^i}}{\mathrm{d}t}\vec{e}_{i} = \frac{\alpha}{\Gamma}\vec{u} - \vec{\beta},\label{eq18}\\
&\frac{\mathrm{d}{u^{\hat i}}}{\mathrm{d}t}\vec{e}_{\hat i} = \frac{\alpha q}{m}\left(\vec{E} + \frac{\vec{u}}{\Gamma}\times\vec{B}\right) + \alpha\Gamma\vec{g}+\alpha\vec{H}\cdot\vec{u}+\vec{F}_{\mathrm{RR}}-\vec{F}_{\mathrm{Coord}}\label{eq19},
\end{align}
where $\vec{g}$ is the gravitational acceleration, $\vec{H}$ is the gravitomagnetic tensor, $\vec{p}=m\Gamma\vec{v}=m\vec{u}$ is the FIDO-measured momentum, $q$ and $m$ are the charge and mass of each macro-particle, $\Gamma=\sqrt{\varepsilon+\vec{u}\cdot\vec{u}}$ is the particle's special relativistic Lorentz factor, and $\varepsilon=0\;\left(=1\right)$ for massless (massive) particles. Equation~\eqref{eq18} shows that the particle velocity is composed of the FIDO-measured velocity, $\alpha\vec{u}/\Gamma$, and the FIDO's velocity, $\vec{\beta}$. Equation~\eqref{eq19} details which forces dictate the particle dynamics. From left to right, we identify the Lorentz, the gravitational, the gravitomagnetic, the classical radiation reaction \citep{Vranic_2016,LL_1975}, and the ``coordinate" forces. The last term comprises the generalized fictitious forces associated with the use of the orthonormal vector basis:
\begin{equation}
    \vec{F}_{\mathrm{Coord}} = u^{\hat i}\frac{\mathrm{d}{\vec{e}_{\hat i}}}{\mathrm{d}t} = \frac{\sqrt{\gamma_{ii}}}{\sqrt{\gamma_{kk}}}\Gamma^i_{kj}u^{\hat k}\frac{\mathrm{d}x^j}{\mathrm{d}t}\vec{e}_{\hat i}=\Gamma^{\hat i}_{\hat{k}j}u^{\hat k}\frac{\mathrm{d}x^j}{\mathrm{d}t}\vec{e}_{\hat i},\;\;k\neq i\label{eq20},
\end{equation}
where $\Gamma^i_{kj}$ are the spatial Christoffel symbols defined in~\ref{AP:A}. We advance in time the contravariant positions and the orthonormal contravariant momenta, i.e. $\vec{r}\equiv r^i\vec{e}_i$ and $\vec{u}\equiv u^{\hat i}\vec{e}_{\hat i}$. To do so, we employ a \color{\colorDO}novel variant of the Strang splitting method to advance in time the equations of motion \citep{Strang68}, which uses leapfrogged positions and velocities. Thus, \color{\colorB}defining the positions centered in integer time steps and the $\vec{u}$ velocities in half-time steps, the discretized equations of motion read:
\begin{align}
    &\vec{u}^{*}=\vec{u}^{n-1/2}+\frac{\Delta t}{2}F_L^n\left(\vec{r}^n,\vec{u}^{n-1/2}\right),\label{eq21}\\
    &\vec{u}^{**}=\vec{u}^{*}+\Delta tF_G^n\left(\vec{r}^n,\vec{u}^{*}\right),\label{eq22}\\
    &\vec{u}^{n+1/2}=\vec{u}^{**}+\frac{\Delta t}{2}F_L^n\left(\vec{r}^n,\vec{u}^{**}\right),\label{eq23}\\
    &\vec{r}^{n+1}=\vec{r}^{n}+\Delta t f^{n+1/2}\left(\vec{r}^n,\vec{u}^{n+1/2}\right),\label{eq24}
\end{align}
with
\begin{align}
    &F_L^n\left(\vec{r}^n,\vec{u}^{n-1/2}\right)=\frac{\alpha\left(\vec{r}^n\right)q}{m}\left(\vec{E}^n\left(\vec{r}^n\right)+\bar{\vec{v}}^n\times \vec{B}^n\left(\vec{r}^n\right)\right),\\
    &F_G^n\left(\vec{r}^n,\vec{u}^{*}\right)=\alpha\left(\vec{r}^n\right)\Gamma\left(\vec{u}^*\right)\vec{g}\left(\vec{r}^n\right)+\alpha\left(\vec{r}^n\right)\vec{H}\left(\vec{r}^n\right)\cdot\vec{u}^*+\nonumber\\
    &\hspace{1.5cm}+\vec{F}_{RR}\left(\vec{r}^n,\vec{u}^{*}\right)-\vec{F}_{Coord}\left(\vec{r}^n,\vec{u}^{*}\right),\\
    &f^{n+1/2}\left(\vec{r}^n,\vec{u}^{n+1/2}\right)=\bar\alpha\vec{u}^{n+1/2}-\bar{\vec{\beta}},
\end{align}
where the terms with a bar highlight the approximated terms. The choice for $\bar{\vec{v}}$ defines if the simulation will be using the Boris \citep{Boris1970}, the Vay \citep{Vay2008}, or the Higuera–Cary (HC) method \citep{Higuera2017} for the electromagnetic force. For the systems we are interested in, the HC method is preferred as it is volume-preserving in the phase space and accurately captures the $E\times B$ drift even when under-resolving the cyclotron frequency \citep{Higuera2017, Ripperda_2018}. The algorithm used to evaluate equations \eqref{eq22} and \eqref{eq24} is the second-order Heun's (or Runge-Kutta) method. 

This generalization of the leapfrog method allows using very efficient methods, such as the ones employed in flat spacetime particle-in-cell codes, while keeping the time centering of the evolution equations \eqref{eq21}-\eqref{eq24}. In each time step, we only need one interpolation of the electromagnetic fields. We check if the relativistic cyclotronic frequency is resolved locally, for each macro-particle, with at least ten temporal steps. If this is not the case, we subtract the particle's perpendicular momenta and add the $E\times B$ drift similarly to \citet{Bacchini_2020}. We perform this verification step when evaluating the Lorentz force. This algorithm to push particles in time is general and can also be applied to model photon dynamics in a fixed background metric (geodesic motion).

In neutron stars, the general relativistic effects in the particle push may be seen as minor corrections to the particle motion very close to the star. In future works, we intend to detail the differences between simulations using a complete GR description and simulations that use a flat spacetime push with the frame-dragging effect in the field solver \draft{\citep{Philippov_2018,Chen_2020,Philippov_2020,Bransgrove_22}}. This study may reveal if a numerically more expensive GR algorithm is necessary for neutron star magnetosphere simulations. Nevertheless, a complete implementation provides versatility to study plasma dynamics around more compact sources such as black holes.  

\color{\colorDO}
\subsection{Current deposition scheme}
\label{Sec: Charge conservation}
\color{\colorB}
The only missing piece to complete the particle-in-cell algorithm is the evaluation of the charged current density in equation \eqref{eq16}. This current term is the source term in the electric field evolution equation and captures the coupling between the electromagnetic fields and the charged particles. In the absence of sources, the Yee algorithm \citep{Yee} ensures that the time-independent Maxwell's equations are always satisfied if initially satisfied. However, in the presence of sources, the only way to satisfy Gauss's law \eqref{eq5} is through the charge conservation condition:
\begin{equation}
    \frac{\partial \rho}{\partial t} + \vec{\nabla} \cdot \vec{J} = 0.\label{eq28}
\end{equation}
An important challenge is then to design an algorithm capable of satisfying this condition to machine precision in an arbitrary curvilinear grid. Such algorithms are readily available for Cartesian grids for any interpolation order \citep{Esirkepov2001}. \color{\colorDT}The usual way to go around this challenge for curvilinear grids is to employ a divergence-cleaning algorithm; {however, these algorithms are significantly less computationally efficient and, from a numerical point of view, their properties are less well known and explored (in a systematic way) }\color{\colorB}. Here, we present the general-relativistic generalization of the current deposition scheme first described in \citet{Cruz_2023} that conserves charge to machine precision in the non-uniform grid in Fig.~\ref{fig:Yee}\color{\colorDO}, even for very high stellar compactness values. It is important to note that this general-relativistic charge conservation scheme is the first of its kind in the literature\color{\colorB}.
\begin{figure}
    \centering
    \resizebox{\hsize}{!}{\includegraphics{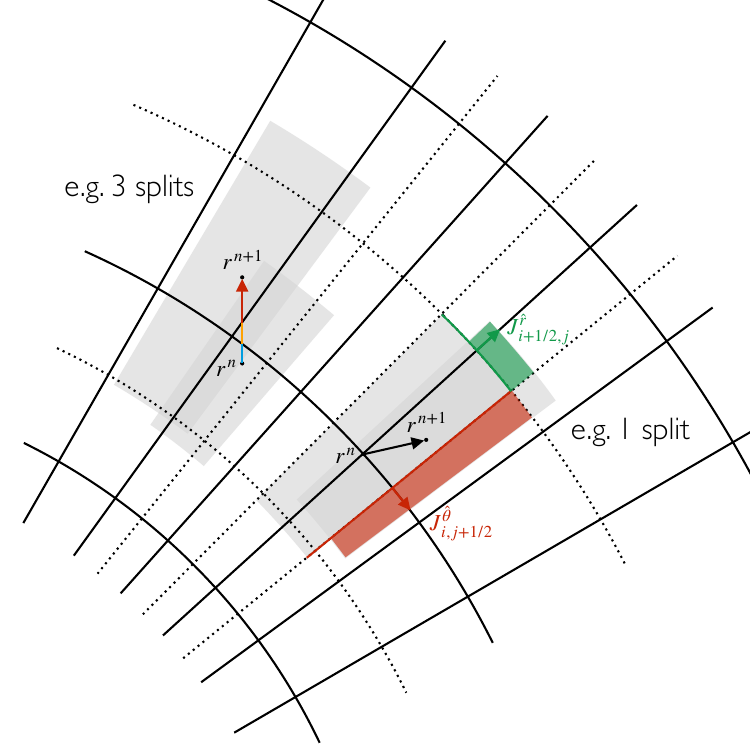}}
    \caption{Schematic of two particle trajectories on a non-uniform grid. The particle to the left has an inter-cell motion of 3 splits, represented in 3 colors. The particle to the right is used as an example of the current deposition scheme for a single split.}
    \label{fig:splits}
\end{figure}

The charge deposition scheme is placed after the particle push and before the field solver. In this way, it has full access to the initial and final positions of each macro-particle, i.e. $\vec{r}^n$ and $\vec {r}^{n+1}$, as well as their half-step velocities, $\vec{u}^{n+1/2}$. The first allows us to deposit the initial and final charge density in the grid nodes, i.e. $\rho_{i,j}^n$ and $\rho_{i,j}^{n+1}$ are known, following~\ref{AP:B}. Reconstructing the charged current density from the half-step velocities and the charge densities to satisfy equation \eqref{eq28} is the challenge we need to overcome. The solution to this problem was inspired in the seminal work by \citet{Villasenor1992} (hereafter VB) that suggested the computation of the charged current density directly from inverting equation \eqref{eq28}, thus enforcing it by construction. As a particle moves from its initial position $\vec{r}^{n}$ to its final position $\vec{r}^{n+1}$, it may cross grid cell boundaries, hence continuously changing its macro-particle shape. To address this, we split the inter-cell motion into several intra-cell trajectories and tackle each split individually, as shown in Fig.\ref{fig:splits}. Without loss of generality, we will describe how to retrieve the charged current density contribution from a single macro-particle given that its motion is bound to a single cell. The algorithm starts by splitting the divergence operator into its radial and meridional components:
\begin{equation}
    \left(\vec{\nabla}\cdot\vec{J}\right) = \left(\vec{\nabla}\cdot\vec{J}\right)^{\hat r} + \left(\vec{\nabla}\cdot\vec{J}\right)^{\hat \theta}.\label{eq29}
\end{equation}
The VB method suggests that each component of \eqref{eq29} can be used to evaluate $J^{\hat r}_{i+1/2,j}$ and $J^{\hat \theta}_{i,j+1/2}$ by computing the temporal derivatives of the charge density when the particle moves purely along the radial (meridional) direction with an average meridional (radial) position. Here, we take a different approach due to the continuous change in the particle shape. We assume that the particle shape changes more significantly along the radial direction; thus, $(\vec{\nabla}\cdot \vec{J})^{\hat \theta}$ cannot be evaluated at a fixed average radius. To overcome this issue, we compute it via
\begin{equation}
    \left(\vec{\nabla}\cdot\vec{J}\right)^{\hat \theta} = \left(\vec{\nabla}\cdot\vec{J}\right) - \left(\vec{\nabla}\cdot\vec{J}\right)^{\hat r},\label{eq30}
\end{equation}
with the right-hand side terms given by
\begin{align}
    \left(\vec{\nabla}\cdot\vec{J}\right)^{n+1/2}_{i,j}&=-\frac{\partial \rho_{i,j}}{\partial t}\Biggr|^{r^{n+1},\theta^{n+1}}_{r^{n},\theta^{n}}=-\frac{\rho_{i,j}\left(r^{n+1},\theta^{n+1}\right)-\rho_{i,j}\left(r^{n},\theta^{n}\right)}{\Delta t},\label{eq31}\\
    \left(\vec{\nabla}\cdot\vec{J}\right)^{\hat r,n+1/2}_{i,j} &=-\frac{\partial \rho_{i,j}}{\partial t}\Biggr|^{r^{n+1},\bar\theta}_{r^{n},\bar\theta}=-\frac{\rho_{i,j}\left(r^{n+1},\bar\theta\right)-\rho_{i,j}\left(r^{n},\bar\theta\right)}{\Delta t}, \label{eq32}
\end{align}
with $\bar\theta=(\theta^{n+1}+\theta^{n})/2$ being the average meridional position of the particle during the split motion. Notice that the poloidal components of the divergence could also be written as
\begin{align}
    \left(\vec{\nabla}\cdot\vec{J}\right)^{\hat r}_{i,j}&=\frac{A^{\hat r}_{i+1/2,j}J^{\hat r}_{i+1/2,j}-A^{\hat r}_{i-1/2,j}J^{\hat r}_{i-1/2,j}}{V_{i,j}}=\frac{A^{\hat r}_{i+1/2,j}J^{\hat r}_{i+1/2,j}}{V_{i,j}},\label{eq33}\\
    \left(\vec{\nabla}\cdot\vec{J}\right)^{\hat \theta}_{i,j}&=\frac{A^{\hat \theta}_{i,j+1/2}J^{\hat \theta}_{i,j+1/2}-A^{\hat \theta}_{i,j-1/2}J^{\hat \theta}_{i,j-1/2}}{V_{i,j}}=\frac{A^{\hat \theta}_{i,j+1/2}J^{\hat \theta}_{i,j+1/2}}{V_{i,j}},\label{eq34}
\end{align}
where $V_{i,j}$ is the volume of the cell evaluated at the cell node, see~\ref{AP:A}. To justify the disappearance of the second terms in equations \eqref{eq33} and \eqref{eq34}, we use the scheme presented in Fig.~\ref{fig:splits}. Note that each split deposits the charged current density in the grid cell its trajectory is bound to. Finally, we obtain the poloidal components of the current by combining equations \eqref{eq32} with \eqref{eq33}, and  \eqref{eq30} with \eqref{eq34}:
\begin{align}
    J_{i+1/2,j}^{\hat r,n+1/2} &=\frac{V_{i,j}}{A_{i+1/2,j}^{\hat r}}\left(\vec{\nabla}\cdot\vec{J}\right)^{\hat r,n+1/2}_{i,j}\\
    J_{i,j+1/2}^{\hat\theta,n+1/2} &=\frac{V_{i,j}}{A_{i,j+1/2}^{\hat \theta}}\left(\vec{\nabla}\cdot\vec{J}\right)^{\hat \theta,n+1/2}_{i,j}.
\end{align}
Due to the axisymmetric condition, equation \eqref{eq28} does not restrict the azimuthal component of the current. Nevertheless, we can construct it via
\begin{equation}
    J_{i,j}^{\hat\phi,n+1/2} = \bar\rho_{i,j}\left(\bar\alpha \frac{u^{\hat\phi,n+1/2}}{\Gamma\left(\vec{u}^{n+1/2}\right)} - \bar{\beta}^{\hat\phi}\right),
\end{equation}
where the bared quantities correspond to a temporal average over the initial and final positions, e.g. 
\begin{equation}
    \bar\rho_{i,j} \equiv \frac{\rho_{i,j}\left(r^n,\theta^n\right)+ \rho_{i,j}\left(r^{n+1},\theta^{n+1}\right)}{2}.
\end{equation}
Also, recall that the total current density is the sum of the contributions of all macro-particles, which, in turn, is a summation over all the splits for each macro-particle.

\color{\colorDO}The current deposit algorithm presented above conserves charge to machine precision independently of the selected coordinates (polar spherical or logical spherical, see \ref{AP:D}). The main difference between both is, in fact, the choice of the cell center position, modifying the effective particle shape. The logical spherical coordinate system possesses a uniform Cartesian-like grid and asymmetric particle shape, thus solving the cell expansion issue characteristic of the symmetric particle shape and non-uniform grid case (i.e., polar spherical case). The latter case requires some particles to deposit charge and current densities in more than two consecutive cells \citep{Cruz_2023}.\color{\colorB}

\subsection{Boundary conditions}
\color{\colorDO}We now discuss the boundary conditions available in the {OSIRIS-GR} particle-in-cell framework. As the OSIRIS code allows for two possible domains (full or half meridional domain), the field and particle boundary conditions change accordingly.\color{\colorB}

\subsubsection{Field boundary conditions}
\label{Sec: Field boundary conditions}
The meridional boundaries are of two kinds: axial or equatorial. The axial boundary condition ensures the axisymmetric condition. Due to the locations of each field component within a spherical Yee grid cell (see Fig.~\ref{fig:Yee}), only the $E^{\hat r}$, $J^{\hat r}$, $B^{\hat \theta}$, $E^{\hat \phi}$ and $J^{\hat \phi}$ components lie in the boundary, hence need to be corrected:
\begin{equation}
    E^{\hat \phi}_{axis} = J^{\hat \phi}_{axis} = B^{\hat \theta}_{axis} = 0.
\end{equation}
All field component values are mirrored and saved in the guard cells for usage in the field evolution equations. For example, $B^{\hat \theta}$ changes direction when crossing the axis, thus changing the sign in the guard cells and being set to zero on the axis, while $B^{\hat r}$ does not invert its direction on the guard cells, hence being copied.

As for the equatorial boundary condition, the physical condition to be satisfied is the magnetospheric up-and-down symmetry. In this boundary, we mirror $B^{\hat r}$, $B^{\hat \phi}$, $E^{\hat \theta}$, and $J^{\hat \theta}$, and copy the rest of the components.

The radial boundaries also take two options: star or open. The star radial boundary condition mimics the surface of a perfect rotating spherical conductor threaded by a dipolar magnetic field. We impose on the interior guard cells and on the surface components (i.e. $E^{\hat \phi}$, $E^{\hat \theta}$ and $B^{\hat r}$) the interior electromagnetic solution of an isolated compact neutron star (equations (144)-(147) in \citet{Torres2023a}). We compute the guard cell values of the current components such that their values are progressively filtered (i.e. with an increasing filter order starting from 0 for the first radial cell to the desired order $n$ for the $(n+1)^{th}$ radial cell \draft{- see~\ref{AP: progressive} for more details}).

The open radial boundary, placed at the outer frontier of the domain, is a Mur-like absorbing boundary condition \citep{Mur_1981} designed to absorb outwardly propagating electromagnetic waves. This condition for an arbitrary field component $\Phi$ in flat spacetime reads:
\begin{equation}
    \frac{\partial \Phi}{\partial_t} + \frac{\partial \Phi}{\partial_r} + \frac{\Phi}{r}\Biggr|_{r=r_{max}} = 0,\label{eq39}
\end{equation}
which is the Sommerfeld radiation condition in spherical coordinates \citep{Novak2004,Espinoza2014}. In~\ref{AP:C}, we present the extension of equation~\eqref{eq39} for curved spacetime and use it to correct the fields that lie in the exterior boundary (i.e. $E^{\hat \phi}$, $E^{\hat \theta}$, and $B^{\hat r}$). This open boundary condition was already used in \citet{Torres2023a} and demonstrated efficient absorption of transient electromagnetic waves launched from a compact neutron star in vacuum. We take the same procedure for the charged current density as in the inner radial boundary.

\subsubsection{Particle boundary conditions}
Particle boundary conditions can be of two types: open or reflecting. In the radial direction, we employ open (or absorbing) boundary conditions on both domain frontiers. Particles that cross these boundaries are subtracted from the simulation. In the meridional direction, we use reflecting (or specular) boundary conditions. Particles are reflected back to the simulation domain with no energy loss.

\subsection{Particle injection mechanism}
\label{Sec: Particle injection mechanism}
OSIRIS allows for several plasma injection mechanisms. Here we limit the discussion to the ones used in the simulations presented in upcoming sections. We distinguish two types of simulations, with or without heuristic pair production \citep{Cruz_2022}. For the latter, we inject pair plasma from the stellar surface at a constant rate with number density $n_{\mathrm{inj}}$ every timestep. Electrons and positrons are initialized in corotation and with a parallel velocity of $0.5~[\mathrm{c}]$. This injection procedure is limited by the local value of the magnetization parameter $\sigma_*$, given by:
\begin{equation}
    \sigma_* = \frac{B_*^2}{4\pi\Gamma \left(n_+ + n_-\right)m_ec^2},
\end{equation}
where $B_*$ is the local amplitude of the magnetic field, $\Gamma$ is the Lorentz factor of the injected particles, and $n_\pm$ is the number density of positrons or electrons. Setting a minimum value for $\sigma_*$ yields a maximum value of the injected plasma density. With the parallel component of the velocity, we can control the injected current provided to the neutron star's magnetosphere. Then, the magnetosphere extracts the sign of the charges required to drive the magnetospheric currents and sustain the global torsion of the magnetic field lines. The goal is to mimic the plasma supply provided by the vacuum work function and the pair-producing cascade close to the stellar surface, i.e. the polar cap gap, while still populating both the closed field line region and the return current. Previous works using global particle-in-cell simulations have already taken this approach as it is a computationally efficient way to populate the magnetosphere without requiring expensive quantum electrodynamics (QED) processes \citep{Cerutti_2015,Philippov_2015b}. Nevertheless, these processes are already implemented in the OSIRIS framework, to model QED cascades in future ultra-intense laser experiments \citep{Grismayer_2016,Grismayer_2017} or in local neutron star polar caps \citep{Cruz_2021}.

The second approach is to sustain a cold corotating atmosphere close to the surface\draft{, similarly to \citet{Hu_2022,Bransgrove_22}}. We characterize the atmosphere by its value at the surface, $n_{0}$, and the scale height of the exponentially decaying profile, $\sigma_{atm}$:
\begin{equation}
    n_{atm}\left(r\right)=n_0e^{-{\left(r-R_*\right)}/{\sigma_{atm}}},\;\; r\leq R_*+k_{atm}\sigma_{atm}.
\end{equation}
We ensure the non-depletion of this atmosphere up to $k_{atm}$ standard deviations from its base by injecting $n_{\mathrm{inj}}$ plasma number density every timestep when required. \draft{In the current study, we consider a plasma composed of electrons and positrons, the injection of ions is deferred to future studies.} This approach differs from the previous one as it allows the magnetosphere to regulate the flux of plasma extracted from the corotating atmosphere. This extraction mechanism would then seed the cascading QED process. To reduce the computational cost of each simulation, we model the cascade with a heuristic pair production \draft{\citep{Philippov_2015a,Philippov_2015b,Philippov_2018,Chen_2020,Guepin_2020,Cruz_2022,Cruz_2023}}. This model for the operation of the vacuum gap defines a threshold energy $\gamma_{thr}$ for a particle to pair produce. Particles that satisfy this condition can pair produce if their radial position corresponds to the range $R_*\leq r\leq R_{\mathrm{PP}}$. Also, this threshold energy has a polar dependence to inhibit pair production close to the polar axis, where the magnetic conversion does not occur due to the infinite magnetic field line curvature. The secondary pair is emitted at the same place and with the same momentum direction as the parent with a Lorentz factor of $\gamma_{pair}/2$. These parameters are user-defined and can be adjusted to study different magnetospheric solutions. This model has shown to be a good model for the polar cap and the outer gaps \draft{\citep{Philippov_2015a,Philippov_2015b,Philippov_2018,Chen_2020,Guepin_2020,Cruz_2022,Cruz_2023}}.

\subsection{Simulation initialization and transient phase}
Simulations start in vacuum with a prescribed general-relativistic dipolar magnetic field \citep{Rezzolla2001a, Torres2023a} threading the neutron star surface. Equivalently, the angular velocity of the star can be either in full rotation or gradually spun to its desired value. For the first scenario, the general-relativistic electric field \citep{Torres2023a} must be prescribed at $t=0$. In the second scenario, the rotation imparted by the stellar boundary condition (see Sect.~\ref{Sec: Field boundary conditions}) generates the self-consistent electric field by launching Alfvénic torsion waves \citep{Cerutti_2015}. At $t=0$, we initiate the plasma injection mechanism described in Sect.~\ref{Sec: Particle injection mechanism}. Due to the high magnetization, the extracted charges from the surface/atmosphere are accelerated along field lines, generating a poloidal current that opens up the field lines that cross the light-cylinder radius, located at $R_{LC}\equiv c/\Omega_*$. Beyond this distance, particles can no longer co-rotate with the field lines anchored to the stellar surface as they would need to co-rotate at speeds faster than the speed of light. These particles then sustain the torsion of the field lines that acquire a toroidal component. In the aligned rotator case, when the magnetic and rotation axes are aligned, the extracted current at the poles is negative due to the predominant outflow of electrons. The magnetosphere self-organizes to close the electric current, which generates the return current. This positive electric current neutralizes the charge density of the neutron star and flows between the closed and open field lines in a Y-shaped structure. Beyond the Y-point, the current sheet sustains the toroidal and poloidal magnetic field reversal at the equator. The magnetospheric solution starts to form in the back of the outwardly propagating Alfvénic torsion wave. We have assumed that after two rotation periods of the star, the magnetospheric solution has converged to the quasi-steady state.

\section{General relativity and the polar cap emission}
\label{Sec: Results sec3}
Coherent radio emission from the polar caps of a neutron star enabled the discovery of pulsars. The pulsed emission profile originates from the cosmic lighthouse effect, which relies on the misalignment angle $\chi$ between the magnetic and rotation axis. Recent works have demonstrated the generation of pulsar radio emission from non-stationary pair plasma discharges \citep{Philippov_2020, Cruz_2021}. Rotation-induced vacuum gaps, regions of an unscreened parallel electric field, potentiate the acceleration of particles to very high energies and subsequent emission of gamma-ray photons via curvature radiation. These hard photons are absorbed in super-strong magnetic fields and create pairs, generating a QED cascade that populates the magnetosphere with a dense electron-positron plasma. The pair production bursts are generated at an angle to the local magnetic field, favouring the induction of electromagnetic modes \citep{Cruz_2021}. However, these models rely on efficient particle acceleration along the polar cap field lines for QED processes to kick in. \citet{Beloborodov_2008} and \citet{Timokhin_2012} addressed the particle accelerator problem at the polar cap. These works identified the $\alpha\equiv j_m/j_{\mathrm{GJ}}$  parameter determinant in defining under which conditions particles accelerate up to relativistic energies and can produce secondary pairs. The motivation behind this parameter, which is the ratio between the electric current extracted from the polar cap and the local co-rotation current $j_{\mathrm{GJ}}=-\Omega_*\cdot B/2\pi$, is that when $0<\alpha<1$, the accelerator is inefficient because the extracted current is enough to sustain the twist of the magnetic field lines near the light cylinder, given by
\begin{equation}
    j_m = \frac{1}{4\pi}\left(\vec\nabla\times\vec{B}_{\mathrm{magnetosphere}}\right)_\parallel.
\end{equation}
When $\alpha>1$ or $\alpha<0$, charges extracted from the surface/atmosphere are insufficient to carry the magnetospheric required current. This current-starved state generates an electric field that accelerates particles. This flow becomes unstable to QED cascading, thus supplying the additional carriers. Under such conditions, the polar cap becomes an efficient accelerator, and pulsar radio emission is viable. However, three-dimensional particle-in-cell simulations of low obliquity rotators (i.e., $\chi\leq40^\circ$) showed weak particle acceleration in the bulk of the polar region {\citep{Philippov_2015b}}, confirming estimates using the $\alpha$ parameter \citep{Bai_2010, Timokhin_2012}. Therefore, low obliquity rotators could not generate pulsar radio emission, which is unsupported by observations. 

The aligned rotator (i.e., $\chi=0^\circ$) constitutes the least favourable configuration for the pulsar mechanism occurrence. We can use this configuration as a robustness diagnostic for any other model that could explain the radio emission. Here we discuss the general-relativistic effects and adopt the aligned rotator configuration.

In the 3+1 formalism, we reformulate the $\alpha$ parameter analysis for the efficiency of the polar cap accelerator as the 4-norm of the current $j_\mu j^\mu$. Equivalently, we have an efficient accelerator for $j_\mu j^\mu>0$ (i.e., spacelike current) or inefficient accelerator for $j_\mu j^\mu<0$ (i.e., timelike current) \citep{Belyaev_2016, Gralla_2016,Huang_2018}. In this way, we can use the spacelike current as an observable for field lines that will sustain pair creation and are viable locations for the pulsar radio emission. \color{\colorDO}Therefore, characterizing the spacelike current region may lead to the direct characterization of the emitted polar radio beam, bridging the gap between numerical models and observations.\color{\colorB}

The wind region of the magnetosphere (the volume of open field lines that extend beyond the light cylinder radius) is somewhat similar for both the dipolar and split-monopolar magnetic field configurations near the rotation axis. The main difference is the existence of a closed field line region at the equator which reduces the number of open field lines crossing the light cylinder. As we are interested in the polar cap of the neutron star magnetosphere, we will neglect this effect and assume, hereafter, that the field configuration resembles that of a split monopole. Taking small angles from the axis, we can approximate the electric current as solely pointing in the radial direction. Hence, the 4-norm of the electric current is given by
\begin{align}
    j_{\mu}j^{\mu} &=  \vec j \cdot \vec j - \left(\rho c\right)^2 \approx  {j^{\hat r}}^2 - \left(\rho c\right)^2 \approx \left(\rho_{\mathrm{GJ}} c\right)^2\left(\left(\frac{B^{\hat \phi}}{E^{\hat \theta}}\right)^2-1\right), \label{eq43}
\end{align}
where we assumed the magnetospheric solution is close to the force-free regime where the charge density closely follows the co-rotation value $\rho\sim\rho_{\mathrm{GJ}}$ and the radial current approximates to the ratio between the azimuthal magnetic field and the polar component of the electric field, as obtained by \citet{Lyutikov_2011}. Although simple, equation \eqref{eq43} is powerful enough to provide us with intuition on the interplay between magnetospheric plasma injection and general-relativistic effects. The first is made explicit through $B^{\hat \phi}$ that measures the torsion of the field lines, which is more efficient in the force-free regime where a strong poloidal current gets extracted from the polar cap. The second point relates to the frame-dragging effect {\citep{Beskin_1990,Muslimov_1992,Philippov_2015a,Torres2023a}}, which reduces the induced electric field due to the mismatch between the stellar and the spacetime rotation. In the force-free regime, the wind region possesses $B^{\hat \phi}\sim E^{\hat \theta}$, corresponding to the null electric current case. This case is not particularly interesting as this would mean that, in flat spacetime, the aligned rotator would not be able to emit in the radio band. However, the frame-dragging effect reduces $E^{\hat \theta}$ raising the ratio above unity and leading to the possibility of pulsar radio emission {\citep{Philippov_2015a}}.

\begin{figure}
  \centering
  \resizebox{\hsize}{!}{\includegraphics{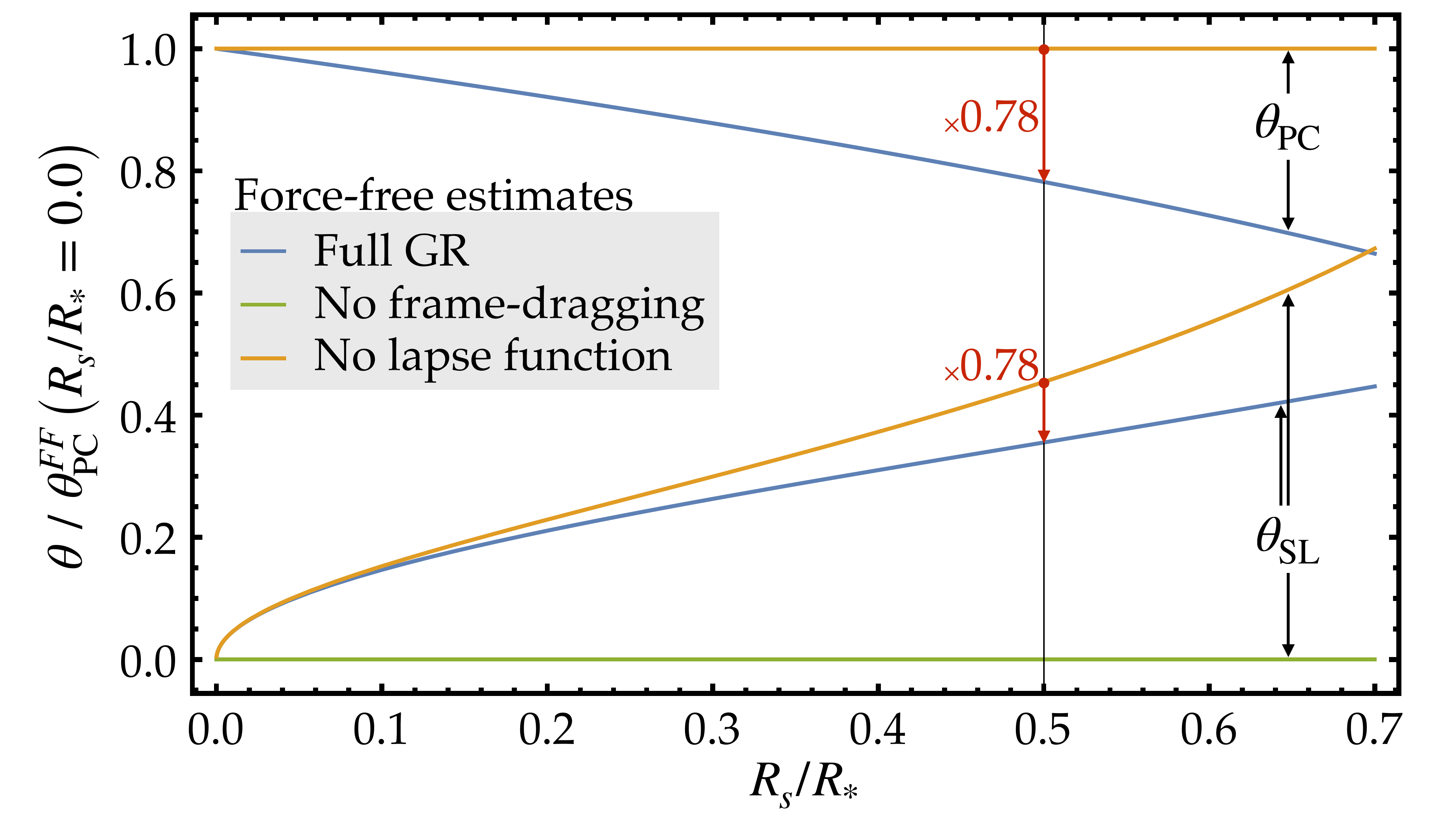}}
  \caption{\color{\colorDTh}Force-free estimates of the maximum polar cap, $\theta_{\mathrm{PC}}$, and spacelike, $\theta_{\mathrm{SL}}$, poloidal extension at the neutron star surface for varying stellar compactness values, $R_s/R_*$. Curves are obtained using the model developed in \citet{Gralla_2016} for a general-relativistic description of force-free electrodynamics. The complete description is presented in blue. Estimates neglecting the frame-dragging contribution and the lapse function are presented in green and orange, respectively. We conclude that not considering the lapse function leads to an overestimate of the open flux tube and effective particle acceleration volume, thus a macroscopic inaccuracy.}
  \label{Figure a}
\end{figure}

\begin{figure*}
  \centering
  \includegraphics[width=17cm]{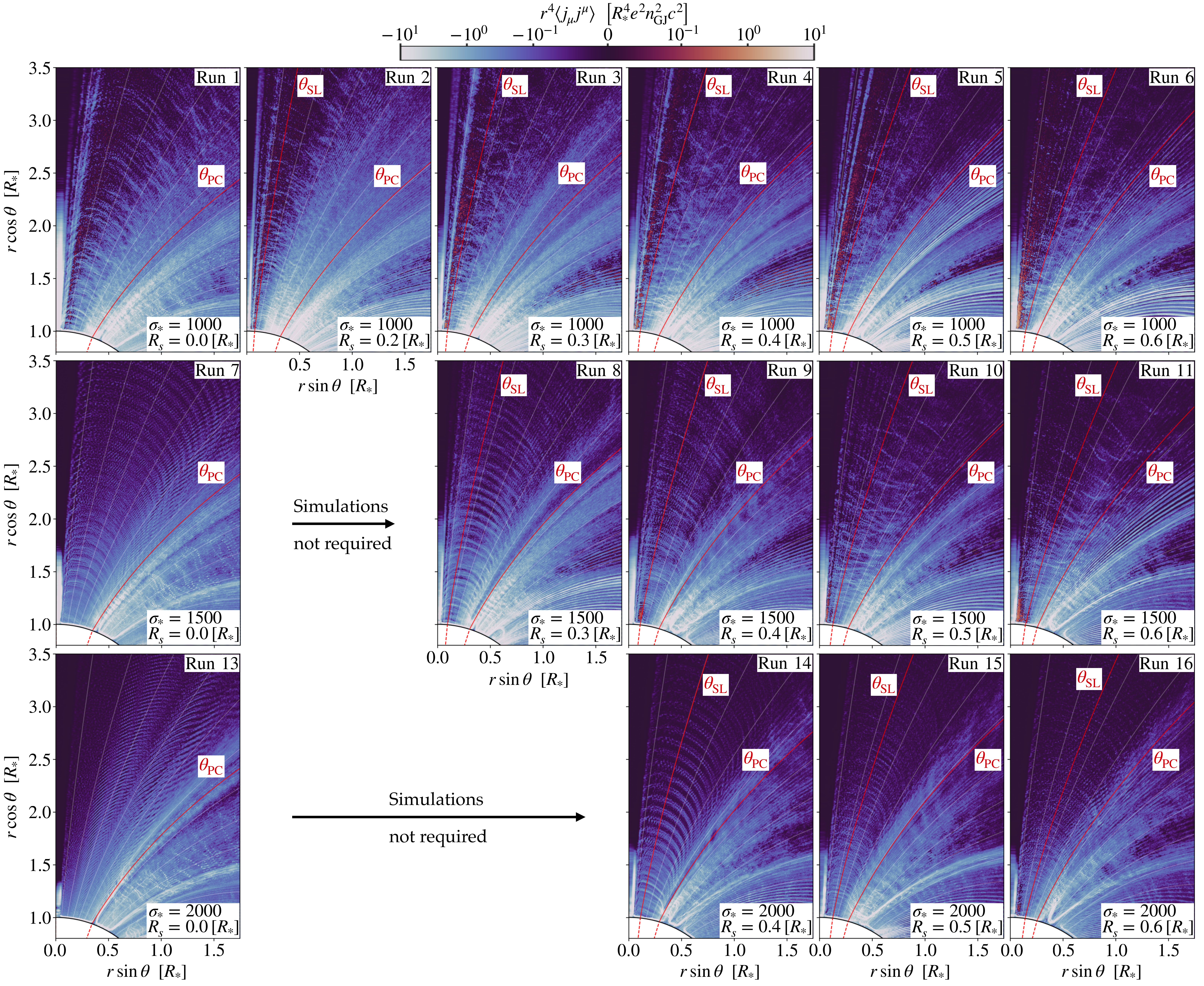}
  \caption{Temporally averaged 4-norm of the electric current for a combination of different stellar compactness and cold magnetization values. This diagnostic allows for the identification of field lines where efficient particle acceleration is expected. The red lines represent the magnetic field lines that start at the maximum predicted angle for the spacelike region ($\theta_{\mathrm{SL}}$) and the polar cap ($\theta_{\mathrm{PC}}$) (as predicted by \citet{Gralla_2016}). The grey lines represent the magnetic field lines. The results show efficient particle acceleration at any compactness for $\sigma_*=1000$, i.e. high plasma supply case. Also, at lower plasma supply, particle acceleration occurs only after a certain value of compactness, e.g. $R_s/R_*\gtrsim0.4$ for $\sigma_*=1500$.}
  \label{Figure 3}
\end{figure*}

To validate this intuitive picture drawn from equation \eqref{eq43}, we simulated the neutron star magnetosphere with no pair production and injected plasma from the surface with parallel velocity as described in Sect.~\ref{Sec: Particle injection mechanism}. This injection method has two advantages: (1) direct control over the extracted poloidal current limited by the cold magnetization parameter; (2) successfully reproduces the force-free magnetospheric solution. Simulations ran in the half-domain setup for $R_s/R_*\in\left[0.0,0.6\right]$ and $\sigma_*\in\left[1000,2000\right]$, presented in Fig.~\ref{Figure 3}. As the maximum meridional extension predicted \color{\colorDO}by force-free models \color{\colorB}for the spacelike current region \citep{Belyaev_2016, Gralla_2016}, $\theta_{\mathrm{SL}}$, is a function of the stellar compactness, the meridional resolution changed accordingly for each simulation to resolve this region with at least 20 numerical cells. This angle also changes with the stellar angular velocity, fixed at $\Omega_*=0.1~[\mathrm{c~rad/R_*}]$ for this set of simulations. Table~\ref{Table 1} contains the detailed list of parameters used. \color{\colorDO}It is important to reiterate that \color{\colorDT}our \color{\colorDO}study takes into account all general-relativistic effects, not just the frame-dragging\color{\colorDT}, unlike previous works on neutron stars {\color{\colorDTh}(e.g., \citet{Philippov_2018, Chen_2020, Bransgrove_22}). In fact, as is shown in Fig.~\ref{Figure a}, this approximation leads to an overestimation of both the polar cap and spacelike angles of $\sim21.8\%$ for $R_s/R_*=0.5$ (typical stellar compactness selected), following the general-relativistic force-free prediction given in \citet{Gralla_2016}}. This overestimation leads to macroscopic magnetospheric changes, including an increased effective open flux tube area, spin-down luminosity, and wider generated radio beam. This point is particularly important when simulating magnetospheric environments of compact neutron stars approaching the Tolman–Oppenheimer–Volkoff mass limit \citep{Kalogera_1996, Romani_2022} {\color{\colorDTh}or long-lived hypermassive neutron stars \citep{Falcke_2014, Chirenti_2019, Chirenti_2023, Selvi_2024}.}\color{\colorB}

\begin{figure}
  \centering
  \resizebox{\hsize}{!}{\includegraphics{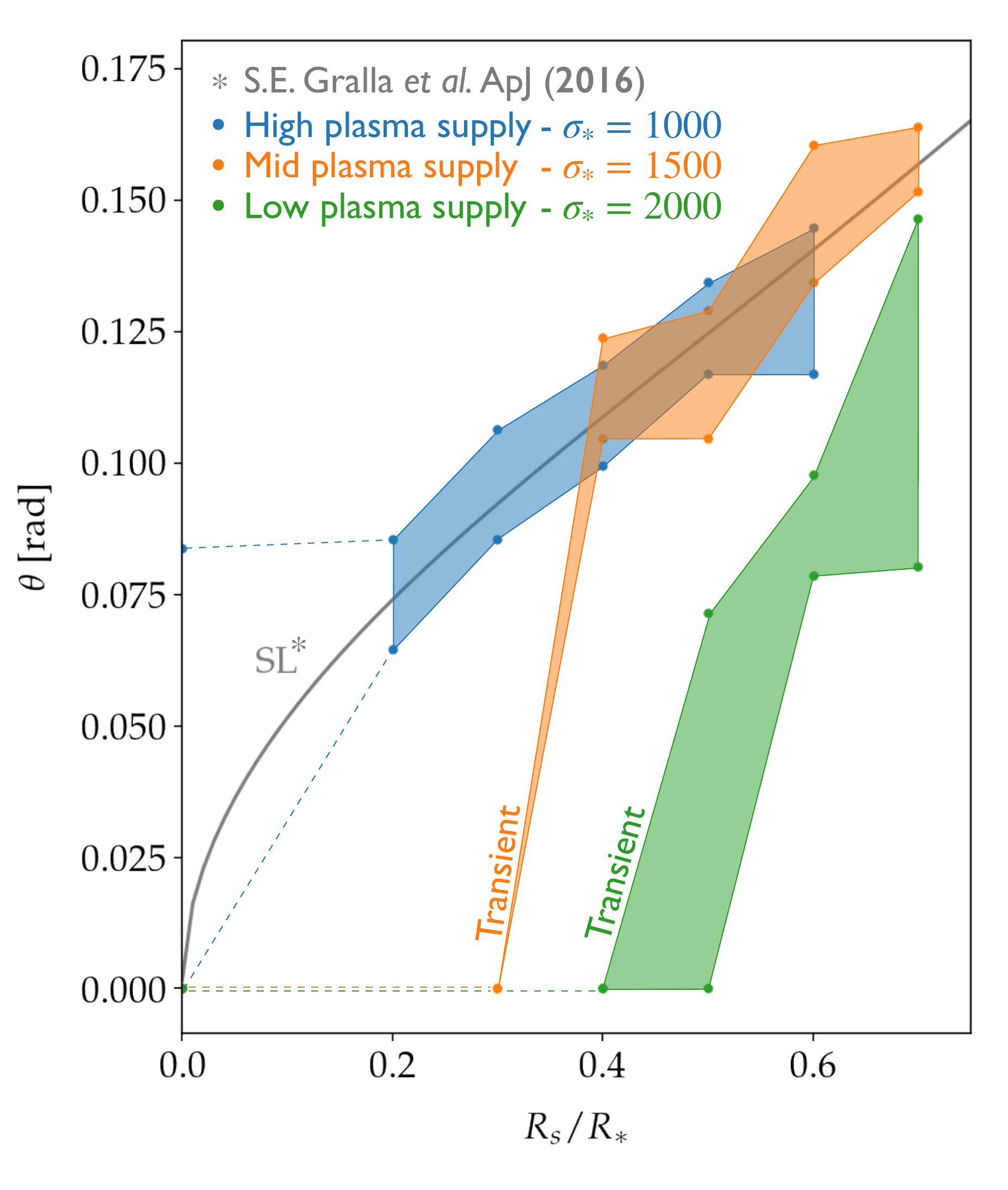}}
  \caption{Maximum poloidal extension of the spacelike region ($\theta_{\mathrm{SL}}$) as function of the stellar compactness $R_s/R_*$. The grey line corresponds to the force-free estimate in \citet{Gralla_2016}. The circles correspond to the maximum and minimum measured angles extracted from numerical simulations presented in Fig.~\ref{Figure 3}. This figure captures the transition from inactive to active polar caps with increasing compactness, e.g. for the $\sigma_*=1500~(2000)$ case, the polar cap activates for $R_s/R_*\gtrsim0.4~(0.5)$.}
  \label{Figure 4}
\end{figure}

The first striking feature in Fig.~\ref{Figure 3} is the inexistence of the spacelike region for some non-zero compactness simulations. The explanation lies in the poloidal current and, consequently, the degree of field line torsion. When $\sigma_*=1000$, the extracted poloidal current is very close to the expected for the force-free split-monopole case, thus ensuring $B^{\hat \phi}\sim E^{\hat \theta}$. In this way, any general-relativistic reduction of the electric field results in a spacelike current, following equation \eqref{eq43}. Lower plasma supply cases represented by higher magnetizations yield a decrease in the poloidal current extracted from the polar cap. The $B^{\hat \phi}$ reduction no longer guarantees the existence of the spacelike region at any compactness, thus explaining the appearance of cut-off conditions as shown in Fig.~\ref{Figure 4}. \color{\colorDO}Figure~\ref{Figure 4} highlights the first direct comparison between global particle-in-cell simulations and force-free model predictions for the meridional extension of the spacelike current volume. Results show a remarkable agreement in the force-free limit but start to deviate for more charge-starved magnetospheric solutions. Nevertheless, the force-free estimates limit the predicted spacelike extension from above, thus it is still a good estimate for the {\color{\colorDTh}maximum} generated radio beam width.\color{\colorB}

From the observational point of view, this shows the robustness of the pulsar radio emission mechanism at any inclination. We have just shown that the spacelike region appears at any compactness for force-free magnetospheric solutions, even for the least favourable inclination configuration (the aligned rotator). Nevertheless, the dependence of the radio beam generation mechanism on the overall magnetospheric solution could explain why some neutron stars are radio-quiet or may present intermittent behaviour. Weak pulsars or pulsars near the death line, where pair production is no longer very efficient, could be strong candidates for this scenario. \color{\colorDO}Also, our model predicts narrower radio beam widths for older \draft{low-obliquity} pulsars, whose magnetospheres are more charge-starved. In some cases, this width reduction can reach half the force-free value, which may be measurable through observations. {\color{\colorDTh}In particular, observations of millisecond pulsars report a systematic tendency for smaller inferred luminosities and narrower emission beams \citep{Kramer_1998}, which support our findings. Thus, the compactness of millisecond pulsar magnetospheres and deviations from force-free solutions may explain the reduction of the angular radius of the generated radio beams.}\color{\colorB}

\begin{table*}[]
\centering
\begin{tabular}{c|ccccccccc}
    \hline
    Run & $\mu_*~\left[\mathrm{m_ec^2R_*^2/e}\right]$ & $\Omega_*~\left[\mathrm{c~rad/R_*}\right]$ & $R_s~\left[\mathrm{R_*}\right]$ & $\sigma_*$ & Injection & $n_{\mathrm{inj}}$ & $v_\parallel~\left[\mathrm{c}\right]$ & Resolution & Spacelike region \\\hline
         1   & 707 & 0.1 & 0.0 & 1000 & Surface & 400 & 0.5 & $2048\times3072$ & \cmark             \\
         2   & 707 & 0.1 & 0.2 & 1000 & Surface & 400 & 0.5 & $2048\times7168$ & \cmark               \\
         3   & 707 & 0.1 & 0.3 & 1000 & Surface & 400 & 0.5 & $2048\times5120$ & \cmark              \\
         4   & 707 & 0.1 & 0.4 & 1000 & Surface & 400 & 0.5 & $3072\times4096$ & \cmark              \\
         5   & 707 & 0.1 & 0.5 & 1000 & Surface & 400 & 0.5 & $2048\times3072$ & \cmark              \\
         6   & 707 & 0.1 & 0.6 & 1000 & Surface & 400 & 0.5 & $2048\times3072$ & \cmark              \\
         7   & 707 & 0.1 & 0.0 & 1500 & Surface & 400 & 0.5 & $2048\times3072$ & \xmark               \\
         8   & 707 & 0.1 & 0.3 & 1500 & Surface & 400 & 0.5 & $2048\times5120$ & \xmark              \\
         9   & 707 & 0.1 & 0.4 & 1500 & Surface & 400 & 0.5 & $2048\times3072$ & \cmark              \\
         10   & 707 & 0.1 & 0.5 & 1500 & Surface & 400 & 0.5 & $2048\times3072$ & \cmark              \\
         11  & 707 & 0.1 & 0.6 & 1500 & Surface & 400 & 0.5 & $2048\times3072$ & \cmark              \\
         12  & 707 & 0.1 & 0.7 & 1500 & Surface & 400 & 0.5 & $2048\times3072$ & \cmark              \\
         13  & 707 & 0.1 & 0.0 & 2000 & Surface & 400 & 0.5 & $2048\times3072$ & \xmark               \\
         14  & 707 & 0.1 & 0.4 & 2000 & Surface & 400 & 0.5 & $2048\times3072$ & \xmark              \\
         15  & 707 & 0.1 & 0.5 & 2000 & Surface & 400 & 0.5 & $2048\times3072$ & \xmark              \\
         16  & 707 & 0.1 & 0.6 & 2000 & Surface & 400 & 0.5 & $2048\times3072$ & \cmark              \\
         17  & 707 & 0.1 & 0.7 & 2000 & Surface & 400 & 0.5 & $2048\times3072$ & \cmark              \\
         18  & 707 & 0.1 & 0.7 & 4000 & Surface & 400 & 0.5 & $2048\times3072$ & \xmark              \\\hline
\end{tabular}
\caption{Simulation parameters used for the first numerical experiment (Sect.~\ref{Sec: Results sec3}), shown in Figs.~\ref{Figure 3} and~\ref{Figure 4}. \draft{These simulations take on average 200k CPU core hours.}}
\label{Table 1}
\end{table*}

\section{Transition from weak to energetic pulsars}
\label{Sec: Results sec4}
In the previous section, we validated the intuitive picture given by equation \eqref{eq43}. However, we adopted a controlled injection methodology that drove the system to the desired solution. In this section, we improve the magnetospheric model by capturing the gap dynamics through heuristic ``on-the-spot" pair production \citep{Cruz_2022}, i.e. in the limit of vanishing photon mean free path. Under these circumstances, solutions become sensitive to the upstream gap plasma supply from the neutron star surface. The charges lifted from the surface form an atmosphere maintained as described in Sect.~\ref{Sec: Particle injection mechanism}, feeding self-consistently the magnetosphere with cold plasma. We follow the PSG non-stationary gap model by sustaining a minimum atmospheric plasma density at the base $n_0\equiv n_+ + n_- \sim 2n_{\mathrm{GJ}}$. A weak parallel component of the electric field persists at the base of some magnetic field lines, providing an initial kick to those particles extracted from the atmosphere. 

\begin{figure*}
  \centering
  \includegraphics[width=17cm]{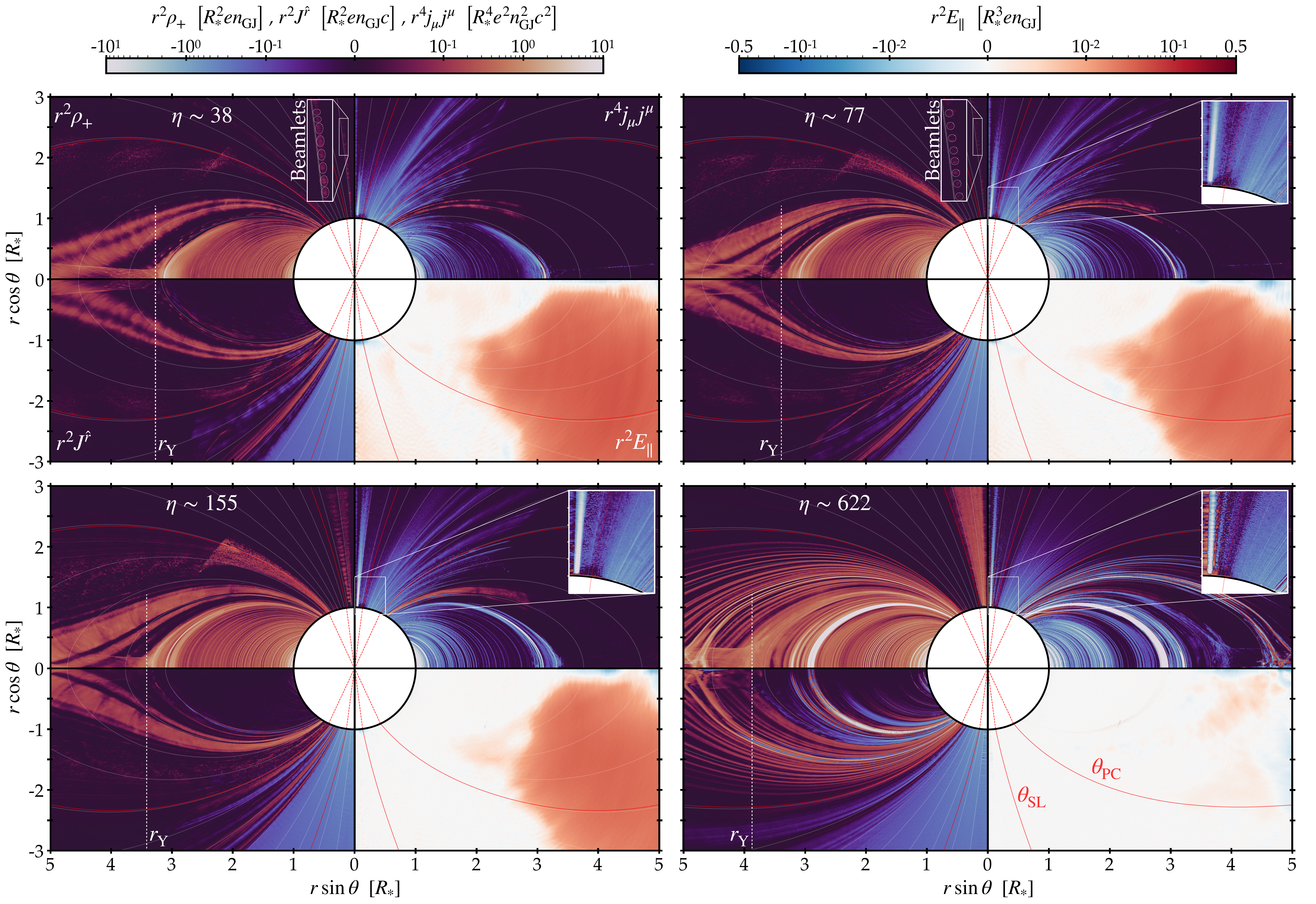}
  \caption{Panels showing the positronic charge density ($\rho_+$), 4-norm of the current ($j_\mu j^\mu$), FIDO-measured radial current ($J^{\hat r}$), and parallel component of the electric field ($E_\parallel$) for increasing pair production efficiency ($\eta$). Field components are multiplied by the radial distance squared ($r^2$) to highlight magnetospheric features further away from the star. Magnetic field lines are in grey and the magnetic field lines that start at the maximum spacelike angle ($\theta_{\mathrm{SL}}$) and polar cap extension angle ($\theta_{\mathrm{PC}}$) are shown in red. This figure captures the transition from a weak to force-free solution, obtained by increasing the stellar magnetic moment and, hence, the pair production efficiency. Consequently, the Y-point \draft{(at $r/R_*\sim r_{\mathrm{Y}}$)} approaches the $R_{\mathrm{LC}}$, the polar cap extends to $\theta_{\mathrm{PC}}$, the parallel electric field component almost vanishes everywhere, and the spacelike current and positrons from pair production appear in the polar cap \draft{(see $j_\mu j^\mu$ zoom insets). The location of the low-multiplicity beamlets is highlighted in the positronic charge density insets close to the poloidal axes.}}
  \label{Figure 5}
\end{figure*}

Figure~\ref{Figure 5} shows the quasi-stationary state of a millisecond neutron star with compactness $R_s/R_*=0.3$ \color{\colorDT}(e.g., \citet{Bhattacharyya_2017}) \color{\colorB}and angular velocity $\Omega_*=0.2~\left[\mathrm{c~rad/R_*}\right]$ for different pair production efficiencies $\eta$ given by:
\begin{equation}
    \eta \equiv \frac{\gamma_{\mathrm{max}}}{\gamma_{\mathrm{thr}}} \in \{ 38,77,155,622\},
\end{equation}
where $\gamma_{\mathrm{max}}\sim B_*\Omega_*^2$ is the estimated maximum achievable particle Lorentz factor. The magnetic field amplitude controls the pair production efficiency by keeping constant the threshold and secondary pair energy at $\gamma_{\mathrm{thr}}=25~[m_ec^2]$ and $\gamma_{\mathrm{pair}}=8~[m_ec^2]$. Also, we restrict pair production to be active only up to three stellar radii (i.e., $R_{\mathrm{PP}}=3R_*$). Table~\ref{Table 2} summarizes the simulation parameters used in this numerical experiment. The low pair production efficiency case, with $\eta\sim38$, highlights the typical solution of a weak pulsar with a persistent outer gap \citep{Shibata_2012, Gruzinov_2012, Gruzinov_2013, Gruzinov_2015, Chen_2020}. This gap is a consequence of the inability of the surface and the pair production in the volumetric return current to supply plasma to the region behind the null surface, i.e. the region where the poloidal magnetic field lines curve towards the equator. As the efficiency increases, the amplitude of the gap's parallel electric field reduces until it becomes almost zero, when the magnetosphere transitions to a force-free solution. The volumetric return current that occupied a portion of the polar cap for low-efficiency values now occupies lower latitudes, increasing the effective polar cap area up to the maximum poloidal extension angle $\theta_{\mathrm{PC}}$\color{\colorDO}, as predicted by force-free models \citep{Gralla_2016,Belyaev_2016}\color{\colorB}. Consequently, the magnetosphere can increase the electric current driven through the open flux tube, leading to a jump in the Poynting flux luminosity. In this model, and for simplicity, we kept the pair production efficiency constant in the magnetosphere. However, different QED processes play a role in distinct magnetospheric regions \citep{Chen_2014}, which may lead to differential pair production efficiencies (deferred for future work). Magnetospheres that continuously oscillate between weak and quasi-force-free states, depending on the morphology of the outer gap, can explain the observed phenomenon of pulsar luminosity intermittency (similar to \citet{Li_2012}).

Several crucial changes accompany the transition from weak to force-free solutions: 1) the Y-point moves closer to the light-cylinder radius due to the more efficient screening of the equatorial superrotation region \citep{Shibata_2012, Hu_2022}; 2) the stronger poloidal current driven through the magnetosphere induces a higher torsion of the magnetic field lines in the toroidal direction. The latter is vital for the activation of the polar cap. At fixed compactness (in this case $R_s/R_*=0.3$), Fig.~\ref{Figure 4} shows that the polar cap may present a spacelike volumetric region only if the magnetosphere drives a stronger current, characteristic of higher plasma supply scenarios. With heuristic pair production, the efficiency controls the plasma supplied from the polar cap gap to the magnetosphere, thus having the same effect as the surface plasma supply from the previous section. In Fig.~\ref{Figure 5}, we observe the appearance of the spacelike volume in the polar cap for efficiencies $\eta\gtrsim155$. However, many processes affect this minimum efficiency value $\eta_{\mathrm{min}}$: 1) the frame-dragging effect for higher compactness neutron stars reduces $\eta_{\mathrm{min}}$ due to the reduction of the poloidal electric field; 2) evident interplay between the outer and inner polar cap gap, more efficient pair production at the Y-point (e.g., including photon-photon processes) pulls the return current away from the polar cap thus increasing the field line torsion of the magnetosphere, which reduces $\eta_{\mathrm{min}}$; 3) a denser neutron star atmosphere also facilitates the launch of a stronger poloidal current, in a similar way as in point 2. In particular, point 2 suggests a correlation between the radio beam and the existence of the outer gap. Intermittent pulsars may also present intermittent radio beams, potential candidates for explaining pulse nullings. \draftt{We have also performed simulations where we inject an atmospheric plasma composed of electrons and ions, where the ions have the same mass and charge as positrons but are not allowed to pair produce \citep{Philippov_2018}. As expected, the injection of ions leads to a lower pair plasma supply to the outer magnetosphere of weak pulsars, thus increasing the minimum efficiency value $\eta_{\mathrm{min}}$. Apart from a shift of the magnetospheric solution towards charge-starvation, we do not observe significant changes in the outer gap dynamics or magnetospheric structure. However, this point may change if ions possess realistic $q/m$, which we differ to a future study. Also, the impact of ion injection on the magnetosphere is more significant in charge-starved pulsar solutions due to the high degree of charge separation, thus this effect is diluted for denser solutions approaching the force-free regime.}

Solutions with very low pair production efficiencies are sustained solely by pair plasma generated at the return current region and progressively tend to the ``dead pulsar" solution (electrosphere), corresponding to the no pair production case \citep{Cruz_2023}. Solutions with even higher pair production efficiencies than the ones presented in this study tend to the force-free regime and possess an active polar cap, i.e. in the accelerator regime. In this sense, we retrieve the same conclusions as in the previous section: low pair production efficiencies (i.e., low plasma supply) require higher stellar compactnesses for the activation of the pulsar mechanism; higher pair production efficiencies (i.e., high plasma supply) relax the compactness restriction. Nevertheless, for low-obliquity rotators, compactness is crucial in the generation phase of the radio beam along the magnetic axis. \color{\colorDO}Also, as observed in the previous section, the width of the radio beam deviates from the force-free estimate depending on the charge-starvation state. \color{\colorDT}Thus, f\color{\colorDO}or older and weaker low-\draft{obliquity} pulsars, the radio beam width should be reduced. 

\draft{Another important feature on the structure of the magnetosphere is the location of the Y-point, which for all simulations presented is located at a significant fraction of the light-cylinder radius (i.e. $r_{\mathrm{Y}}/R_{\mathrm{LC}}\sim 0.6-0.8$), consistent with previous kinetic models of neutron star magnetospheres \citep{Chen_2020,Guepin_2020,Cruz_2023,Hakobyan_2023}. The spatial restriction imposed on the heuristic pair production ($R_{\mathrm{PP}}=3R_*$) limits the supply of plasma to the outer magnetosphere, in particular, to regions close to the Y-point. This weak supply of plasma beyond $R_{\mathrm{LC}}$ explains the formation of the persistent outer gap, volumetric return current and wide equatorial current sheet. This is in contrast to the typical ideal force-free magnetosphere with abundant plasma supply, clear and localized Y-shaped return current and thin equatorial current sheet. Wider current sheets dissipate energy at a much slower rate, thus we expect our novel reduced-domain model (i.e., imposing equatorial symmetry) to be accurate in modelling weak pulsar magnetospheres in the aligned rotator configuration. We have also verified the convergence of the presented results by doubling the number of particles per cell injected every timestep.}

\color{\colorDT}Particularly interesting is the systematic formation of core component-associated beamlets at lower pair production efficiencies \draft{(see positronic charge density for $\eta=38-77$ simulations in Fig.~\ref{Figure 5}). These beamlets are generated by the cyclic opening of a fragile gap near the magnetic pole of the star. As pair production is prohibited close to the axis due to the small curvature of the magnetic field lines, some electrons from the magnetosphere are reversed back to the star, thus intermittently shorting out the accelerating gap from above \citep{Lyubarsky_2009}. This cyclic behaviour allows the magnetosphere to drive the required quasi-steady-state polar current that supports the magnetic field line torsion in the outer magnetosphere.} We conjecture that these beamlets are responsible for the narrow dwarf pulses observed during ordinary pulse nullings of PSR B2111+46 in \citet{XChen_2023}, \draft{referred to} as ``particle raindrops" by the authors.\color{\colorB}

\begin{table*}[]
\centering
\begin{tabular}{c|cccccccccc}
    \hline
    Run & $\mu_*~\left[\mathrm{m_ec^2R_*^2/e}\right]$ & $\Omega_*~\left[\mathrm{c~rad/R_*}\right]$ & $R_s~\left[\mathrm{R_*}\right]$ & Injection & $n_{\mathrm{inj}}$ & $n_0$ & $v_\parallel~\left[\mathrm{c}\right]$ & $\sigma_{\mathrm{atm}}$ & $k_{\mathrm{atm}}$ & Resolution \\\hline
         1   & 11312 & 0.2 & 0.3 & Atmosphere & 400 & $2n_{\mathrm{GJ}}$ & 0.0 & $R_*/80$ & 4 & $2048\times3072$           \\
         2   & 22624 & 0.2 & 0.3 & Atmosphere & 800 & $2n_{\mathrm{GJ}}$ & 0.0 & $R_*/80$ & 4 & $2048\times3072$           \\
         3   & 45248 & 0.2 & 0.3 & Atmosphere & 1600 & $2n_{\mathrm{GJ}}$ & 0.0 & $R_*/80$ & 4 & $2048\times3072$          \\
         4   & 180992 & 0.2 & 0.3 & Atmosphere & 6400 & $2n_{\mathrm{GJ}}$ & 0.0 & $R_*/80$ & 4 & $2048\times3072$        \\\hline
\end{tabular}
\caption{Simulation parameters used for the second numerical experiment (Sect.~\ref{Sec: Results sec4}), shown in Fig.~\ref{Figure 5}. \draft{These simulations take on average 500k CPU core hours.}}
\label{Table 2}
\end{table*}

\section{Conclusions}
\label{Sec: Conclusions}
Understanding neutron star magnetosphere dynamics is of utmost importance to reveal the locations of efficient particle acceleration and radiation generation. The connection between the rich pulsar observations and theoretical models is only possible through global ab initio simulations. In particular, due to the extreme nature of such astrophysical environments, kinetic models must capture self-consistent QED processes and general relativistic effects. This paper introduces a new general-relativistic module for the particle-in-cell code OSIRIS \citep{Osiris}, intended to study plasma dynamics in intense electromagnetic and gravitational fields, capturing self-consistently GR effects in all components of the PIC algorithm (field solver, particle pusher, charge and current deposit scheme).

We designed two numerical experiments to gain a deeper insight into the role of general relativity in activating the polar cap, focusing on the conditions for efficient particle acceleration. Therefore, we used the 4-norm of the current to probe under what circumstances the emission of a coherent radio beam along the magnetic axis is viable. In the first experiment, we injected plasma at $0.5~\left[\mathrm{c}\right]$ along the magnetic field lines, controlled by the local magnetization, to reproduce different quasi-stationary magnetospheric states. This setup allowed for a structured parametric scan over the plasma supply and stellar compactness. Without the general-relativistic effects, a low obliquity rotator would not be able to accelerate particles efficiently enough to potentiate the radio beam generation independently of the plasma supply. The general relativistic frame-dragging effect reduces the poloidal electric field, thus increasing the current amplitude. We show that this GR effect activates the polar cap of young and energetic neutron stars at any non-zero compactness, thus providing a robust mechanism for radio beam generation. For older neutron stars, it depends on both the compactness and plasma supply. Nevertheless, we show that GR could enable radio emission even for weak pulsars that drive weaker poloidal currents. \color{\colorDO}Therefore, weak pulsars may be observable in the radio spectrum beyond their expected death line, some of which can present themselves with narrower beams concerning force-free estimates. \color{\colorDT}{We note, however, that measuring such deviations is challenging }due to other competing effects, such as \draft{the observer's line of sight limitation or the misalignment angle $\chi$, requiring a full three-dimensional model extension. Previous kinetic studies of pulsar magnetospheres at arbitrary inclinations and low plasma injection rates demonstrated a similar behaviour of the magnetospheric solution \citep{Kalapotharakos_2018,Brambilla_2018}, with decreased Poynting flux and an outer gap beyond the null surface and around the separatrix region. Consequently, the volumetric field-aligned magnetospheric current and the torsion of the magnetic field lines decrease for these weak pulsar solutions. Following the conclusions from this work, we expect a reduction of the spacelike volume ($j_\mu j^\mu>0$ or $\alpha>1$) in the outflowing region of the polar cap \citep{Bai_2010,Kramer_2022}, where pair production and the generation of the radio beam are expected. However, this effect should be more prominent in low-obliquity or very charge-starved pulsars, where this deviation from the force-free solution produces clear differences in the expected beam width. O}{\color{\colorDTh}ur findings \draft{and conjecture} seem to agree with observations of millisecond pulsars, which do not follow the scaling predicted from a canonical pulsar model for the inferred luminosity and angular radius of the generated radio beam \citep{Kramer_1998}. The compactness of their magnetospheres and deviations from the force-free solution may explain such systematic tendencies.}\color{\colorB}

In the second experiment (Sect.~\ref{Sec: Results sec4}), we improve the accuracy of the previous model by resolving the polar cap gap dynamics using a heuristic ``on-the-spot" pair production model. Here, we allow pair production to populate only regions of efficient particle acceleration. The magnetosphere self-regulates to drive the required currents that sustain the magnetic field line torsion. By varying the pair production efficiency, i.e. through an increase in the magnetic field amplitude, we observe a clear transition between the weak and force-free pulsar solutions at fixed compactness (fiducial value of $R_s/R_*=0.3$). Once again, young and energetic neutron stars support strong poloidal currents, characteristic of a denser force-free magnetospheric solution, and possess active polar caps at any non-zero stellar compactnesses. Older pulsars, i.e. pulsars approaching their death line, transition to weak pulsar solutions and must rely on both the pair production efficiency (plasma supply) and stellar compactness for the presence of efficient particle acceleration and consequent radio beam. In Fig.~\ref{Figure 5}, we demonstrate the existence of a minimum pair production efficiency for which the spacelike current develops in the bulk of the polar cap. The previous numerical experiment also captured this transient behaviour of the spacelike current via varying the plasma supply at fixed compactness (see Fig.~\ref{Figure 4}). Also, simulations highlight the poloidal current-mediated interplay between the polar cap and outer gaps. Magnetospheres with such null surface gaps develop a volumetric return current, constricting the polar cap area and limiting the poloidal current in the open magnetic field line bundle. Consequently, equation~\eqref{eq43} requires a lower poloidal electric field to activate the polar cap discharge, achieved by increasing the stellar compactness. Magnetospheres with pair production near the Y-point populate these exterior regions with plasma, relaxing the polar cap poloidal extension and pair production efficiency at the poles. This interconnection between outer and inner gaps may explain the intermittency of certain observed pulsars or even justify the existence of pulse nullings. \color{\colorDT}In particular, weak pulsar solutions generate low-multiplicity plasma beamlets near the magnetic axis, compatible with the observed narrow-width dwarf pulses in \citet{XChen_2023}.\color{\colorB}

\draft{In this work, we have employed a novel reduced-domain approach to model the global magnetosphere of a neutron star. This reduced model allows for simulations with higher spatiotemporal resolution and better macro-particle statistics, at the expense of introducing an up-down equatorial symmetry. Consequently, by construction, this model does not capture the development of the kink instability in the equatorial current sheet \citep{Cerutti_2015,Philippov_2015a}. In that sense, we do not expect our method to model correctly the dissipation properties of thin current sheets, characterized by frequent magnetic reconnection events and plasmoid formation \citep{Hakobyan_2019,Hakobyan_2023,Schoeffler_2023}. However, we expect good agreement for weak pulsar solutions with no pair production near $R_{\mathrm{LC}}$, which possess a volumetric return-current and wide equatorial current sheet (with no significant dissipation/reconnection events occurring). Also, the location of the Y-point is consistent with similar full-domain particle-in-cell simulations \citep{Chen_2020,Guepin_2020,Cruz_2023,Hakobyan_2023}, indicating that this reduced model preserves the main elements of a neutron star magnetosphere in the aligned rotator configuration.}

Future works with more accurate pair production models, i.e. heuristic models including non-zero photon mean free path and geodesic propagation, or even modelling QED processes more accurately with Monte Carlo techniques \citep{Cruz_2021}, should explore this global outer-to-inner gap dynamics. Also, the variation of the upstream plasma supply from the stellar surface/atmosphere may affect the dynamics of the polar cap gap and its dependency on the driven magnetospheric current, requiring further investigation. \draft{In particular, with the surface/atmospheric injection of electron-ion plasma instead of pair plasma, which} \draftt{affects} \draft{the discharge dynamics on the return-current, modifying the plasma supply to the outer magnetosphere.} Also, three-dimensional models of weak pulsar magnetospheres for low-obliquity rotators may unveil the importance of GR in the pulsar observational appearance.

\section*{Acknowledgements}
This work is partially supported by the European Research Council (ERC-2015-AdG Grant 695088) and FCT (Foundation for Science and Technology, Portugal) under the project X-MASER No. 2022.02230.PTDC. RT is supported by FCT (Portugal) (Grant PD/BD/142971/2018) in the framework of the Advanced Program in Plasma Science and Engineering (APPLAuSE, FCT Grant PD/00505/2018). We acknowledge EuroHPC for granting access to LUMI (Large Unified Modern Infrastructure, Kajaani, Finland) within the EuroHPC-JU project EHPC-REG-2021R0038, where the simulations presented in this work were performed. \draft{The authors thank the anonymous referee for the thoughtful and detailed comments that have significantly improved the manuscript.}

\appendix

\section{Metric dependent quantities}
\label{AP:A}
We show expressions of the line and areal elements that allow the determination of the curl terms in equations \eqref{eq9}-\eqref{eq14} for the logical spherical coordinate system ($\tilde r$,$\tilde \theta$,$\phi$), i.e. ($\tilde r$,$\tilde \theta$,$\phi$)=($\log r$,$-\cos\theta$,$\phi$) with the metric line element given in eq.~\eqref{eqlineele}:
\begin{align}
    &l^{\hat r}_{i+1/2} = \int_{\tilde r_i}^{\tilde r_{i+1}} \sqrt{\gamma_{\tilde r\tilde r}}\mathrm{d}\tilde r'=\left[\alpha e^{\tilde r'}+\frac{R_s}{2}\log{\frac{2e^{\tilde r'}\left(1+\alpha\right)-R_s}{R_s}}\right]_{\tilde r_i}^{\tilde r_{i+1}},\\
    &l^{\hat \theta}_{i,j+1/2} = \int_{\tilde \theta_j}^{\tilde \theta_{j+1}} \sqrt{\gamma_{\tilde \theta\tilde \theta}}\mathrm{d}\tilde \theta'=e^{\tilde r_i}\left[\arcsin\tilde\theta'\right]_{\tilde \theta_j}^{\tilde \theta_{j+1}},
\end{align}
\begin{align}
    &l^{\hat \phi}_{i,j} = \int_{\phi}^{\phi+\Delta \phi} \sqrt{\gamma_{\phi\phi}}\mathrm{d}\phi'=e^{\tilde r_i}\sqrt{1-\tilde\theta_j^2}\Delta\phi,\\
    &A^{\hat r}_{i,j+1/2} = \int_{\tilde \theta_j}^{\tilde \theta_{j+1}}\int_{\phi}^{\phi+\Delta \phi} \sqrt{\gamma_{\tilde \theta\tilde \theta}\gamma_{\phi\phi}}\mathrm{d}\tilde \theta'\mathrm{d}\phi'=e^{2\tilde r_i}\left[\tilde\theta'\right]_{\tilde \theta_j}^{\tilde \theta_{j+1}}\Delta\phi,\\
    &A^{\hat \theta}_{i+1/2,j} = \int_{\tilde r_i}^{\tilde r_{i+1}}\int_{\phi}^{\phi+\Delta \phi} \sqrt{\gamma_{\tilde r\tilde r}\gamma_{\phi\phi}}\mathrm{d}\tilde r'\mathrm{d}\phi'=\nonumber\\
    &\sqrt{1-\tilde\theta_j^2}\Delta\phi\left[\frac{\alpha e^{\tilde r'}}{4}\left(2e^{\tilde r'}+3R_s\right)+\frac{3R_s^2}{8}\log{\frac{2e^{\tilde r'}\left(1+\alpha\right)-R_s}{R_s}}\right]_{\tilde r_i}^{\tilde r_{i+1}},\\
    &A^{\hat \phi}_{i+1/2,j+1/2} = \int_{\tilde r_i}^{\tilde r_{i+1}} \int_{\tilde \theta_j}^{\tilde \theta_{j+1}} \sqrt{\gamma_{\tilde r\tilde r}\gamma_{\tilde \theta\tilde \theta}}\mathrm{d}\tilde r'\mathrm{d}\tilde \theta'=\nonumber\\
    &\left[\frac{\alpha e^{\tilde r'}}{4}\left(2e^{\tilde r'}+3R_s\right)+\frac{3R_s^2}{8}\log{\frac{2e^{\tilde r'}\left(1+\alpha\right)-R_s}{R_s}}\right]_{\tilde r_i}^{\tilde r_{i+1}}\left[\arcsin\tilde\theta'\right]_{\tilde \theta_j}^{\tilde \theta_{j+1}}.
\end{align}

The volume element can also be evaluated via
\begin{align}
    V_{i+1/2,j+1/2}&=\int_{\tilde r_i}^{\tilde r_{i+1}} \int_{\tilde \theta_j}^{\tilde \theta_{j+1}}\int_{\phi}^{\phi+\Delta \phi}\sqrt{\gamma_{\tilde r\tilde r}\gamma_{\tilde \theta\tilde \theta}\gamma_{\phi\phi}}\mathrm{d}\tilde r'\mathrm{d}\tilde \theta'\mathrm{d} \phi'=\nonumber\\
    &=I^{\hat r}_{i+1/2}I^{\hat\theta}_{j+1/2}I^{\hat\phi}=I^{\hat r}_{i+1/2}\left[\tilde\theta'\right]_{\tilde \theta_j}^{\tilde \theta_{j+1}}\Delta\phi,\label{eqA7}
\end{align}
which is used in the charge conservation scheme of Sect.~\ref{Sec: Charge conservation}. The radial part of the volume integral is given by
\begin{align}
    I^{\hat r}_{i+1/2}=&\left[\frac{\alpha e^{\tilde r'}}{24}\left(8e^{2\tilde r'}+10e^{\tilde r'}R_s+15R_s^2\right)+\right.\nonumber\\
    &+\left.\frac{5R_s^3}{16}\log{\frac{2e^{\tilde r'}\left(1+\alpha\right)-R_s}{R_s}}\right]_{\tilde r_i}^{\tilde r_{i+1}}.\label{eqA8}
\end{align}

The same procedure can be repeated for any diagonal spatial metric to retrieve the metric elements of other metric or coordinate systems.

In Sect.~\ref{Sec: Particle Pusher}, the spatial Christoffel symbols used in equation \eqref{eq20} for the logical spherical coordinate system are given by:
\begin{align}
    \Gamma^{\hat{\tilde r}}_{\hat{i}j}&= \begin{bmatrix}
\frac{3}{2}-\frac{1}{2\alpha^2} & 0 & 0 \\
0 & -\frac{\alpha}{\sqrt{1-\tilde\theta^2}} & 0 \\
0 & 0 & -\alpha\sqrt{1-\tilde\theta^2}\end{bmatrix},\\ 
\Gamma^{\hat{\tilde \theta}}_{\hat{i}j}&= \begin{bmatrix}
0 & \frac{\alpha}{\sqrt{1-\tilde\theta^2}} & 0 \\
1 & \frac{\tilde\theta}{{1-\tilde\theta^2}} & 0 \\
0 & 0 & \tilde\theta\end{bmatrix},\\ 
\Gamma^{\hat{\phi}}_{\hat{i}j}&= \begin{bmatrix}
0 & 0 & \alpha\sqrt{1-\tilde\theta^2} \\
0 & 0 & -\tilde\theta \\
1 & -\frac{\tilde\theta}{{1-\tilde\theta^2}} & 0\end{bmatrix}.
\end{align}

\section{Charge deposition scheme}
\label{AP:B}
In the 3+1 formalism, we can retrieve the charge conservation condition via the divergence of the Amp\'{e}re's law: 
\begin{equation}
    \vec{\nabla}\cdot\left[\vec{\nabla}\times\left(\alpha\vec{B}-\vec{\beta}\times\vec{E}\right)\right]=\vec{\nabla}\cdot\left[\frac{\partial\vec{E}}{\partial t} + 4\pi\left(\alpha\vec{j}-\rho\vec{\beta}\right)\right],
\end{equation}
which simplifies to
\begin{equation}
    0= \frac{\partial \rho}{\partial t} + \vec{\nabla} \cdot \left(\alpha\vec{j}-\rho\vec{\beta}\right)= \frac{\partial \rho}{\partial t} + \vec{\nabla} \cdot \vec{J}
\end{equation}
using Gauss's law \eqref{eq5} and the vector calculus property that states that the divergence of a curl of a vector field is always zero.

For the sake of clarification, we start by defining that each macro-particle, located at $(\tilde r_p, \tilde \theta_p )$, has the same shape as the cell it is in, which is centered at $(\tilde r_{i+1/2}, \tilde \theta_{j+1/2} )$. Each macro-particle occupies a volume given by the particle shape function $S(\tilde r, \tilde \theta, \tilde r_p, \tilde \theta_p)$, which we will define later. The number of particles in each macro-particle, $N_p$, is given by
\begin{equation}
    \int_V n\left(\tilde r, \tilde \theta\right)~dV = N_p,\label{eqB3}
\end{equation}
where $n(\tilde r, \tilde \theta)$ is the particle number density, which we will assume to be constant within each cell (i.e. waterbag-like). In this way, we can obtain the particle number density in each grid cell:
\begin{equation}
    n\left(\tilde r_{i+1/2}, \tilde \theta_{j+1/2}\right)= \frac{N_p}{V_{i+1/2,j+1/2}}=\frac{N_p}{I^{\hat r}_{i+1/2}I^{\hat \theta}_{j+1/2}2\pi},\label{eqB4}
\end{equation}
using equations \eqref{eqA7} and \eqref{eqA8}. The extension of this expression to an arbitrary particle position taking the continuous limit reads
\begin{equation}
    n\left(\tilde r_p, \tilde \theta_p\right)=\frac{N_p}{I^{\hat r}_{p}I^{\hat \theta}_{p}2\pi},\label{eqB5}
\end{equation}
with
\begin{align}
    I^{\hat r}_{p} =&\left[\frac{\alpha e^{\tilde r'}}{24}\left(8e^{2\tilde r'}+10e^{\tilde r'}R_s+15R_s^2\right)+\right.\nonumber\\
    &+\left.\frac{5R_s^3}{16}\log{\frac{2e^{\tilde r'}\left(1+\alpha\right)-R_s}{R_s}}\right]_{\tilde r_{min}}^{\tilde r_{max}},\\
    I^{\hat \theta}_{p} =& \left[\tilde\theta'\right]_{\tilde\theta_{min}}^{\tilde\theta_{max}}=\tilde\theta_{max}-\tilde\theta_{min}=\Delta\tilde\theta,
\end{align}
where $\tilde r_{min}=\tilde r_p-\Delta \tilde r/2$, $\tilde r_{max}=\tilde r_p+\Delta \tilde r/2$, $\tilde\theta_{min}=\tilde \theta_p-\Delta \tilde \theta/2$ and $\tilde\theta_{max}=\tilde \theta_p+\Delta \tilde \theta/2$ are the poloidal limits of the macro-particle, and $\Delta \tilde r$ and $\Delta \tilde \theta$ are the spatial dimensions of the corresponding grid cell. The continuous limit ensures that the particle shape is a smooth function of the particle position that satisfies exactly equations \eqref{eqB3} and \eqref{eqB4} when $(\tilde r_p, \tilde \theta_p )=(\tilde r_{i+1/2}, \tilde \theta_{j+1/2} )$. The particle shape function is inferred from equation \eqref{eqB5} as
\begin{equation}
    S\left(\tilde r,\tilde \theta,\tilde r_p,\tilde \theta_p\right) = \left[I^{\hat r}_{p}\left(\tilde r_p\right) I^{\hat \theta}_{p}\left(\tilde \theta_p\right)2\pi\right]^{-1} b_0\left(\frac{\tilde r-\tilde r_p}{\Delta\tilde r}\right) b_0\left(\frac{\tilde \theta-\tilde \theta_p}{\Delta\tilde \theta}\right),\label{eqB8}
\end{equation}
where $b_0$ is the zeroth order b-spline function given by $b_0(x)=1$ if $|x|<0.5$ and $b_0(x)=0$ otherwise. The charge density at any position due to a macro-particle located at $(\tilde r_p,\tilde\theta_p)$ with $N_p$ particles of charge $q_p$ reads
\begin{equation}
    \rho_p\left(\tilde r,\tilde \theta,\tilde r_p,\tilde \theta_p\right)=q_pN_pS\left(\tilde r,\tilde \theta,\tilde r_p,\tilde \theta_p\right).
\end{equation}
Consequently, the contribution of the macro-particle's charge density to the charge density evaluated at the grid nodes is defined through the volume weighting technique as
\begin{align}
    \rho_{i,j}\left(\tilde r_p,\tilde \theta_p\right) &= \frac{\int_{V_{i,j}}\rho_p\left(\tilde r,\tilde \theta,\tilde r_p,\tilde \theta_p\right) dV_{i,j}}{V_{i,j}}=\nonumber\\
    &=\frac{2\pi\int_{\tilde r_{i-1/2}}^{\tilde r_{i+1/2}} \int_{\tilde \theta_{j-1/2}}^{\tilde \theta_{j+1/2}}\rho_p\left(\tilde r,\tilde \theta,\tilde r_p,\tilde \theta_p\right)\sqrt{\gamma_{\tilde r\tilde r}\gamma_{\tilde \theta\tilde \theta}\gamma_{\phi\phi}}\mathrm{d}\tilde r\mathrm{d}\tilde \theta}{V_{i,j}}=\nonumber\\
    &=2\pi q_p N_p \mathcal{S}^{\hat r}\left(\tilde r_p\right) \mathcal{S}^{\hat \theta}\left(\tilde \theta_p\right),\label{eqB10}
\end{align}
where $\mathcal{S}^{\hat r}(\tilde r_p)$ and $\mathcal{S}^{\hat \theta}(\tilde \theta_p)$ are spline functions given by:
\begin{align}
    \mathcal{S}^{\hat r}\left(\tilde r_p\right)&=\frac{I^{\hat r}_{<p>}\left[\tilde r_{<}\rightarrow\tilde r_{>}\right]}{I^{\hat r}_p\left[\tilde r_{min}\rightarrow\tilde r_{max}\right]I^{\hat r}_{i}\left[\tilde r_{i-1/2}\rightarrow\tilde r_{i+1/2}\right]},\\
    \mathcal{S}^{\hat \theta}\left(\tilde \theta_p\right)&=\frac{I^{\hat \theta}_{<p>}\left[\tilde \theta_{<}\rightarrow\tilde \theta_{>}\right]}{I^{\hat \theta}_p\left[\tilde \theta_{min}\rightarrow\tilde \theta_{max}\right]I^{\hat \theta}_{j}\left[\tilde \theta_{j-1/2}\rightarrow\tilde \theta_{j+1/2}\right]},
\end{align}
specifying the limits taken for each integral function between square brackets. The new integration limits $\tilde r_<=\mathrm{max}(\tilde r_{min},\tilde r_{i-1/2})$, $\tilde \theta_<=\mathrm{max}(\tilde \theta_{min},\tilde \theta_{i-1/2})$, $\tilde r_>=\mathrm{min}(\tilde r_{max},\tilde r_{i+1/2})$, and $\tilde \theta_>=\mathrm{max}(\tilde \theta_{max},\tilde \theta_{i+1/2})$ arise from the integral of the $b_0$ functions present in equation \eqref{eqB8}.

The total charge density at the grid position $(i,j)$ is the sum of \eqref{eqB10} over all the particles that contribute, i.e. 
\begin{equation}
    \rho_{i,j}=\sum_p \rho_{i,j}\left(\tilde r_p,\tilde \theta_p\right).
\end{equation}

\color{\colorDO}
\section{Charge conservation}
\label{AP:D}
\begin{figure*}
  \centering
  \includegraphics[width=17cm]{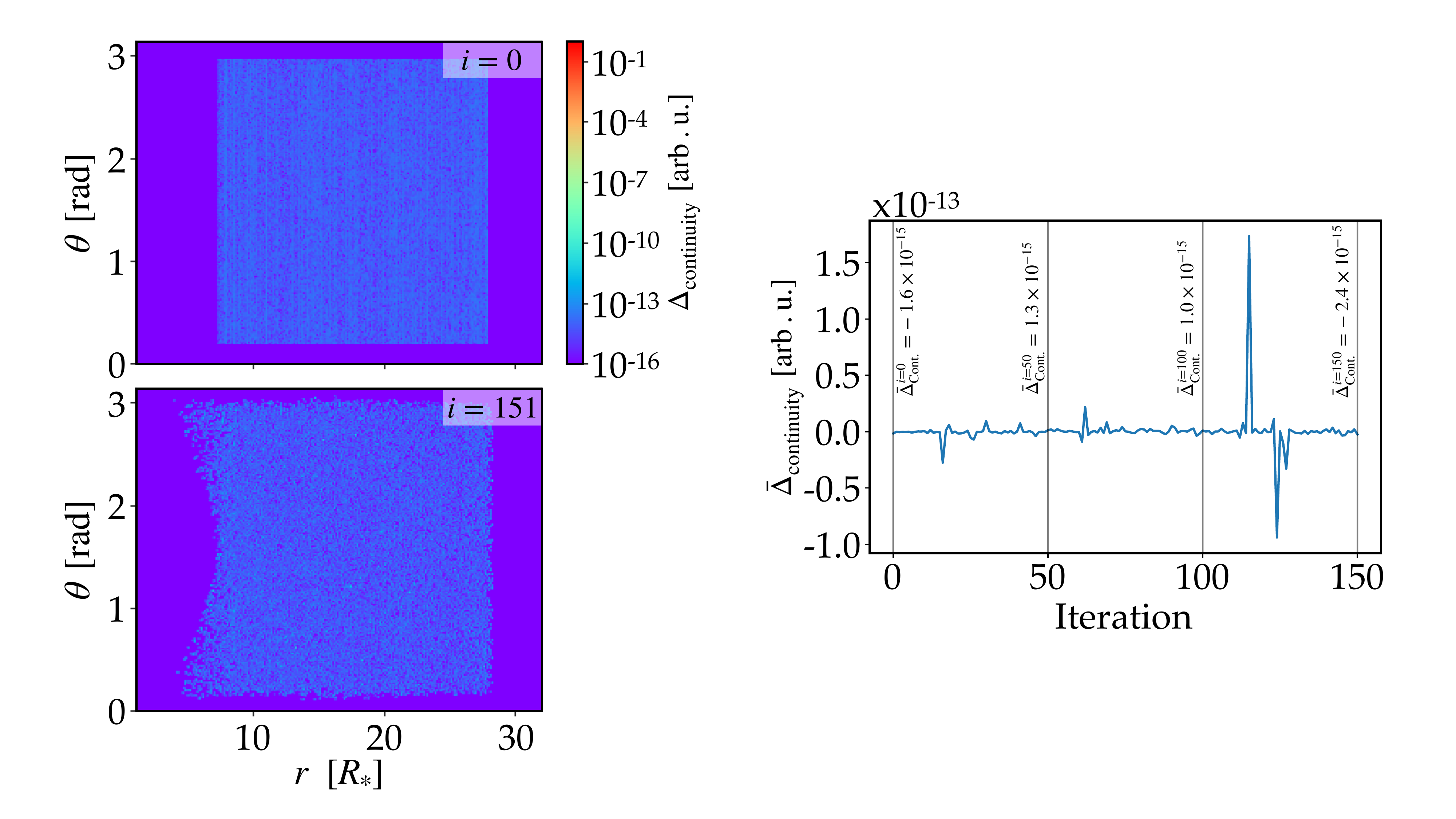}
  \caption{Charge conservation diagnostic for a randomized sample of electrons initialized with thermal velocities. This setup uses nonlinear grid spacings in both $r$ (logarithmic) and $\theta$ (equal area) to highlight that the proposed scheme works even when not using logical coordinate systems. Left panels show the initial and final states, the right panel displays the temporal evolution of the spatial averaged diagnostic.}
  \label{fig: ChargeCons_SPH}
\end{figure*}

\begin{figure*}
  \centering
  \includegraphics[width=17cm]{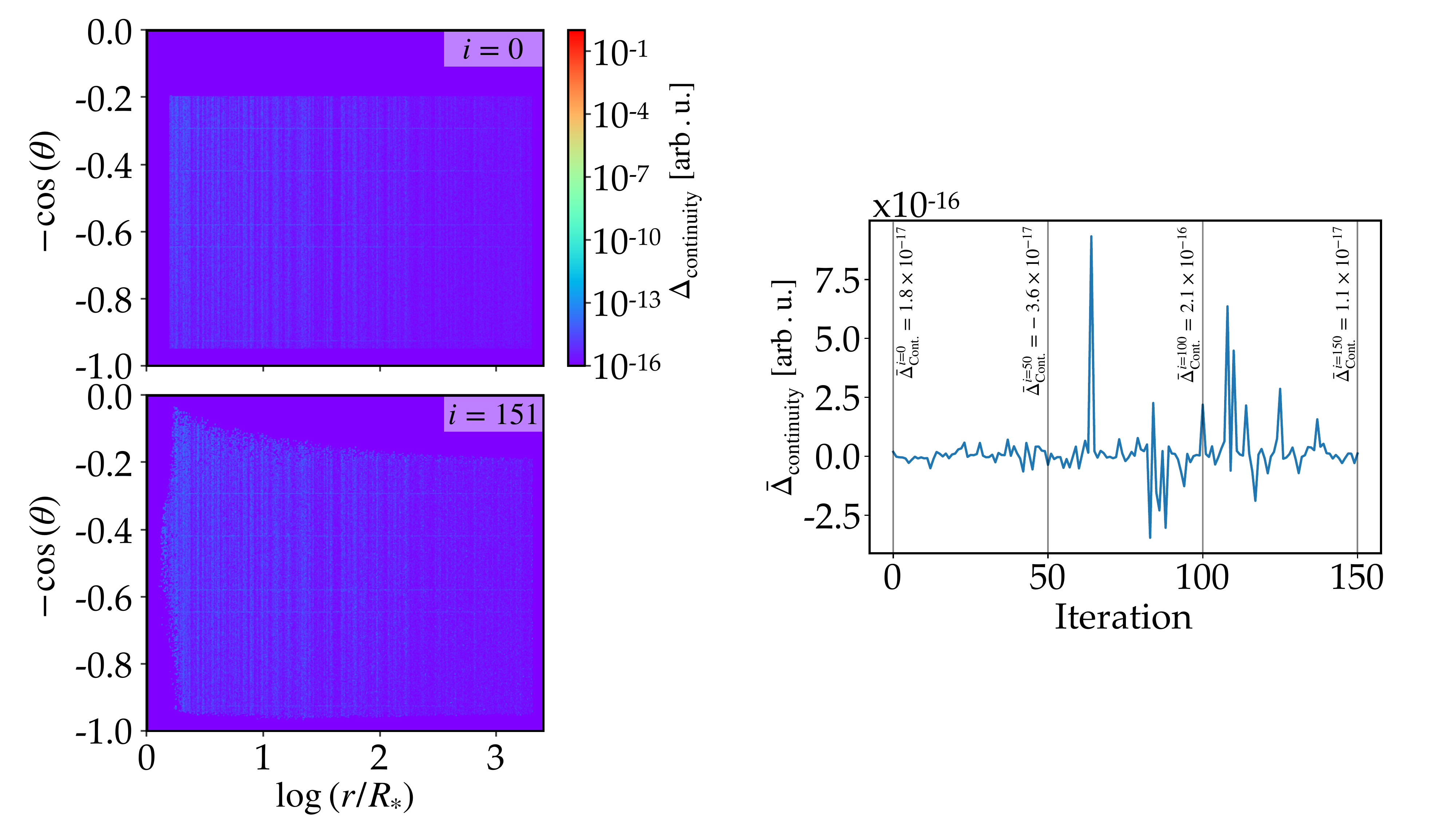}
  \caption{Charge conservation diagnostic for a randomized sample of electrons initialized with thermal velocities. This setup uses uniform grid spacings in both $\tilde{r}$ and $\tilde{\theta}$, i.e. using the logical coordinate system. Left panels show the initial and final states, the right panel displays the temporal evolution of the spatial averaged diagnostic.}
  \label{fig: ChargeCons_Logical}
\end{figure*}

We demonstrate that the code conserves charge to machine precision independently of the selected coordinate system. To exemplify, we use the standard spherical coordinate system with a non-uniform grid spacing in both radial and meridional directions, i.e. uniform spacing in $\log r$ and $-\cos\theta$, and initialize particles with a thermal distribution of velocities in all directions, i.e. $\vec{u}_{\mathrm{thermal}}/c=(2.0,2.0,2.0)$, in the vicinity of a rotating neutron star with dipolar magnetic field $B_*=1000~[m_ec^2/eR_*]$ and $\Omega_*=0.1~[c~\mathrm{rad}/R_*]$, see upper-left panel in Fig.~\ref{fig: ChargeCons_SPH}. Also, we selected a compactness parameter $R_s/R_*=0.5$ to show that the scheme is valid for significant curved spacetimes. Every iteration, we compute the deviation from the continuity equation given by
\begin{align}
    &\Delta_{\mathrm{continuity}}\equiv\frac{\partial\rho}{\partial t} + \vec{\nabla}\cdot\vec{J},\\ &\Delta^{n+1/2}_{\mathrm{cont.}~i,j}\equiv\frac{\rho_{i,j}^{n+1}-\rho_{i,j}^{n}}{\Delta t} + \left(\vec{\nabla}\cdot\vec{J}\right)^{\hat r, n+1/2}_{i,j}+\left(\vec{\nabla}\cdot\vec{J}\right)^{\hat \theta, n+1/2}_{i,j},
\end{align}
and evaluate the grid average value of that quantity, shown in the right panel of Fig.~\ref{fig: ChargeCons_SPH}. Under the same conditions, but now using the logical coordinate system, i.e. using a uniform grid in $(\tilde r,\tilde \theta)=(\log r,-\cos\theta)$, yields the results in Fig.~\ref{fig: ChargeCons_Logical}. Apart from small numerical fluctuations, the grid averaged deviation of the continuity equation stays close to machine precision over the test duration for both cases.

\section{Progressive filter}
\label{AP: progressive}

\draft{We explain in more detail the implementation of the generic progressive filter applied to the first and last radial domain cells. In particular, this progressive filter is used to filter the current density grid components, in this appendix designated $f_i^k$ with $i$ and $k$ being the cell index and the filtering order, respectively. Outside the simulation domain, the current density is unknown, thus invalidating high-order smoothing near the boundaries. Instead of not filtering the boundary cells or filling the guard cells with unphysical field values, we compute the exterior current component values such that only the first and last cells remain unfiltered.} 

\draft{Within the OSIRIS framework, smoothing any field component to order $n$ requires a single-passage kernel with $2n+1$ entries. For example, the single-passage first-, second-, and third-order kernels read}
\begin{align}
    &\mathcal{K}^1 = \left[K^1_1\;\;K^1_2\;\;K^1_3\right],\label{eqE1}\\
    &\mathcal{K}^2 = \left[K^2_1\;\;K^2_2\;\;K^2_3\;\;K^2_4\;\;K^2_5\right],\label{eqE2}\\
    &\mathcal{K}^3 = \left[K^3_1\;\;K^3_2\;\;K^3_3\;\;K^3_4\;\;K^3_5\;\;K^3_6\;\;K^3_7\right],\label{eqE3}
\end{align}
\draft{with $K$ being the weight given to each neighbour. The third-order filtered value of the first four cells yields}
\begin{align}
    &\color{black}f^3_1 = K^3_1\color{red}f^0_{-2}\color{black} + K^3_2\color{red}f^0_{-1}\color{black} + K^3_3\color{red}f^0_{0}\color{black} + K^3_4f^0_{1} +K^3_5f^0_{2} +K^3_6f^0_{3} +K^3_7f^0_{4},\label{eqE4}\\
    &\color{black}f^3_2 = K^3_1\color{red}f^0_{-1}\color{black} + K^3_2\color{red}f^0_{0}\color{black} + K^3_3f^0_{1} + K^3_4f^0_{2} +K^3_5f^0_{3} +K^3_6f^0_{4} +K^3_7f^0_{5},\\
    &\color{black}f^3_3 = K^3_1\color{red}f^0_{0}\color{black} + K^3_2f^0_{1} + K^3_3f^0_{2} + K^3_4f^0_{3} +K^3_5f^0_{4} +K^3_6f^0_{5} +K^3_7f^0_{6},\label{eqE6}\\
    &\color{black}f^3_4 = K^3_1f^0_{1} + K^3_2f^0_{2} + K^3_3f^0_{3} + K^3_4f^0_{4} +K^3_5f^0_{5} +K^3_6f^0_{6} +K^3_7f^0_{7},\label{eqE7}
\end{align}
\draft{where the terms in red are the unknown field values inside the guard cells. Equations~\eqref{eqE4}-\eqref{eqE7} highlight that we can filter the fourth cell using the third-order single passage filter in eq.~\eqref{eqE3}, the third cell using the second-order single passage filter in eq.~\eqref{eqE2}, the second cell using the first-order single-passage filter in eq.~\eqref{eqE1}, and the first cell remains unfiltered, i.e. }
\begin{align}
    &f^3_3 = f_3^2 = K^2_1f^0_{1} + K^2_2f^0_{2} + K^2_3f^0_{3} +K^2_4f^0_{4} +K^2_5f^0_{5},\label{eqE8}\\
    &f^3_2 = f_2^1 = K^1_1f^0_{1} + K^1_2f^0_{2} +K^1_3f^0_{3},\\
    &f^3_1 = f_1^0.\label{eqE10}
\end{align}
\draft{Inserting eqs.~\eqref{eqE4}-\eqref{eqE6} into eqs.~\eqref{eqE8}-\eqref{eqE10} determines the three unknown field components defined in the guard cells, i.e.}
\begin{align}
    &f_0^0 = \frac{1}{K_1^3}\left[\left(K_1^2-K_2^3\right)f^0_1+\left(K_2^2-K_3^3\right)f^0_2+\right.\nonumber\\
    &+\left.\left(K_3^2-K_4^3\right)f^0_3+\left(K_4^2-K_5^3\right)f^0_4+\left(K_5^2-K_6^3\right)f^0_5-K_7^3f^0_6\right],\\
    &f_{-1}^0 = \frac{1}{K_1^3}\left[-K_2^3f^0_0+\left(K_1^1-K_3^3\right)f^0_1+\left(K_2^1-K_4^3\right)f^0_2+\right.\nonumber\\
    &+\left.\left(K_3^1-K_5^3\right)f^0_3-K_6^3f^0_4-K_7^3f^0_5\right],\\
    &f_{-2}^0 = \frac{1}{K_1^3}\left[-K_2^3f_{-1}^0-K_3^3f_{0}^0+\left(1-K_4^3\right)f_{1}^0-K_5^3f_{2}^0-\right.\nonumber\\
    &-\left.K_6^3f_{3}^0-K_7^3f_{4}^0\right].
\end{align}
\draft{All other field values are filtered using the third-order single passage kernel.}

\color{\colorB}
\section{Mur open boundary}
\label{AP:C}
We discuss the implementation of the exterior radial boundary condition that mimics an open radiation boundary. We define the generic wave equation in spherical coordinates
\begin{equation}
    \square\Phi\left(t,r,\theta,\phi\right)=\sigma\left(t,r,\theta,\phi\right),
\end{equation}
where $\sigma$ is the source term and $\square$ is the d'Alembertian operator. Since we place this absorbing boundary at the extreme of the domain, i.e. $r=R_{max}$ with $R_{max} \gg R_*$, we can neglect the effect of the source term. Also, electromagnetic perturbations that propagate radially outwards are dominantly radial due to the geometry and dynamics of the system, i.e. $\partial_\theta \Phi\approx0$. With axisymmetry ($\partial_\phi \Phi=0$), we can neglect the d'Alembertian operator for the angular derivatives:
\begin{equation}
    \square\Phi\left(t,r,\theta,\phi\right)\approx\left(\partial^2_t-\partial^2_r-\frac{2}{r}\partial _r\right)\Phi=0.\label{eqC2}
\end{equation}
The radiation condition impinging on a spherical shell placed far from the compact point source derived in \citep{Sommerfeld1949} reads
\begin{equation}
    \lim_{r\rightarrow\infty}\left(\partial_r+\partial_t\right)\left(r\Phi\right)=0,
\end{equation}
describes only the outward propagating component of equation~\eqref{eqC2} and is hereafter designated as the Sommerfeld condition. At a finite distance from the source, it approximates to \citep{Novak2004, Espinoza2014}
\begin{equation}
    \partial_t \Phi + \partial_r \Phi +\frac{\Phi}{r}\Biggr|_{r=R_{max}}=0,\label{eqC4}
\end{equation}
which is exact for pure monopolar waves. 

In General Relativity, using the 3+1 formalism in spherical coordinates, the modification of the differential operators leads to the generalization of equation~\eqref{eqC4} as
\begin{equation}
    \alpha^{-1} \partial_t \Phi + \alpha\partial_r \Phi +\frac{\Phi}{r}\Biggr|_{r=R_{max}}=0,
\end{equation}
which we discretize adopting a spatiotemporal centering at $(r,t)=(r_{N+1/2},t^{n+1/2})$:
\begin{equation}
    \frac{\Phi^{n+1}-\Phi^{n}}{\alpha \Delta t}\Biggr|_{r_{N+1/2}}+ \frac{\alpha_{N+1}\Phi_{N+1}-\alpha_{N}\Phi_{N}}{\Delta r}\Biggr|^{n+1/2}+\frac{\Phi^{n+1/2}}{r}\Biggr|_{r_{N+1/2}}=0, \label{eqC6}
\end{equation}
where $N$ and $N+1$ are the radial indices of the last cell of the domain and the first guard cell outside of it, respectively. The field components that lie in the outer grid frontier, i.e. at $r=r_{N+1}$, are $E^{\hat \phi}$, $E^{\hat \theta}$, and $B^{\hat r}$, which means that equation \eqref{eqC6} needs to be adapted for each of these components. Recalling that we are using the leapfrog method to push the electromagnetic field components in time, we see that we have the electric field components at times $n$ and $n+1$, while the magnetic field components are available at times $n$ and $n+1/2$, for the first half push, and $n+1/2$ and $n+1$, for the second half push. Consequently, the algorithm to compute the correct field value outside the domain, i.e. $\Phi^{n+1}_{N+1}$, differs for the electric and magnetic field components. 

\subsection{Electric field}
Here we discuss the algorithm with $\Phi$ representing $E^{\hat \phi}$ and $E^{\hat \theta}$. The first term in equation~\eqref{eqC6} is already in the right temporal positions but requires the correct spatial centering:
\begin{equation}
    \frac{\Phi^{n+1}-\Phi^{n}}{\alpha \Delta t}\Biggr|_{r_{N+1/2}} = \frac{1}{\Delta t}\left[\frac{1}{2}\left(\frac{\Phi^{n+1}_{N+1}}{\alpha_{N+1}}+\frac{\Phi^{n+1}_N}{\alpha_{N}}\right)-\frac{1}{2}\left(\frac{\Phi^{n}_{N+1}}{\alpha_{N+1}}+\frac{\Phi^{n}_N}{\alpha_{N}}\right)\right].\label{eqC7}
\end{equation}
In opposition, the second term is in the right spatial position and requires the correct temporal centering:
\begin{align}
    \frac{\alpha_{N+1}\Phi_{N+1}-\alpha_{N}\Phi_{N}}{\Delta r}\Biggr|^{n+1/2} = \frac{1}{\Delta r}&\left[\frac{\alpha_{N+1}}{2}\left(\Phi^{n+1}_{N+1}+\Phi^{n}_{N+1}\right)-\right.\nonumber\\
    &-\left.\frac{\alpha_{N}}{2}\left(\Phi^{n+1}_{N}+\Phi^{n}_{N}\right)\right].\label{eqC8}
\end{align}
The last term in equation~\eqref{eqC6} is more complicated because we need to center it in time and space:
\begin{align}
    \frac{\Phi^{n+1/2}}{r}\Biggr|_{r_{N+1/2}} &= \frac{1}{2}\left[\frac{\Phi^{n+1/2}_{N+1}}{r_{N+1}}+\frac{\Phi^{n+1/2}_{N}}{r_{N}}\right]=\nonumber\\
    &=\frac{1}{4}\left[\frac{\Phi^{n+1}_{N+1}+\Phi^{n}_{N+1}}{r_{N+1}}+\frac{\Phi^{n+1}_{N}+\Phi^{n}_{N}}{r_{N}}\right].\label{eqC9}
\end{align}
Inputting equations \eqref{eqC7}-\eqref{eqC9} into \eqref{eqC6} and solving for the corrected field outside the domain, we get:
\begin{align}
    \Phi^{n+1}_{N+1} &= \left(\frac{1}{\alpha_{N+1}\Delta t}+\frac{\alpha_{N+1}}{\Delta r}+\frac{1}{2r_{N+1}}\right)^{-1}\times \nonumber\\
    &\times\left[\Phi^{n}_{N+1}\left(\frac{1}{\alpha_{N+1}\Delta t}-\frac{\alpha_{N+1}}{\Delta r}-\frac{1}{2r_{N+1}}\right)-\right.\nonumber\\
    &\hspace{.15cm}\left.-\Phi^{n+1}_{N}\left(\frac{1}{\alpha_{N}\Delta t}-\frac{\alpha_{N}}{\Delta r}+\frac{1}{2r_{N}}\right)+\Phi^{n}_{N}\left(\frac{1}{\alpha_{N}\Delta t}+\frac{\alpha_{N}}{\Delta r}-\frac{1}{2r_{N}}\right)\right]
\end{align}
Notice that to compute this, we require knowledge of the term $\Phi^{n+1}_{N}$ that is available on the grid, and the knowledge of the terms $\Phi^{n}_{N+1}$ and $\Phi^{n}_{N}$ that need to be saved in memory as, otherwise, they would be overwritten on the grid.

\subsection{Magnetic field}
The approach taken to correct the out-of-bound $B^{\hat r}$ is very similar but it must be split into two steps, as in the magnetic evolution algorithm. Therefore, we need to compute $\Phi^{n+1/2}_{N+1}$ ($\Phi^{n+1}_{N+1}$) after the first (second) half-step. To obtain these expressions, we need to repeat the same procedure as for the electric field with the extra step of adding and subtracting the half-step value, i.e. equation \eqref{eqC7} reads
\begin{align}
    \frac{\Phi^{n+1}-\Phi^{n}}{\alpha \Delta t}\Biggr|_{r_{N+1/2}} = &\frac{1}{2\Delta t}\left[\left(\frac{\Phi^{n+1/2}_{N+1}}{\alpha_{N+1}}+\frac{\Phi^{n+1/2}_N}{\alpha_{N}}\right)-\left(\frac{\Phi^{n}_{N+1}}{\alpha_{N+1}}+\frac{\Phi^{n}_N}{\alpha_{N}}\right)\right] + \nonumber\\
    +&\frac{1}{2\Delta t}\left[\left(\frac{\Phi^{n+1}_{N+1}}{\alpha_{N+1}}+\frac{\Phi^{n+1}_N}{\alpha_{N}}\right)-\left(\frac{\Phi^{n+1/2}_{N+1}}{\alpha_{N+1}}+\frac{\Phi^{n+1/2}_N}{\alpha_{N}}\right)\right],
\end{align}
which we repeat for equations \eqref{eqC8} and \eqref{eqC9}. Grouping terms that depend on the temporal indices $n$ and $n+1/2$ yields the first half-step correction:
\begin{align}
    \Phi^{n+1/2}_{N+1} = -\frac{\alpha_{N+1}}{\alpha_N}\Phi^{n+1/2}_{N}+\Phi^{n}_{N+1}\left(1-\frac{\alpha^2_{N+1}\Delta t}{\Delta r}-\frac{\alpha_{N+1}\Delta t}{2 r_{N+1}}\right)+\nonumber\\
    +\Phi^{n}_{N}\left(\frac{\alpha_{N+1}}{\alpha_{N}}+\frac{\alpha_{N+1}\alpha_{N}\Delta t}{\Delta r}-\frac{\alpha_{N+1}\Delta t}{2r_N}\right).\label{eqC12}
\end{align}
We obtain the second half-step correction through the terms that depend on the temporal indices $n+1$ and $n+1/2$, giving
\begin{align}
    &\Phi^{n+1}_{N+1} = \left(\frac{1}{\alpha_{N+1}}+\frac{\alpha_{N+1}\Delta t}{\Delta r}+\frac{\Delta t}{2r_{N+1}}\right)^{-1}\times\nonumber\\
    &\times\left[\frac{1}{\alpha_{N+1}}\Phi^{n+1/2}_{N+1}+\frac{1}{\alpha_{N}}\Phi^{n+1/2}_{N}-\Phi^{n+1}_{N}\left(\frac{1}{\alpha_N}-\frac{\alpha_N\Delta t}{\Delta r}+\frac{\Delta t}{2r_N}\right)\right].\label{eqC13}
\end{align}
In this case, we need to save in memory the values of $\Phi_{N}^{n}$ and $\Phi_{N+1}^{n}$ for equation \eqref{eqC12}, and $\Phi_{N}^{n+1/2}$ and $\Phi_{N+1}^{n+1/2}$ for equation \eqref{eqC13}. The other components are readily available on the grid.

\bibliographystyle{elsarticle-harv} 
\bibliography{elsdoc}

\begin{thebibliography}{112}
\expandafter\ifx\csname natexlab\endcsname\relax\def\natexlab#1{#1}\fi
\providecommand{\url}[1]{\texttt{#1}}
\providecommand{\href}[2]{#2}
\providecommand{\path}[1]{#1}
\providecommand{\DOIprefix}{doi:}
\providecommand{\ArXivprefix}{arXiv:}
\providecommand{\URLprefix}{URL: }
\providecommand{\Pubmedprefix}{pmid:}
\providecommand{\doi}[1]{\href{http://dx.doi.org/#1}{\path{#1}}}
\providecommand{\Pubmed}[1]{\href{pmid:#1}{\path{#1}}}
\providecommand{\bibinfo}[2]{#2}
\ifx\xfnm\relax \def\xfnm[#1]{\unskip,\space#1}\fi
\bibitem[{{Arons}(1983)}]{Arons_1983}
\bibinfo{author}{{Arons}, J.}, \bibinfo{year}{1983}.
\newblock \bibinfo{title}{{Pair creation above pulsar polar caps : geometrical
  structure and energetics of slot gaps.}}
\newblock \bibinfo{journal}{\apj} \bibinfo{volume}{266},
  \bibinfo{pages}{215--241}.
\newblock \DOIprefix\doi{10.1086/160771}.
\bibitem[{{Arons} and {Scharlemann}(1979)}]{Arons_1979}
\bibinfo{author}{{Arons}, J.}, \bibinfo{author}{{Scharlemann}, E.T.},
  \bibinfo{year}{1979}.
\newblock \bibinfo{title}{{Pair formation above pulsar polar caps: structure of
  the low altitude acceleration zone.}}
\newblock \bibinfo{journal}{\apj} \bibinfo{volume}{231},
  \bibinfo{pages}{854--879}.
\newblock \DOIprefix\doi{10.1086/157250}.
\bibitem[{Bacchini et~al.(2020)Bacchini, Ripperda, Philippov and
  Parfrey}]{Bacchini_2020}
\bibinfo{author}{Bacchini, F.}, \bibinfo{author}{Ripperda, B.},
  \bibinfo{author}{Philippov, A.A.}, \bibinfo{author}{Parfrey, K.},
  \bibinfo{year}{2020}.
\newblock \bibinfo{title}{A coupled guiding center{\textendash}boris particle
  pusher for magnetized plasmas in compact-object magnetospheres}.
\newblock \bibinfo{journal}{The Astrophysical Journal Supplement Series}
  \bibinfo{volume}{251}, \bibinfo{pages}{10}.
\newblock \URLprefix \url{https://doi.org/10.3847/1538-4365/abb604},
  \DOIprefix\doi{10.3847/1538-4365/abb604}.
\bibitem[{Bai and Spitkovsky(2010)}]{Bai_2010}
\bibinfo{author}{Bai, X.N.}, \bibinfo{author}{Spitkovsky, A.},
  \bibinfo{year}{2010}.
\newblock \bibinfo{title}{Modeling of gamma-ray pulsar light curves using the
  force-free magnetic field}.
\newblock \bibinfo{journal}{The Astrophysical Journal} \bibinfo{volume}{715},
  \bibinfo{pages}{1282}.
\newblock \URLprefix \url{https://dx.doi.org/10.1088/0004-637X/715/2/1282},
  \DOIprefix\doi{10.1088/0004-637X/715/2/1282}.
\bibitem[{Basu et~al.(2018)Basu, Mitra, Melikidze and Skrzypczak}]{Basu_2018}
\bibinfo{author}{Basu, R.}, \bibinfo{author}{Mitra, D.},
  \bibinfo{author}{Melikidze, G.I.}, \bibinfo{author}{Skrzypczak, A.},
  \bibinfo{year}{2018}.
\newblock \bibinfo{title}{{Classification of subpulse drifting in pulsars}}.
\newblock \bibinfo{journal}{Monthly Notices of the Royal Astronomical Society}
  \bibinfo{volume}{482}, \bibinfo{pages}{3757--3788}.
\newblock \URLprefix \url{https://doi.org/10.1093/mnras/sty2846},
  \DOIprefix\doi{10.1093/mnras/sty2846},
  \href{http://arxiv.org/abs/https://academic.oup.com/mnras/article-pdf/482/3/3757/26694419/sty2846.pdf}{{\tt
  arXiv:https://academic.oup.com/mnras/article-pdf/482/3/3757/26694419/sty2846.pdf}}.
\bibitem[{Beloborodov(2008)}]{Beloborodov_2008}
\bibinfo{author}{Beloborodov, A.M.}, \bibinfo{year}{2008}.
\newblock \bibinfo{title}{Polar-cap accelerator and radio emission from
  pulsars}.
\newblock \bibinfo{journal}{The Astrophysical Journal} \bibinfo{volume}{683},
  \bibinfo{pages}{L41--L44}.
\newblock \URLprefix \url{https://doi.org/10.1086/590079},
  \DOIprefix\doi{10.1086/590079}.
\bibitem[{Belyaev(2015a)}]{Belyaev_2015b}
\bibinfo{author}{Belyaev, M.A.}, \bibinfo{year}{2015}a.
\newblock \bibinfo{title}{{Dissipation, energy transfer, and spin-down
  luminosity in 2.5D PIC simulations of the pulsar magnetosphere}}.
\newblock \bibinfo{journal}{Monthly Notices of the Royal Astronomical Society}
  \bibinfo{volume}{449}, \bibinfo{pages}{2759--2767}.
\newblock \URLprefix \url{https://doi.org/10.1093/mnras/stv468},
  \DOIprefix\doi{10.1093/mnras/stv468},
  \href{http://arxiv.org/abs/https://academic.oup.com/mnras/article-pdf/449/3/2759/9386867/stv468.pdf}{{\tt
  arXiv:https://academic.oup.com/mnras/article-pdf/449/3/2759/9386867/stv468.pdf}}.
\bibitem[{Belyaev(2015b)}]{Belyaev_2015a}
\bibinfo{author}{Belyaev, M.A.}, \bibinfo{year}{2015}b.
\newblock \bibinfo{title}{Picsar: A 2.5d axisymmetric, relativistic,
  electromagnetic, particle in cell code with a radiation absorbing boundary}.
\newblock \bibinfo{journal}{New Astronomy} \bibinfo{volume}{36},
  \bibinfo{pages}{37--49}.
\newblock \URLprefix
  \url{https://www.sciencedirect.com/science/article/pii/S1384107614001407},
  \DOIprefix\doi{https://doi.org/10.1016/j.newast.2014.09.006}.
\bibitem[{Belyaev and Parfrey(2016)}]{Belyaev_2016}
\bibinfo{author}{Belyaev, M.A.}, \bibinfo{author}{Parfrey, K.},
  \bibinfo{year}{2016}.
\newblock \bibinfo{title}{{SPATIAL} {DISTRIBUTION} {OF} {PAIR} {PRODUCTION}
  {OVER} {THE} {PULSAR} {POLAR} {CAP}}.
\newblock \bibinfo{journal}{The Astrophysical Journal} \bibinfo{volume}{830},
  \bibinfo{pages}{119}.
\newblock \URLprefix \url{https://doi.org/10.3847/0004-637x/830/2/119},
  \DOIprefix\doi{10.3847/0004-637x/830/2/119}.
\bibitem[{{Beskin}(1990)}]{Beskin_1990}
\bibinfo{author}{{Beskin}, V.S.}, \bibinfo{year}{1990}.
\newblock \bibinfo{title}{{General Relativity Effects on Electrodynamic
  Processes in Radio Pulsars}}.
\newblock \bibinfo{journal}{Soviet Astronomy Letters} \bibinfo{volume}{16},
  \bibinfo{pages}{286}.
\bibitem[{Bhattacharyya et~al.(2017)Bhattacharyya, Bombaci, Bandyopadhyay,
  Thampan and Logoteta}]{Bhattacharyya_2017}
\bibinfo{author}{Bhattacharyya, S.}, \bibinfo{author}{Bombaci, I.},
  \bibinfo{author}{Bandyopadhyay, D.}, \bibinfo{author}{Thampan, A.V.},
  \bibinfo{author}{Logoteta, D.}, \bibinfo{year}{2017}.
\newblock \bibinfo{title}{Millisecond radio pulsars with known masses:
  Parameter values and equation of state models}.
\newblock \bibinfo{journal}{New Astronomy} \bibinfo{volume}{54},
  \bibinfo{pages}{61--71}.
\newblock \URLprefix
  \url{https://www.sciencedirect.com/science/article/pii/S1384107616301798},
  \DOIprefix\doi{https://doi.org/10.1016/j.newast.2017.01.008}.
\bibitem[{Boris and Shanny(1972)}]{Boris1970}
\bibinfo{author}{Boris, J.P.}, \bibinfo{author}{Shanny, R.A.},
  \bibinfo{year}{1972}.
\newblock \bibinfo{title}{Proceedings: Fourth Conference on Numerical
  Simulation of Plasmas, November 2, 3, 1970}.
\newblock \bibinfo{publisher}{Naval Research Laboratory}.
\bibitem[{Brambilla et~al.(2018)Brambilla, Kalapotharakos, Timokhin, Harding
  and Kazanas}]{Brambilla_2018}
\bibinfo{author}{Brambilla, G.}, \bibinfo{author}{Kalapotharakos, C.},
  \bibinfo{author}{Timokhin, A.N.}, \bibinfo{author}{Harding, A.K.},
  \bibinfo{author}{Kazanas, D.}, \bibinfo{year}{2018}.
\newblock \bibinfo{title}{Electron–positron pair flow and current composition
  in the pulsar magnetosphere}.
\newblock \bibinfo{journal}{The Astrophysical Journal} \bibinfo{volume}{858},
  \bibinfo{pages}{81}.
\newblock \URLprefix \url{https://dx.doi.org/10.3847/1538-4357/aab3e1},
  \DOIprefix\doi{10.3847/1538-4357/aab3e1}.
\bibitem[{Bransgrove et~al.(2022)Bransgrove, Beloborodov and
  Levin}]{Bransgrove_22}
\bibinfo{author}{Bransgrove, A.}, \bibinfo{author}{Beloborodov, A.M.},
  \bibinfo{author}{Levin, Y.}, \bibinfo{year}{2022}.
\newblock \bibinfo{title}{Radio emission and electric gaps in pulsar
  magnetospheres}.
\newblock \URLprefix \url{https://arxiv.org/abs/2209.11362},
  \DOIprefix\doi{10.48550/ARXIV.2209.11362}.
\bibitem[{Cerutti and Beloborodov(2016)}]{Cerutti_2016b}
\bibinfo{author}{Cerutti, B.}, \bibinfo{author}{Beloborodov, A.M.},
  \bibinfo{year}{2016}.
\newblock \bibinfo{title}{Electrodynamics of pulsar magnetospheres}.
\newblock \bibinfo{journal}{Space Science Reviews} \bibinfo{volume}{207},
  \bibinfo{pages}{111–136}.
\newblock \URLprefix \url{http://dx.doi.org/10.1007/s11214-016-0315-7},
  \DOIprefix\doi{10.1007/s11214-016-0315-7}.
\bibitem[{Cerutti et~al.(2015)Cerutti, Philippov, Parfrey and
  Spitkovsky}]{Cerutti_2015}
\bibinfo{author}{Cerutti, B.}, \bibinfo{author}{Philippov, A.},
  \bibinfo{author}{Parfrey, K.}, \bibinfo{author}{Spitkovsky, A.},
  \bibinfo{year}{2015}.
\newblock \bibinfo{title}{{ Particle acceleration in axisymmetric pulsar
  current sheets}}.
\newblock \bibinfo{journal}{Monthly Notices of the Royal Astronomical Society}
  \bibinfo{volume}{448}, \bibinfo{pages}{606--619}.
\newblock \URLprefix \url{https://doi.org/10.1093/mnras/stv042},
  \DOIprefix\doi{10.1093/mnras/stv042},
  \href{http://arxiv.org/abs/https://academic.oup.com/mnras/article-pdf/448/1/606/9388852/stv042.pdf}{{\tt
  arXiv:https://academic.oup.com/mnras/article-pdf/448/1/606/9388852/stv042.pdf}}.
\bibitem[{Cerutti et~al.(2013)Cerutti, Werner, Uzdensky and
  Begelman}]{Cerutti_2013}
\bibinfo{author}{Cerutti, B.}, \bibinfo{author}{Werner, G.R.},
  \bibinfo{author}{Uzdensky, D.A.}, \bibinfo{author}{Begelman, M.C.},
  \bibinfo{year}{2013}.
\newblock \bibinfo{title}{Simulations of particle acceleration beyond the
  classical synchrotron burnoff limit in magnetic reconnection: An explanation
  of the crab flares}.
\newblock \bibinfo{journal}{The Astrophysical Journal} \bibinfo{volume}{770},
  \bibinfo{pages}{147}.
\newblock \URLprefix \url{https://dx.doi.org/10.1088/0004-637X/770/2/147},
  \DOIprefix\doi{10.1088/0004-637X/770/2/147}.
\bibitem[{{Cerutti, B.} and {Philippov, A. A.}(2017)}]{Cerutti_2017}
\bibinfo{author}{{Cerutti, B.}}, \bibinfo{author}{{Philippov, A. A.}},
  \bibinfo{year}{2017}.
\newblock \bibinfo{title}{Dissipation of the striped pulsar wind}.
\newblock \bibinfo{journal}{A\&A} \bibinfo{volume}{607}, \bibinfo{pages}{A134}.
\newblock \URLprefix \url{https://doi.org/10.1051/0004-6361/201731680},
  \DOIprefix\doi{10.1051/0004-6361/201731680}.
\bibitem[{{Cerutti, Beno\^{\i}t} et~al.(2020){Cerutti, Beno\^{\i}t},
  {Philippov, Alexander A.} and {Dubus, Guillaume}}]{Cerutti_2020}
\bibinfo{author}{{Cerutti, Beno\^{\i}t}}, \bibinfo{author}{{Philippov,
  Alexander A.}}, \bibinfo{author}{{Dubus, Guillaume}}, \bibinfo{year}{2020}.
\newblock \bibinfo{title}{Dissipation of the striped pulsar wind and
  non-thermal particle acceleration: 3d pic simulations}.
\newblock \bibinfo{journal}{A\&A} \bibinfo{volume}{642}, \bibinfo{pages}{A204}.
\newblock \URLprefix \url{https://doi.org/10.1051/0004-6361/202038618},
  \DOIprefix\doi{10.1051/0004-6361/202038618}.
\bibitem[{Chen and Beloborodov(2014)}]{Chen_2014}
\bibinfo{author}{Chen, A.Y.}, \bibinfo{author}{Beloborodov, A.M.},
  \bibinfo{year}{2014}.
\newblock \bibinfo{title}{{ELECTRODYNAMICS} {OF} {AXISYMMETRIC} {PULSAR}
  {MAGNETOSPHERE} {WITH} {ELECTRON}-{POSITRON} {DISCHARGE}: A {NUMERICAL}
  {EXPERIMENT}}.
\newblock \bibinfo{journal}{The Astrophysical Journal} \bibinfo{volume}{795},
  \bibinfo{pages}{L22}.
\newblock \URLprefix \url{https://doi.org/10.1088/2041-8205/795/1/l22},
  \DOIprefix\doi{10.1088/2041-8205/795/1/l22}.
\bibitem[{Chen et~al.(2020)Chen, Cruz and Spitkovsky}]{Chen_2020}
\bibinfo{author}{Chen, A.Y.}, \bibinfo{author}{Cruz, F.},
  \bibinfo{author}{Spitkovsky, A.}, \bibinfo{year}{2020}.
\newblock \bibinfo{title}{Filling the magnetospheres of weak pulsars}.
\newblock \bibinfo{journal}{The Astrophysical Journal} \bibinfo{volume}{889},
  \bibinfo{pages}{69}.
\newblock \URLprefix \url{https://dx.doi.org/10.3847/1538-4357/ab5c20},
  \DOIprefix\doi{10.3847/1538-4357/ab5c20}.
\bibitem[{Chen et~al.(2023)Chen, Yan, Han, Wang, Wang, Jing, Lee, Zhang, Xu,
  Wang, Yang, Su, Cai, Wang, Qiao, Xu and Zhou}]{XChen_2023}
\bibinfo{author}{Chen, X.}, \bibinfo{author}{Yan, Y.}, \bibinfo{author}{Han,
  J.L.}, \bibinfo{author}{Wang, C.}, \bibinfo{author}{Wang, P.F.},
  \bibinfo{author}{Jing, W.C.}, \bibinfo{author}{Lee, K.J.},
  \bibinfo{author}{Zhang, B.}, \bibinfo{author}{Xu, R.X.},
  \bibinfo{author}{Wang, T.}, \bibinfo{author}{Yang, Z.L.},
  \bibinfo{author}{Su, W.Q.}, \bibinfo{author}{Cai, N.N.},
  \bibinfo{author}{Wang, W.Y.}, \bibinfo{author}{Qiao, G.J.},
  \bibinfo{author}{Xu, J.}, \bibinfo{author}{Zhou, D.J.}, \bibinfo{year}{2023}.
\newblock \bibinfo{title}{Strong and weak pulsar radio emission due to
  thunderstorms and raindrops of particles in the magnetosphere}.
\newblock \bibinfo{journal}{Nature Astronomy} \bibinfo{volume}{7},
  \bibinfo{pages}{1235--1244}.
\newblock \URLprefix \url{https://doi.org/10.1038/s41550-023-02056-z},
  \DOIprefix\doi{10.1038/s41550-023-02056-z}.
\bibitem[{{Cheng} et~al.(1986){Cheng}, {Ho} and {Ruderman}}]{Cheng_1986}
\bibinfo{author}{{Cheng}, K.S.}, \bibinfo{author}{{Ho}, C.},
  \bibinfo{author}{{Ruderman}, M.}, \bibinfo{year}{1986}.
\newblock \bibinfo{title}{{Energetic Radiation from Rapidly Spinning Pulsars.
  I. Outer Magnetosphere Gaps}}.
\newblock \bibinfo{journal}{\apj} \bibinfo{volume}{300}, \bibinfo{pages}{500}.
\newblock \DOIprefix\doi{10.1086/163829}.
\bibitem[{Chirenti et~al.(2023)Chirenti, Dichiara, Lien, Miller and
  Preece}]{Chirenti_2023}
\bibinfo{author}{Chirenti, C.}, \bibinfo{author}{Dichiara, S.},
  \bibinfo{author}{Lien, A.}, \bibinfo{author}{Miller, M.C.},
  \bibinfo{author}{Preece, R.}, \bibinfo{year}{2023}.
\newblock \bibinfo{title}{Kilohertz quasiperiodic oscillations in short
  gamma-ray bursts}.
\newblock \bibinfo{journal}{Nature} \bibinfo{volume}{613},
  \bibinfo{pages}{253--256}.
\newblock \URLprefix \url{https://doi.org/10.1038/s41586-022-05497-0},
  \DOIprefix\doi{10.1038/s41586-022-05497-0}.
\bibitem[{{Chirenti} et~al.(2019){Chirenti}, {Miller}, {Strohmayer} and
  {Camp}}]{Chirenti_2019}
\bibinfo{author}{{Chirenti}, C.}, \bibinfo{author}{{Miller}, M.C.},
  \bibinfo{author}{{Strohmayer}, T.}, \bibinfo{author}{{Camp}, J.},
  \bibinfo{year}{2019}.
\newblock \bibinfo{title}{{Searching for Hypermassive Neutron Stars with Short
  Gamma-Ray Bursts}}.
\newblock \bibinfo{journal}{\apjl} \bibinfo{volume}{884}, \bibinfo{pages}{L16}.
\newblock \DOIprefix\doi{10.3847/2041-8213/ab43e0},
  \href{http://arxiv.org/abs/1906.09647}{{\tt arXiv:1906.09647}}.
\bibitem[{Contopoulos et~al.(1999)Contopoulos, Kazanas and
  Fendt}]{Contopoulos_1999}
\bibinfo{author}{Contopoulos, I.}, \bibinfo{author}{Kazanas, D.},
  \bibinfo{author}{Fendt, C.}, \bibinfo{year}{1999}.
\newblock \bibinfo{title}{The axisymmetric pulsar magnetosphere}.
\newblock \bibinfo{journal}{The Astrophysical Journal} \bibinfo{volume}{511},
  \bibinfo{pages}{351}.
\newblock \URLprefix \url{https://dx.doi.org/10.1086/306652},
  \DOIprefix\doi{10.1086/306652}.
\bibitem[{Cruz et~al.(2023)Cruz, Grismayer, Chen, Spitkovsky, Fonseca and
  Silva}]{Cruz_2023}
\bibinfo{author}{Cruz, F.}, \bibinfo{author}{Grismayer, T.},
  \bibinfo{author}{Chen, A.Y.}, \bibinfo{author}{Spitkovsky, A.},
  \bibinfo{author}{Fonseca, R.A.}, \bibinfo{author}{Silva, L.O.},
  \bibinfo{year}{2023}.
\newblock \bibinfo{title}{Particle-in-cell simulations of pulsar
  magnetospheres: transition between electrosphere and force-free regimes}.
\newblock \href{http://arxiv.org/abs/2309.04834}{{\tt arXiv:2309.04834}}.
\bibitem[{Cruz et~al.(2021)Cruz, Grismayer, Chen, Spitkovsky and
  Silva}]{Cruz_2021}
\bibinfo{author}{Cruz, F.}, \bibinfo{author}{Grismayer, T.},
  \bibinfo{author}{Chen, A.Y.}, \bibinfo{author}{Spitkovsky, A.},
  \bibinfo{author}{Silva, L.O.}, \bibinfo{year}{2021}.
\newblock \bibinfo{title}{Coherent emission from qed cascades in pulsar polar
  caps}.
\newblock \bibinfo{journal}{The Astrophysical Journal Letters}
  \bibinfo{volume}{919}, \bibinfo{pages}{L4}.
\newblock \URLprefix \url{https://dx.doi.org/10.3847/2041-8213/ac2157},
  \DOIprefix\doi{10.3847/2041-8213/ac2157}.
\bibitem[{Cruz et~al.(2022)Cruz, Grismayer, Iteanu, Tortone and
  Silva}]{Cruz_2022}
\bibinfo{author}{Cruz, F.}, \bibinfo{author}{Grismayer, T.},
  \bibinfo{author}{Iteanu, S.}, \bibinfo{author}{Tortone, P.},
  \bibinfo{author}{Silva, L.O.}, \bibinfo{year}{2022}.
\newblock \bibinfo{title}{{Model of pulsar pair cascades in non-uniform
  electric fields: Growth rate, density profile, and screening time}}.
\newblock \bibinfo{journal}{Physics of Plasmas} \bibinfo{volume}{29},
  \bibinfo{pages}{052902}.
\newblock \URLprefix \url{https://doi.org/10.1063/5.0085847},
  \DOIprefix\doi{10.1063/5.0085847},
  \href{http://arxiv.org/abs/https://pubs.aip.org/aip/pop/article-pdf/doi/10.1063/5.0085847/16596951/052902\_1\_online.pdf}{{\tt
  arXiv:https://pubs.aip.org/aip/pop/article-pdf/doi/10.1063/5.0085847/16596951/052902\_1\_online.pdf}}.
\bibitem[{{Daugherty} and {Harding}(1982)}]{Daugherty_1982}
\bibinfo{author}{{Daugherty}, J.K.}, \bibinfo{author}{{Harding}, A.K.},
  \bibinfo{year}{1982}.
\newblock \bibinfo{title}{{Electromagnetic cascades in pulsars.}}
\newblock \bibinfo{journal}{\apj} \bibinfo{volume}{252},
  \bibinfo{pages}{337--347}.
\newblock \DOIprefix\doi{10.1086/159561}.
\bibitem[{{Daugherty} and {Harding}(1996)}]{Daugherty_1996}
\bibinfo{author}{{Daugherty}, J.K.}, \bibinfo{author}{{Harding}, A.K.},
  \bibinfo{year}{1996}.
\newblock \bibinfo{title}{{Gamma-Ray Pulsars: Emission from Extended Polar CAP
  Cascades}}.
\newblock \bibinfo{journal}{\apj} \bibinfo{volume}{458}, \bibinfo{pages}{278}.
\newblock \DOIprefix\doi{10.1086/176811},
  \href{http://arxiv.org/abs/astro-ph/9508155}{{\tt arXiv:astro-ph/9508155}}.
\bibitem[{Deshpande and Rankin(1999)}]{Deshpande_1999}
\bibinfo{author}{Deshpande, A.A.}, \bibinfo{author}{Rankin, J.M.},
  \bibinfo{year}{1999}.
\newblock \bibinfo{title}{Pulsar magnetospheric emission mapping: Images and
  implications of polar cap weather}.
\newblock \bibinfo{journal}{The Astrophysical Journal} \bibinfo{volume}{524},
  \bibinfo{pages}{1008}.
\newblock \URLprefix \url{https://dx.doi.org/10.1086/307862},
  \DOIprefix\doi{10.1086/307862}.
\bibitem[{Drake and Craft(1968)}]{Drake_1968}
\bibinfo{author}{Drake, F.D.}, \bibinfo{author}{Craft, H.D.},
  \bibinfo{year}{1968}.
\newblock \bibinfo{title}{Second periodic pulsation in pulsars}.
\newblock \bibinfo{journal}{Nature} \bibinfo{volume}{220},
  \bibinfo{pages}{231--235}.
\newblock \URLprefix \url{https://doi.org/10.1038/220231a0},
  \DOIprefix\doi{10.1038/220231a0}.
\bibitem[{Esirkepov(2001)}]{Esirkepov2001}
\bibinfo{author}{Esirkepov, T.}, \bibinfo{year}{2001}.
\newblock \bibinfo{title}{Exact charge conservation scheme for particle-in-cell
  simulation with an arbitrary form-factor}.
\newblock \bibinfo{journal}{Computer Physics Communications}
  \bibinfo{volume}{135}, \bibinfo{pages}{144--153}.
\newblock \URLprefix
  \url{https://www.sciencedirect.com/science/article/pii/S0010465500002289},
  \DOIprefix\doi{https://doi.org/10.1016/S0010-4655(00)00228-9}.
\bibitem[{Espinoza et~al.(2014)Espinoza, Codina and Badia}]{Espinoza2014}
\bibinfo{author}{Espinoza, H.}, \bibinfo{author}{Codina, R.},
  \bibinfo{author}{Badia, S.}, \bibinfo{year}{2014}.
\newblock \bibinfo{title}{A sommerfeld non-reflecting boundary condition for
  the wave equation in mixed form}.
\newblock \bibinfo{journal}{Computer Methods in Applied Mechanics and
  Engineering} \bibinfo{volume}{276}, \bibinfo{pages}{122--148}.
\newblock \URLprefix
  \url{https://www.sciencedirect.com/science/article/pii/S0045782514001017},
  \DOIprefix\doi{https://doi.org/10.1016/j.cma.2014.03.015}.
\bibitem[{{Falcke} and {Rezzolla}(2014)}]{Falcke_2014}
\bibinfo{author}{{Falcke}, H.}, \bibinfo{author}{{Rezzolla}, L.},
  \bibinfo{year}{2014}.
\newblock \bibinfo{title}{{Fast radio bursts: the last sign of supramassive
  neutron stars}}.
\newblock \bibinfo{journal}{\aap} \bibinfo{volume}{562}, \bibinfo{pages}{A137}.
\newblock \DOIprefix\doi{10.1051/0004-6361/201321996},
  \href{http://arxiv.org/abs/1307.1409}{{\tt arXiv:1307.1409}}.
\bibitem[{Fonseca et~al.(2002)Fonseca, Silva, Tsung, Decyk, Lu, Ren, Mori,
  Deng, Lee, Katsouleas and Adam}]{Osiris}
\bibinfo{author}{Fonseca, R.A.}, \bibinfo{author}{Silva, L.O.},
  \bibinfo{author}{Tsung, F.S.}, \bibinfo{author}{Decyk, V.K.},
  \bibinfo{author}{Lu, W.}, \bibinfo{author}{Ren, C.}, \bibinfo{author}{Mori,
  W.B.}, \bibinfo{author}{Deng, S.}, \bibinfo{author}{Lee, S.},
  \bibinfo{author}{Katsouleas, T.}, \bibinfo{author}{Adam, J.C.},
  \bibinfo{year}{2002}.
\newblock \bibinfo{title}{Osiris: A three-dimensional, fully relativistic
  particle in cell code for modeling plasma based accelerators}, in:
  \bibinfo{editor}{Sloot, P.M.A.}, \bibinfo{editor}{Hoekstra, A.G.},
  \bibinfo{editor}{Tan, C.J.K.}, \bibinfo{editor}{Dongarra, J.J.} (Eds.),
  \bibinfo{booktitle}{Computational Science --- ICCS 2002},
  \bibinfo{publisher}{Springer Berlin Heidelberg}, \bibinfo{address}{Berlin,
  Heidelberg}. pp. \bibinfo{pages}{342--351}.
\bibitem[{{Gil, J.} et~al.(2003){Gil, J.}, {Melikidze, G. I.} and {Geppert,
  U.}}]{Gil_2003}
\bibinfo{author}{{Gil, J.}}, \bibinfo{author}{{Melikidze, G. I.}},
  \bibinfo{author}{{Geppert, U.}}, \bibinfo{year}{2003}.
\newblock \bibinfo{title}{Drifting subpulses and inner acceleration regions in
  radio pulsars*}.
\newblock \bibinfo{journal}{A\&A} \bibinfo{volume}{407},
  \bibinfo{pages}{315--324}.
\newblock \URLprefix \url{https://doi.org/10.1051/0004-6361:20030854},
  \DOIprefix\doi{10.1051/0004-6361:20030854}.
\bibitem[{{Goldreich} and {Julian}(1969)}]{Goldreich_1969}
\bibinfo{author}{{Goldreich}, P.}, \bibinfo{author}{{Julian}, W.H.},
  \bibinfo{year}{1969}.
\newblock \bibinfo{title}{{Pulsar Electrodynamics}}.
\newblock \bibinfo{journal}{\apj} \bibinfo{volume}{157}, \bibinfo{pages}{869}.
\newblock \DOIprefix\doi{10.1086/150119}.
\bibitem[{Gralla et~al.(2016)Gralla, Lupsasca and Philippov}]{Gralla_2016}
\bibinfo{author}{Gralla, S.E.}, \bibinfo{author}{Lupsasca, A.},
  \bibinfo{author}{Philippov, A.}, \bibinfo{year}{2016}.
\newblock \bibinfo{title}{{PULSAR} {MAGNETOSPHERES}: {BEYOND} {THE} {FLAT}
  {SPACETIME} {DIPOLE}}.
\newblock \bibinfo{journal}{The Astrophysical Journal} \bibinfo{volume}{833},
  \bibinfo{pages}{258}.
\newblock \URLprefix \url{https://doi.org/10.3847/1538-4357/833/2/258},
  \DOIprefix\doi{10.3847/1538-4357/833/2/258}.
\bibitem[{Grismayer et~al.(2016)Grismayer, Vranic, Martins, Fonseca and
  Silva}]{Grismayer_2016}
\bibinfo{author}{Grismayer, T.}, \bibinfo{author}{Vranic, M.},
  \bibinfo{author}{Martins, J.L.}, \bibinfo{author}{Fonseca, R.A.},
  \bibinfo{author}{Silva, L.O.}, \bibinfo{year}{2016}.
\newblock \bibinfo{title}{{Laser absorption via quantum electrodynamics
  cascades in counter propagating laser pulses}}.
\newblock \bibinfo{journal}{Physics of Plasmas} \bibinfo{volume}{23},
  \bibinfo{pages}{056706}.
\newblock \URLprefix \url{https://doi.org/10.1063/1.4950841},
  \DOIprefix\doi{10.1063/1.4950841},
  \href{http://arxiv.org/abs/https://pubs.aip.org/aip/pop/article-pdf/doi/10.1063/1.4950841/15946479/056706\_1\_online.pdf}{{\tt
  arXiv:https://pubs.aip.org/aip/pop/article-pdf/doi/10.1063/1.4950841/15946479/056706\_1\_online.pdf}}.
\bibitem[{Grismayer et~al.(2017)Grismayer, Vranic, Martins, Fonseca and
  Silva}]{Grismayer_2017}
\bibinfo{author}{Grismayer, T.}, \bibinfo{author}{Vranic, M.},
  \bibinfo{author}{Martins, J.L.}, \bibinfo{author}{Fonseca, R.A.},
  \bibinfo{author}{Silva, L.O.}, \bibinfo{year}{2017}.
\newblock \bibinfo{title}{Seeded qed cascades in counterpropagating laser
  pulses}.
\newblock \bibinfo{journal}{Phys. Rev. E} \bibinfo{volume}{95},
  \bibinfo{pages}{023210}.
\newblock \URLprefix \url{https://link.aps.org/doi/10.1103/PhysRevE.95.023210},
  \DOIprefix\doi{10.1103/PhysRevE.95.023210}.
\bibitem[{Gruzinov(2005)}]{Gruzinov_2005}
\bibinfo{author}{Gruzinov, A.}, \bibinfo{year}{2005}.
\newblock \bibinfo{title}{Power of an axisymmetric pulsar}.
\newblock \bibinfo{journal}{Phys. Rev. Lett.} \bibinfo{volume}{94},
  \bibinfo{pages}{021101}.
\newblock \URLprefix
  \url{https://link.aps.org/doi/10.1103/PhysRevLett.94.021101},
  \DOIprefix\doi{10.1103/PhysRevLett.94.021101}.
\bibitem[{{Gruzinov}(2012)}]{Gruzinov_2012}
\bibinfo{author}{{Gruzinov}, A.}, \bibinfo{year}{2012}.
\newblock \bibinfo{title}{{Pulsar Efficiency}}.
\newblock \bibinfo{journal}{arXiv e-prints} ,
  \bibinfo{pages}{arXiv:1209.5121}\DOIprefix\doi{10.48550/arXiv.1209.5121},
  \href{http://arxiv.org/abs/1209.5121}{{\tt arXiv:1209.5121}}.
\bibitem[{{Gruzinov}(2013)}]{Gruzinov_2013}
\bibinfo{author}{{Gruzinov}, A.}, \bibinfo{year}{2013}.
\newblock \bibinfo{title}{{Aristotelian Electrodynamics solves the Pulsar:
  Lower Efficiency of Strong Pulsars}}.
\newblock \bibinfo{journal}{arXiv e-prints} ,
  \bibinfo{pages}{arXiv:1303.4094}\DOIprefix\doi{10.48550/arXiv.1303.4094},
  \href{http://arxiv.org/abs/1303.4094}{{\tt arXiv:1303.4094}}.
\bibitem[{{Gruzinov}(2015)}]{Gruzinov_2015}
\bibinfo{author}{{Gruzinov}, A.}, \bibinfo{year}{2015}.
\newblock \bibinfo{title}{{No-Hair Theorem for Weak Pulsar}}.
\newblock \bibinfo{journal}{arXiv e-prints} ,
  \bibinfo{pages}{arXiv:1503.05158}\DOIprefix\doi{10.48550/arXiv.1503.05158},
  \href{http://arxiv.org/abs/1503.05158}{{\tt arXiv:1503.05158}}.
\bibitem[{{Gu{\'e}pin} et~al.(2020){Gu{\'e}pin}, {Cerutti} and
  {Kotera}}]{Guepin_2020}
\bibinfo{author}{{Gu{\'e}pin}, C.}, \bibinfo{author}{{Cerutti}, B.},
  \bibinfo{author}{{Kotera}, K.}, \bibinfo{year}{2020}.
\newblock \bibinfo{title}{{Proton acceleration in pulsar magnetospheres}}.
\newblock \bibinfo{journal}{\aap} \bibinfo{volume}{635}, \bibinfo{pages}{A138}.
\newblock \DOIprefix\doi{10.1051/0004-6361/201936816},
  \href{http://arxiv.org/abs/1910.11387}{{\tt arXiv:1910.11387}}.
\bibitem[{Hakobyan et~al.(2019)Hakobyan, Philippov and
  Spitkovsky}]{Hakobyan_2019}
\bibinfo{author}{Hakobyan, H.}, \bibinfo{author}{Philippov, A.},
  \bibinfo{author}{Spitkovsky, A.}, \bibinfo{year}{2019}.
\newblock \bibinfo{title}{Effects of synchrotron cooling and pair production on
  collisionless relativistic reconnection}.
\newblock \bibinfo{journal}{The Astrophysical Journal} \bibinfo{volume}{877},
  \bibinfo{pages}{53}.
\newblock \URLprefix \url{https://dx.doi.org/10.3847/1538-4357/ab191b},
  \DOIprefix\doi{10.3847/1538-4357/ab191b}.
\bibitem[{Hakobyan et~al.(2023)Hakobyan, Philippov and
  Spitkovsky}]{Hakobyan_2023}
\bibinfo{author}{Hakobyan, H.}, \bibinfo{author}{Philippov, A.},
  \bibinfo{author}{Spitkovsky, A.}, \bibinfo{year}{2023}.
\newblock \bibinfo{title}{Magnetic energy dissipation and $\gamma$-ray emission
  in energetic pulsars}.
\newblock \bibinfo{journal}{The Astrophysical Journal} \bibinfo{volume}{943},
  \bibinfo{pages}{105}.
\newblock \URLprefix \url{https://dx.doi.org/10.3847/1538-4357/acab05},
  \DOIprefix\doi{10.3847/1538-4357/acab05}.
\bibitem[{{Hartle}(1967)}]{Hartle1967}
\bibinfo{author}{{Hartle}, J.B.}, \bibinfo{year}{1967}.
\newblock \bibinfo{title}{{Slowly Rotating Relativistic Stars. I. Equations of
  Structure}}.
\newblock \bibinfo{journal}{\apj} \bibinfo{volume}{150}, \bibinfo{pages}{1005}.
\newblock \DOIprefix\doi{10.1086/149400}.
\bibitem[{{Hartle} and {Thorne}(1968)}]{Hartle1968}
\bibinfo{author}{{Hartle}, J.B.}, \bibinfo{author}{{Thorne}, K.S.},
  \bibinfo{year}{1968}.
\newblock \bibinfo{title}{{Slowly Rotating Relativistic Stars. II. Models for
  Neutron Stars and Supermassive Stars}}.
\newblock \bibinfo{journal}{\apj} \bibinfo{volume}{153}, \bibinfo{pages}{807}.
\newblock \DOIprefix\doi{10.1086/149707}.
\bibitem[{Hewish et~al.(1968)Hewish, BELL, PILKINGTON, SCOTT and
  COLLINS}]{Hewish_1968}
\bibinfo{author}{Hewish, A.}, \bibinfo{author}{BELL, S.J.},
  \bibinfo{author}{PILKINGTON, J.D.H.}, \bibinfo{author}{SCOTT, P.F.},
  \bibinfo{author}{COLLINS, R.A.}, \bibinfo{year}{1968}.
\newblock \bibinfo{title}{Observation of a rapidly pulsating radio source}.
\newblock \bibinfo{journal}{Nature} \bibinfo{volume}{217},
  \bibinfo{pages}{709--713}.
\newblock \URLprefix \url{https://doi.org/10.1038/217709a0},
  \DOIprefix\doi{10.1038/217709a0}.
\bibitem[{Higuera and Cary(2017)}]{Higuera2017}
\bibinfo{author}{Higuera, A.V.}, \bibinfo{author}{Cary, J.R.},
  \bibinfo{year}{2017}.
\newblock \bibinfo{title}{{Structure-preserving second-order integration of
  relativistic charged particle trajectories in electromagnetic fields}}.
\newblock \bibinfo{journal}{Physics of Plasmas} \bibinfo{volume}{24}.
\newblock \URLprefix \url{https://doi.org/10.1063/1.4979989},
  \DOIprefix\doi{10.1063/1.4979989}. \bibinfo{note}{052104}.
\bibitem[{Hirotani(2008)}]{Hirotani_2008}
\bibinfo{author}{Hirotani, K.}, \bibinfo{year}{2008}.
\newblock \bibinfo{title}{Outer-gap versus slot-gap models for pulsar
  high-energy emissions: The case of the crab pulsar}.
\newblock \bibinfo{journal}{The Astrophysical Journal} \bibinfo{volume}{688},
  \bibinfo{pages}{L25}.
\newblock \URLprefix \url{https://dx.doi.org/10.1086/595000},
  \DOIprefix\doi{10.1086/595000}.
\bibitem[{Hu and Beloborodov(2022)}]{Hu_2022}
\bibinfo{author}{Hu, R.}, \bibinfo{author}{Beloborodov, A.M.},
  \bibinfo{year}{2022}.
\newblock \bibinfo{title}{Axisymmetric pulsar magnetosphere revisited}.
\newblock \bibinfo{journal}{The Astrophysical Journal} \bibinfo{volume}{939},
  \bibinfo{pages}{42}.
\newblock \URLprefix \url{https://dx.doi.org/10.3847/1538-4357/ac961d},
  \DOIprefix\doi{10.3847/1538-4357/ac961d}.
\bibitem[{Huang et~al.(2018)Huang, Pan and Yu}]{Huang_2018}
\bibinfo{author}{Huang, L.}, \bibinfo{author}{Pan, Z.}, \bibinfo{author}{Yu,
  C.}, \bibinfo{year}{2018}.
\newblock \bibinfo{title}{{An improved algorithm for crossing curved light
  surfaces: rapidly rotating pulsar magnetospheres in curved space–time}}.
\newblock \bibinfo{journal}{Monthly Notices of the Royal Astronomical Society}
  \bibinfo{volume}{479}, \bibinfo{pages}{4499--4508}.
\newblock \URLprefix \url{https://doi.org/10.1093/mnras/sty1761},
  \DOIprefix\doi{10.1093/mnras/sty1761},
  \href{http://arxiv.org/abs/https://academic.oup.com/mnras/article-pdf/479/4/4499/25180510/sty1761.pdf}{{\tt
  arXiv:https://academic.oup.com/mnras/article-pdf/479/4/4499/25180510/sty1761.pdf}}.
\bibitem[{Jun-Tao et~al.(2021)Jun-Tao, Xin, Quan-Wei, Qi-Jun, Lun-Hua,
  Yuan-Jie, You-Li, Shi-Jun, Shuang-Qiang, Ru-Shuang, Ai-Jun and
  Guo-Jun}]{Tao_2021}
\bibinfo{author}{Jun-Tao, B.}, \bibinfo{author}{Xin, X.},
  \bibinfo{author}{Quan-Wei, L.}, \bibinfo{author}{Qi-Jun, Z.},
  \bibinfo{author}{Lun-Hua, S.}, \bibinfo{author}{Yuan-Jie, D.},
  \bibinfo{author}{You-Li, T.}, \bibinfo{author}{Shi-Jun, D.},
  \bibinfo{author}{Shuang-Qiang, W.}, \bibinfo{author}{Ru-Shuang, Z.},
  \bibinfo{author}{Ai-Jun, D.}, \bibinfo{author}{Guo-Jun, Q.},
  \bibinfo{year}{2021}.
\newblock \bibinfo{title}{Modeling multiband x-ray light curves of the crab
  pulsar with the annular and core gap models}.
\newblock \bibinfo{journal}{New Astronomy} \bibinfo{volume}{83},
  \bibinfo{pages}{101480}.
\newblock \URLprefix
  \url{https://www.sciencedirect.com/science/article/pii/S138410762030227X},
  \DOIprefix\doi{https://doi.org/10.1016/j.newast.2020.101480}.
\bibitem[{Kalapotharakos et~al.(2018)Kalapotharakos, Brambilla, Timokhin,
  Harding and Kazanas}]{Kalapotharakos_2018}
\bibinfo{author}{Kalapotharakos, C.}, \bibinfo{author}{Brambilla, G.},
  \bibinfo{author}{Timokhin, A.}, \bibinfo{author}{Harding, A.K.},
  \bibinfo{author}{Kazanas, D.}, \bibinfo{year}{2018}.
\newblock \bibinfo{title}{Three-dimensional kinetic pulsar magnetosphere
  models: Connecting to gamma-ray observations}.
\newblock \bibinfo{journal}{The Astrophysical Journal} \bibinfo{volume}{857},
  \bibinfo{pages}{44}.
\newblock \URLprefix \url{https://dx.doi.org/10.3847/1538-4357/aab550},
  \DOIprefix\doi{10.3847/1538-4357/aab550}.
\bibitem[{{Kalapotharakos, C.} and {Contopoulos,
  I.}(2009)}]{Kalapotharakos_2009}
\bibinfo{author}{{Kalapotharakos, C.}}, \bibinfo{author}{{Contopoulos, I.}},
  \bibinfo{year}{2009}.
\newblock \bibinfo{title}{Three-dimensional numerical simulations of the pulsar
  magnetosphere: preliminary results}.
\newblock \bibinfo{journal}{A\&A} \bibinfo{volume}{496},
  \bibinfo{pages}{495--502}.
\newblock \URLprefix \url{https://doi.org/10.1051/0004-6361:200810281},
  \DOIprefix\doi{10.1051/0004-6361:200810281}.
\bibitem[{Kalogera and Baym(1996)}]{Kalogera_1996}
\bibinfo{author}{Kalogera, V.}, \bibinfo{author}{Baym, G.},
  \bibinfo{year}{1996}.
\newblock \bibinfo{title}{The maximum mass of a neutron star}.
\newblock \bibinfo{journal}{The Astrophysical Journal} \bibinfo{volume}{470},
  \bibinfo{pages}{L61}.
\newblock \URLprefix \url{https://dx.doi.org/10.1086/310296},
  \DOIprefix\doi{10.1086/310296}.
\bibitem[{Kirk et~al.(2009)Kirk, Lyubarsky and Petri}]{Kirk_2009}
\bibinfo{author}{Kirk, J.G.}, \bibinfo{author}{Lyubarsky, Y.},
  \bibinfo{author}{Petri, J.}, \bibinfo{year}{2009}.
\newblock \bibinfo{title}{The Theory of Pulsar Winds and Nebulae}.
  \bibinfo{publisher}{Springer Berlin Heidelberg}, \bibinfo{address}{Berlin,
  Heidelberg}.
\newblock pp. \bibinfo{pages}{421--450}.
\newblock \URLprefix \url{https://doi.org/10.1007/978-3-540-76965-1$\_$16},
  \DOIprefix\doi{10.1007/978-3-540-76965-1$\_$16}.
\bibitem[{Komissarov(2011)}]{Komissarov2011}
\bibinfo{author}{Komissarov, S.S.}, \bibinfo{year}{2011}.
\newblock \bibinfo{title}{{3+1 magnetodynamics}}.
\newblock \bibinfo{journal}{Monthly Notices of the Royal Astronomical Society:
  Letters} \bibinfo{volume}{418}, \bibinfo{pages}{L94--L98}.
\newblock \URLprefix \url{https://doi.org/10.1111/j.1745-3933.2011.01150.x},
  \DOIprefix\doi{10.1111/j.1745-3933.2011.01150.x},
  \href{http://arxiv.org/abs/https://academic.oup.com/mnrasl/article-pdf/418/1/L94/6418894/418-1-L94.pdf}{{\tt
  arXiv:https://academic.oup.com/mnrasl/article-pdf/418/1/L94/6418894/418-1-L94.pdf}}.
\bibitem[{Kramer et~al.(1998)Kramer, Xilouris, Lorimer, Doroshenko, Jessner,
  Wielebinski, Wolszczan and Camilo}]{Kramer_1998}
\bibinfo{author}{Kramer, M.}, \bibinfo{author}{Xilouris, K.M.},
  \bibinfo{author}{Lorimer, D.R.}, \bibinfo{author}{Doroshenko, O.},
  \bibinfo{author}{Jessner, A.}, \bibinfo{author}{Wielebinski, R.},
  \bibinfo{author}{Wolszczan, A.}, \bibinfo{author}{Camilo, F.},
  \bibinfo{year}{1998}.
\newblock \bibinfo{title}{The characteristics of millisecond pulsar emission.
  i. spectra, pulse shapes, and the beaming fraction}.
\newblock \bibinfo{journal}{The Astrophysical Journal} \bibinfo{volume}{501},
  \bibinfo{pages}{270}.
\newblock \URLprefix \url{https://dx.doi.org/10.1086/305790},
  \DOIprefix\doi{10.1086/305790}.
\bibitem[{Landau and Lifschitz(1975)}]{LL_1975}
\bibinfo{author}{Landau, L.D.}, \bibinfo{author}{Lifschitz, E.M.},
  \bibinfo{year}{1975}.
\newblock \bibinfo{title}{{The Classical Theory of Fields}}. volume
  \bibinfo{volume}{Volume 2} of \textit{\bibinfo{series}{Course of Theoretical
  Physics}}.
\newblock \bibinfo{publisher}{Pergamon Press}, \bibinfo{address}{Oxford}.
\bibitem[{Li et~al.(2012)Li, Spitkovsky and Tchekhovskoy}]{Li_2012}
\bibinfo{author}{Li, J.}, \bibinfo{author}{Spitkovsky, A.},
  \bibinfo{author}{Tchekhovskoy, A.}, \bibinfo{year}{2012}.
\newblock \bibinfo{title}{On the spin-down of intermittent pulsars}.
\newblock \bibinfo{journal}{The Astrophysical Journal Letters}
  \bibinfo{volume}{746}, \bibinfo{pages}{L24}.
\newblock \URLprefix \url{https://dx.doi.org/10.1088/2041-8205/746/2/L24},
  \DOIprefix\doi{10.1088/2041-8205/746/2/L24}.
\bibitem[{{Lyubarsky}(2009)}]{Lyubarsky_2009}
\bibinfo{author}{{Lyubarsky}, Y.}, \bibinfo{year}{2009}.
\newblock \bibinfo{title}{{Adjustment of the Electric Charge and Current in
  Pulsar Magnetospheres}}.
\newblock \bibinfo{journal}{\apj} \bibinfo{volume}{696},
  \bibinfo{pages}{320--327}.
\newblock \DOIprefix\doi{10.1088/0004-637X/696/1/320},
  \href{http://arxiv.org/abs/0904.2446}{{\tt arXiv:0904.2446}}.
\bibitem[{Lyutikov(2011)}]{Lyutikov_2011}
\bibinfo{author}{Lyutikov, M.}, \bibinfo{year}{2011}.
\newblock \bibinfo{title}{Electromagnetic power of merging and collapsing
  compact objects}.
\newblock \bibinfo{journal}{Phys. Rev. D} \bibinfo{volume}{83},
  \bibinfo{pages}{124035}.
\newblock \URLprefix \url{https://link.aps.org/doi/10.1103/PhysRevD.83.124035},
  \DOIprefix\doi{10.1103/PhysRevD.83.124035}.
\bibitem[{Medin and Lai(2010)}]{Medin_2010}
\bibinfo{author}{Medin, Z.}, \bibinfo{author}{Lai, D.}, \bibinfo{year}{2010}.
\newblock \bibinfo{title}{{Pair cascades in the magnetospheres of strongly
  magnetized neutron stars}}.
\newblock \bibinfo{journal}{Monthly Notices of the Royal Astronomical Society}
  \bibinfo{volume}{406}, \bibinfo{pages}{1379--1404}.
\newblock \URLprefix \url{https://doi.org/10.1111/j.1365-2966.2010.16776.x},
  \DOIprefix\doi{10.1111/j.1365-2966.2010.16776.x},
  \href{http://arxiv.org/abs/https://academic.oup.com/mnras/article-pdf/406/2/1379/18718642/mnr0406-1379.pdf}{{\tt
  arXiv:https://academic.oup.com/mnras/article-pdf/406/2/1379/18718642/mnr0406-1379.pdf}}.
\bibitem[{{Michel}(1973)}]{Michel_1973}
\bibinfo{author}{{Michel}, F.C.}, \bibinfo{year}{1973}.
\newblock \bibinfo{title}{{Rotating Magnetospheres: an Exact 3-D Solution}}.
\newblock \bibinfo{journal}{\apjl} \bibinfo{volume}{180},
  \bibinfo{pages}{L133}.
\newblock \DOIprefix\doi{10.1086/181169}.
\bibitem[{Mitchell and Gelfand(2022)}]{Mitchell_2022}
\bibinfo{author}{Mitchell, A.M.W.}, \bibinfo{author}{Gelfand, J.},
  \bibinfo{year}{2022}.
\newblock \bibinfo{title}{Pulsar Wind Nebulae}. \bibinfo{publisher}{Springer
  Nature Singapore}, \bibinfo{address}{Singapore}.
\newblock pp. \bibinfo{pages}{1--52}.
\newblock \URLprefix \url{https://doi.org/10.1007/978-981-16-4544-0$\_$157-1},
  \DOIprefix\doi{10.1007/978-981-16-4544-0$\_$157-1}.
\bibitem[{Mur(1981)}]{Mur_1981}
\bibinfo{author}{Mur, G.}, \bibinfo{year}{1981}.
\newblock \bibinfo{title}{Absorbing boundary conditions for the
  finite-difference approximation of the time-domain electromagnetic-field
  equations}.
\newblock \bibinfo{journal}{IEEE Transactions on Electromagnetic Compatibility}
  \bibinfo{volume}{EMC-23}, \bibinfo{pages}{377--382}.
\newblock \DOIprefix\doi{10.1109/TEMC.1981.303970}.
\bibitem[{Muslimov and Harding(2003)}]{Muslimov_2003}
\bibinfo{author}{Muslimov, A.G.}, \bibinfo{author}{Harding, A.K.},
  \bibinfo{year}{2003}.
\newblock \bibinfo{title}{Extended acceleration in slot gaps and pulsar
  high-energy emission}.
\newblock \bibinfo{journal}{The Astrophysical Journal} \bibinfo{volume}{588},
  \bibinfo{pages}{430}.
\newblock \URLprefix \url{https://dx.doi.org/10.1086/368162},
  \DOIprefix\doi{10.1086/368162}.
\bibitem[{Muslimov and Harding(2004)}]{Muslimov_2004}
\bibinfo{author}{Muslimov, A.G.}, \bibinfo{author}{Harding, A.K.},
  \bibinfo{year}{2004}.
\newblock \bibinfo{title}{High-altitude particle acceleration and radiation in
  pulsar slot gaps}.
\newblock \bibinfo{journal}{The Astrophysical Journal} \bibinfo{volume}{606},
  \bibinfo{pages}{1143}.
\newblock \URLprefix \url{https://dx.doi.org/10.1086/383079},
  \DOIprefix\doi{10.1086/383079}.
\bibitem[{Muslimov and Tsygan(1992)}]{Muslimov_1992}
\bibinfo{author}{Muslimov, A.G.}, \bibinfo{author}{Tsygan, A.I.},
  \bibinfo{year}{1992}.
\newblock \bibinfo{title}{{General relativistic electric potential drops above
  pulsar polar caps}}.
\newblock \bibinfo{journal}{Monthly Notices of the Royal Astronomical Society}
  \bibinfo{volume}{255}, \bibinfo{pages}{61--70}.
\newblock \URLprefix \url{https://doi.org/10.1093/mnras/255.1.61},
  \DOIprefix\doi{10.1093/mnras/255.1.61},
  \href{http://arxiv.org/abs/https://academic.oup.com/mnras/article-pdf/255/1/61/18523363/mnras255-0061.pdf}{{\tt
  arXiv:https://academic.oup.com/mnras/article-pdf/255/1/61/18523363/mnras255-0061.pdf}}.
\bibitem[{Novak and Bonazzola(2004)}]{Novak2004}
\bibinfo{author}{Novak, J.}, \bibinfo{author}{Bonazzola, S.},
  \bibinfo{year}{2004}.
\newblock \bibinfo{title}{Absorbing boundary conditions for simulation of
  gravitational waves with spectral methods in spherical coordinates}.
\newblock \bibinfo{journal}{Journal of Computational Physics}
  \bibinfo{volume}{197}, \bibinfo{pages}{186--196}.
\newblock \URLprefix
  \url{https://www.sciencedirect.com/science/article/pii/S0021999103006259},
  \DOIprefix\doi{https://doi.org/10.1016/j.jcp.2003.11.027}.
\bibitem[{Parfrey et~al.(2019)Parfrey, Philippov and Cerutti}]{Parfrey_2019}
\bibinfo{author}{Parfrey, K.}, \bibinfo{author}{Philippov, A.},
  \bibinfo{author}{Cerutti, B.}, \bibinfo{year}{2019}.
\newblock \bibinfo{title}{First-principles plasma simulations of black-hole jet
  launching}.
\newblock \bibinfo{journal}{Phys. Rev. Lett.} \bibinfo{volume}{122},
  \bibinfo{pages}{035101}.
\newblock \URLprefix
  \url{https://link.aps.org/doi/10.1103/PhysRevLett.122.035101},
  \DOIprefix\doi{10.1103/PhysRevLett.122.035101}.
\bibitem[{{Philippov} and {Kramer}(2022)}]{Kramer_2022}
\bibinfo{author}{{Philippov}, A.}, \bibinfo{author}{{Kramer}, M.},
  \bibinfo{year}{2022}.
\newblock \bibinfo{title}{{Pulsar Magnetospheres and Their Radiation}}.
\newblock \bibinfo{journal}{\araa} \bibinfo{volume}{60},
  \bibinfo{pages}{495--558}.
\newblock \DOIprefix\doi{10.1146/annurev-astro-052920-112338}.
\bibitem[{Philippov et~al.(2020)Philippov, Timokhin and
  Spitkovsky}]{Philippov_2020}
\bibinfo{author}{Philippov, A.}, \bibinfo{author}{Timokhin, A.},
  \bibinfo{author}{Spitkovsky, A.}, \bibinfo{year}{2020}.
\newblock \bibinfo{title}{Origin of pulsar radio emission}.
\newblock \bibinfo{journal}{Phys. Rev. Lett.} \bibinfo{volume}{124},
  \bibinfo{pages}{245101}.
\newblock \URLprefix
  \url{https://link.aps.org/doi/10.1103/PhysRevLett.124.245101},
  \DOIprefix\doi{10.1103/PhysRevLett.124.245101}.
\bibitem[{Philippov et~al.(2015a)Philippov, Cerutti, Tchekhovskoy and
  Spitkovsky}]{Philippov_2015a}
\bibinfo{author}{Philippov, A.A.}, \bibinfo{author}{Cerutti, B.},
  \bibinfo{author}{Tchekhovskoy, A.}, \bibinfo{author}{Spitkovsky, A.},
  \bibinfo{year}{2015}a.
\newblock \bibinfo{title}{{AB} {INITIO} {PULSAR} {MAGNETOSPHERE}: {THE} {ROLE}
  {OF} {GENERAL} {RELATIVITY}}.
\newblock \bibinfo{journal}{The Astrophysical Journal} \bibinfo{volume}{815},
  \bibinfo{pages}{L19}.
\newblock \URLprefix \url{https://doi.org/10.1088/2041-8205/815/2/l19},
  \DOIprefix\doi{10.1088/2041-8205/815/2/l19}.
\bibitem[{Philippov and Spitkovsky(2014)}]{Philippov_2014a}
\bibinfo{author}{Philippov, A.A.}, \bibinfo{author}{Spitkovsky, A.},
  \bibinfo{year}{2014}.
\newblock \bibinfo{title}{{AB} {INITIO} {PULSAR} {MAGNETOSPHERE}:
  {THREE}-{DIMENSIONAL} {PARTICLE}-{IN}-{CELL} {SIMULATIONS} {OF}
  {AXISYMMETRIC} {PULSARS}}.
\newblock \bibinfo{journal}{The Astrophysical Journal} \bibinfo{volume}{785},
  \bibinfo{pages}{L33}.
\newblock \URLprefix \url{https://doi.org/10.1088/2041-8205/785/2/l33},
  \DOIprefix\doi{10.1088/2041-8205/785/2/l33}.
\bibitem[{Philippov and Spitkovsky(2018)}]{Philippov_2018}
\bibinfo{author}{Philippov, A.A.}, \bibinfo{author}{Spitkovsky, A.},
  \bibinfo{year}{2018}.
\newblock \bibinfo{title}{Ab-initio pulsar magnetosphere: Particle acceleration
  in oblique rotators and high-energy emission modeling}.
\newblock \bibinfo{journal}{The Astrophysical Journal} \bibinfo{volume}{855},
  \bibinfo{pages}{94}.
\newblock \URLprefix \url{https://doi.org/10.3847/1538-4357/aaabbc},
  \DOIprefix\doi{10.3847/1538-4357/aaabbc}.
\bibitem[{Philippov et~al.(2015b)Philippov, Spitkovsky and
  Cerutti}]{Philippov_2015b}
\bibinfo{author}{Philippov, A.A.}, \bibinfo{author}{Spitkovsky, A.},
  \bibinfo{author}{Cerutti, B.}, \bibinfo{year}{2015}b.
\newblock \bibinfo{title}{{AB} {INITIO} {PULSAR} {MAGNETOSPHERE}:
  {THREE}-{DIMENSIONAL} {PARTICLE}-{IN}-{CELL} {SIMULATIONS} {OF} {OBLIQUE}
  {PULSARS}}.
\newblock \bibinfo{journal}{The Astrophysical Journal} \bibinfo{volume}{801},
  \bibinfo{pages}{L19}.
\newblock \URLprefix \url{https://doi.org/10.1088/2041-8205/801/1/l19},
  \DOIprefix\doi{10.1088/2041-8205/801/1/l19}.
\bibitem[{Pétri(2012)}]{Petri_2012}
\bibinfo{author}{Pétri, J.}, \bibinfo{year}{2012}.
\newblock \bibinfo{title}{{The pulsar force-free magnetosphere linked to its
  striped wind: time-dependent pseudo-spectral simulations}}.
\newblock \bibinfo{journal}{Monthly Notices of the Royal Astronomical Society}
  \bibinfo{volume}{424}, \bibinfo{pages}{605--619}.
\newblock \URLprefix \url{https://doi.org/10.1111/j.1365-2966.2012.21238.x},
  \DOIprefix\doi{10.1111/j.1365-2966.2012.21238.x},
  \href{http://arxiv.org/abs/https://academic.oup.com/mnras/article-pdf/424/1/605/3290757/mnras0424-0605.pdf}{{\tt
  arXiv:https://academic.oup.com/mnras/article-pdf/424/1/605/3290757/mnras0424-0605.pdf}}.
\bibitem[{Qiao et~al.(2004)Qiao, Lee, Zhang, Xu and Wang}]{Qiao_2004}
\bibinfo{author}{Qiao, G.J.}, \bibinfo{author}{Lee, K.J.},
  \bibinfo{author}{Zhang, B.}, \bibinfo{author}{Xu, R.X.},
  \bibinfo{author}{Wang, H.G.}, \bibinfo{year}{2004}.
\newblock \bibinfo{title}{A model for the challenging “bi-drifting”
  phenomenon in psr j0815+09}.
\newblock \bibinfo{journal}{The Astrophysical Journal} \bibinfo{volume}{616},
  \bibinfo{pages}{L127}.
\newblock \URLprefix \url{https://dx.doi.org/10.1086/426862},
  \DOIprefix\doi{10.1086/426862}.
\bibitem[{{Ravenhall} and {Pethick}(1994)}]{Ravenhall1994}
\bibinfo{author}{{Ravenhall}, D.G.}, \bibinfo{author}{{Pethick}, C.J.},
  \bibinfo{year}{1994}.
\newblock \bibinfo{title}{{Neutron Star Moments of Inertia}}.
\newblock \bibinfo{journal}{\apj} \bibinfo{volume}{424}, \bibinfo{pages}{846}.
\newblock \DOIprefix\doi{10.1086/173935}.
\bibitem[{Ren-xin and Guo-jun(1998)}]{Xu_1998}
\bibinfo{author}{Ren-xin, X.}, \bibinfo{author}{Guo-jun, Q.},
  \bibinfo{year}{1998}.
\newblock \bibinfo{title}{`bare' strange stars might not be bare}.
\newblock \bibinfo{journal}{Chinese Physics Letters} \bibinfo{volume}{15},
  \bibinfo{pages}{934}.
\newblock \URLprefix \url{https://dx.doi.org/10.1088/0256-307X/15/12/026},
  \DOIprefix\doi{10.1088/0256-307X/15/12/026}.
\bibitem[{Rezzolla et~al.(2001)Rezzolla, Ahmedov and Miller}]{Rezzolla2001a}
\bibinfo{author}{Rezzolla, L.}, \bibinfo{author}{Ahmedov, B.J.},
  \bibinfo{author}{Miller, J.C.}, \bibinfo{year}{2001}.
\newblock \bibinfo{title}{{General relativistic electromagnetic fields of a
  slowly rotating magnetized neutron star — I. Formulation of the
  equations}}.
\newblock \bibinfo{journal}{Monthly Notices of the Royal Astronomical Society}
  \bibinfo{volume}{322}, \bibinfo{pages}{723--740}.
\newblock \URLprefix \url{https://doi.org/10.1046/j.1365-8711.2001.04161.x},
  \DOIprefix\doi{10.1046/j.1365-8711.2001.04161.x},
  \href{http://arxiv.org/abs/https://academic.oup.com/mnras/article-pdf/322/4/723/3323790/322-4-723.pdf}{{\tt
  arXiv:https://academic.oup.com/mnras/article-pdf/322/4/723/3323790/322-4-723.pdf}}.
\bibitem[{Ripperda et~al.(2018)Ripperda, Bacchini, Teunissen, Xia, Porth,
  Sironi, Lapenta and Keppens}]{Ripperda_2018}
\bibinfo{author}{Ripperda, B.}, \bibinfo{author}{Bacchini, F.},
  \bibinfo{author}{Teunissen, J.}, \bibinfo{author}{Xia, C.},
  \bibinfo{author}{Porth, O.}, \bibinfo{author}{Sironi, L.},
  \bibinfo{author}{Lapenta, G.}, \bibinfo{author}{Keppens, R.},
  \bibinfo{year}{2018}.
\newblock \bibinfo{title}{A comprehensive comparison of relativistic particle
  integrators}.
\newblock \bibinfo{journal}{The Astrophysical Journal Supplement Series}
  \bibinfo{volume}{235}, \bibinfo{pages}{21}.
\newblock \URLprefix \url{https://dx.doi.org/10.3847/1538-4365/aab114},
  \DOIprefix\doi{10.3847/1538-4365/aab114}.
\bibitem[{{Romani}(1996)}]{Romani_1996}
\bibinfo{author}{{Romani}, R.W.}, \bibinfo{year}{1996}.
\newblock \bibinfo{title}{{Gamma-Ray Pulsars: Radiation Processes in the Outer
  Magnetosphere}}.
\newblock \bibinfo{journal}{\apj} \bibinfo{volume}{470}, \bibinfo{pages}{469}.
\newblock \DOIprefix\doi{10.1086/177878}.
\bibitem[{Romani et~al.(2022)Romani, Kandel, Filippenko, Brink and
  Zheng}]{Romani_2022}
\bibinfo{author}{Romani, R.W.}, \bibinfo{author}{Kandel, D.},
  \bibinfo{author}{Filippenko, A.V.}, \bibinfo{author}{Brink, T.G.},
  \bibinfo{author}{Zheng, W.}, \bibinfo{year}{2022}.
\newblock \bibinfo{title}{Psr j0952-0607: The fastest and heaviest known
  galactic neutron star}.
\newblock \bibinfo{journal}{The Astrophysical Journal Letters}
  \bibinfo{volume}{934}, \bibinfo{pages}{L17}.
\newblock \URLprefix \url{https://dx.doi.org/10.3847/2041-8213/ac8007},
  \DOIprefix\doi{10.3847/2041-8213/ac8007}.
\bibitem[{{Ruderman} and {Sutherland}(1975)}]{Ruderman_1975}
\bibinfo{author}{{Ruderman}, M.A.}, \bibinfo{author}{{Sutherland}, P.G.},
  \bibinfo{year}{1975}.
\newblock \bibinfo{title}{{Theory of pulsars: polar gaps, sparks, and coherent
  microwave radiation.}}
\newblock \bibinfo{journal}{\apj} \bibinfo{volume}{196},
  \bibinfo{pages}{51--72}.
\newblock \DOIprefix\doi{10.1086/153393}.
\bibitem[{{Scharlemann} and {Wagoner}(1973)}]{Scharlemann_1973}
\bibinfo{author}{{Scharlemann}, E.T.}, \bibinfo{author}{{Wagoner}, R.V.},
  \bibinfo{year}{1973}.
\newblock \bibinfo{title}{{Aligned Rotating Magnetospheres. General Analysis}}.
\newblock \bibinfo{journal}{\apj} \bibinfo{volume}{182},
  \bibinfo{pages}{951--960}.
\newblock \DOIprefix\doi{10.1086/152195}.
\bibitem[{Schoeffler et~al.(2023)Schoeffler, Grismayer, Uzdensky and
  Silva}]{Schoeffler_2023}
\bibinfo{author}{Schoeffler, K.M.}, \bibinfo{author}{Grismayer, T.},
  \bibinfo{author}{Uzdensky, D.}, \bibinfo{author}{Silva, L.O.},
  \bibinfo{year}{2023}.
\newblock \bibinfo{title}{{High-energy synchrotron flares powered by strongly
  radiative relativistic magnetic reconnection: 2D and 3D PIC simulations}}.
\newblock \bibinfo{journal}{Monthly Notices of the Royal Astronomical Society}
  \bibinfo{volume}{523}, \bibinfo{pages}{3812--3839}.
\newblock \URLprefix \url{https://doi.org/10.1093/mnras/stad1588},
  \DOIprefix\doi{10.1093/mnras/stad1588},
  \href{http://arxiv.org/abs/https://academic.oup.com/mnras/article-pdf/523/3/3812/50604710/stad1588.pdf}{{\tt
  arXiv:https://academic.oup.com/mnras/article-pdf/523/3/3812/50604710/stad1588.pdf}}.
\bibitem[{{Selvi} et~al.(2024){Selvi}, {Porth}, {Ripperda} and
  {Sironi}}]{Selvi_2024}
\bibinfo{author}{{Selvi}, S.}, \bibinfo{author}{{Porth}, O.},
  \bibinfo{author}{{Ripperda}, B.}, \bibinfo{author}{{Sironi}, L.},
  \bibinfo{year}{2024}.
\newblock \bibinfo{title}{{Current sheet alignment in oblique black hole
  magnetospheres -- a black hole pulsar?}}
\newblock \bibinfo{journal}{arXiv e-prints} ,
  \bibinfo{pages}{arXiv:2402.16055}\DOIprefix\doi{10.48550/arXiv.2402.16055},
  \href{http://arxiv.org/abs/2402.16055}{{\tt arXiv:2402.16055}}.
\bibitem[{Shinya and Shinpei(2012)}]{Shibata_2012}
\bibinfo{author}{Shinya, Y.}, \bibinfo{author}{Shinpei, S.},
  \bibinfo{year}{2012}.
\newblock \bibinfo{title}{{A Particle Simulation for the Pulsar Magnetosphere:
  Relationship of Polar Cap, Slot Gap, and Outer Gap}}.
\newblock \bibinfo{journal}{Publications of the Astronomical Society of Japan}
  \bibinfo{volume}{64}, \bibinfo{pages}{43}.
\newblock \URLprefix \url{https://doi.org/10.1093/pasj/64.3.43},
  \DOIprefix\doi{10.1093/pasj/64.3.43},
  \href{http://arxiv.org/abs/https://academic.oup.com/pasj/article-pdf/64/3/43/54690438/pasj\_64\_3\_43.pdf}{{\tt
  arXiv:https://academic.oup.com/pasj/article-pdf/64/3/43/54690438/pasj\_64\_3\_43.pdf}}.
\bibitem[{Sommerfeld(1949)}]{Sommerfeld1949}
\bibinfo{author}{Sommerfeld, A.}, \bibinfo{year}{1949}.
\newblock \bibinfo{title}{Partial Differential Equations in Physics}.
\newblock \bibinfo{publisher}{Elsevier}.
\newblock \URLprefix \url{https://doi.org/10.1016/b978-0-12-654658-3.x5001-0},
  \DOIprefix\doi{10.1016/b978-0-12-654658-3.x5001-0}.
\bibitem[{{Spitkovsky}(2006)}]{Spitkovsky_2006}
\bibinfo{author}{{Spitkovsky}, A.}, \bibinfo{year}{2006}.
\newblock \bibinfo{title}{{Time-dependent Force-free Pulsar Magnetospheres:
  Axisymmetric and Oblique Rotators}}.
\newblock \bibinfo{journal}{\apjl} \bibinfo{volume}{648},
  \bibinfo{pages}{L51--L54}.
\newblock \DOIprefix\doi{10.1086/507518},
  \href{http://arxiv.org/abs/astro-ph/0603147}{{\tt arXiv:astro-ph/0603147}}.
\bibitem[{Strang(1968)}]{Strang68}
\bibinfo{author}{Strang, G.}, \bibinfo{year}{1968}.
\newblock \bibinfo{title}{On the construction and comparison of difference
  schemes}.
\newblock \bibinfo{journal}{SIAM Journal on Numerical Analysis}
  \bibinfo{volume}{5}, \bibinfo{pages}{506--517}.
\newblock \URLprefix \url{https://doi.org/10.1137/0705041},
  \DOIprefix\doi{10.1137/0705041},
  \href{http://arxiv.org/abs/https://doi.org/10.1137/0705041}{{\tt
  arXiv:https://doi.org/10.1137/0705041}}.
\bibitem[{Takata et~al.(2004)Takata, Shibata and Hirotani}]{Takata_2004}
\bibinfo{author}{Takata, J.}, \bibinfo{author}{Shibata, S.},
  \bibinfo{author}{Hirotani, K.}, \bibinfo{year}{2004}.
\newblock \bibinfo{title}{{A pulsar outer gap model with trans‐field
  structure}}.
\newblock \bibinfo{journal}{Monthly Notices of the Royal Astronomical Society}
  \bibinfo{volume}{354}, \bibinfo{pages}{1120--1132}.
\newblock \URLprefix \url{https://doi.org/10.1111/j.1365-2966.2004.08270.x},
  \DOIprefix\doi{10.1111/j.1365-2966.2004.08270.x},
  \href{http://arxiv.org/abs/https://academic.oup.com/mnras/article-pdf/354/4/1120/3586427/354-4-1120.pdf}{{\tt
  arXiv:https://academic.oup.com/mnras/article-pdf/354/4/1120/3586427/354-4-1120.pdf}}.
\bibitem[{{Thorne} and {MacDonald}(1982)}]{Thorne1982}
\bibinfo{author}{{Thorne}, K.S.}, \bibinfo{author}{{MacDonald}, D.},
  \bibinfo{year}{1982}.
\newblock \bibinfo{title}{{Electrodynamics in Curved Spacetime - 3+1
  Formulation}}.
\newblock \bibinfo{journal}{\mnras} \bibinfo{volume}{198},
  \bibinfo{pages}{339}.
\newblock \DOIprefix\doi{10.1093/mnras/198.2.339}.
\bibitem[{Timokhin(2007)}]{Timokhin_2007}
\bibinfo{author}{Timokhin, A.N.}, \bibinfo{year}{2007}.
\newblock \bibinfo{title}{Force-free magnetosphere of an aligned rotator with
  differential rotation of open magnetic field lines}.
\newblock \bibinfo{journal}{Astrophysics and Space Science}
  \bibinfo{volume}{308}, \bibinfo{pages}{575--579}.
\newblock \URLprefix \url{https://doi.org/10.1007/s10509-007-9348-4},
  \DOIprefix\doi{10.1007/s10509-007-9348-4}.
\bibitem[{Timokhin(2010)}]{Timokhin_2010}
\bibinfo{author}{Timokhin, A.N.}, \bibinfo{year}{2010}.
\newblock \bibinfo{title}{{Time-dependent pair cascades in magnetospheres of
  neutron stars – I. Dynamics of the polar cap cascade with no particle
  supply from the neutron star surface}}.
\newblock \bibinfo{journal}{Monthly Notices of the Royal Astronomical Society}
  \bibinfo{volume}{408}, \bibinfo{pages}{2092--2114}.
\newblock \URLprefix \url{https://doi.org/10.1111/j.1365-2966.2010.17286.x},
  \DOIprefix\doi{10.1111/j.1365-2966.2010.17286.x},
  \href{http://arxiv.org/abs/https://academic.oup.com/mnras/article-pdf/408/4/2092/4220523/mnras0408-2092.pdf}{{\tt
  arXiv:https://academic.oup.com/mnras/article-pdf/408/4/2092/4220523/mnras0408-2092.pdf}}.
\bibitem[{Timokhin and Arons(2012)}]{Timokhin_2012}
\bibinfo{author}{Timokhin, A.N.}, \bibinfo{author}{Arons, J.},
  \bibinfo{year}{2012}.
\newblock \bibinfo{title}{{Current flow and pair creation at low altitude in
  rotation-powered pulsars’ force-free magnetospheres: space charge limited
  flow}}.
\newblock \bibinfo{journal}{Monthly Notices of the Royal Astronomical Society}
  \bibinfo{volume}{429}, \bibinfo{pages}{20--54}.
\newblock \URLprefix \url{https://doi.org/10.1093/mnras/sts298},
  \DOIprefix\doi{10.1093/mnras/sts298},
  \href{http://arxiv.org/abs/https://academic.oup.com/mnras/article-pdf/429/1/20/18684664/sts298.pdf}{{\tt
  arXiv:https://academic.oup.com/mnras/article-pdf/429/1/20/18684664/sts298.pdf}}.
\bibitem[{Torres et~al.(2023)Torres, Grismayer, Cruz and Silva}]{Torres2023a}
\bibinfo{author}{Torres, R.}, \bibinfo{author}{Grismayer, T.},
  \bibinfo{author}{Cruz, F.}, \bibinfo{author}{Silva, L.O.},
  \bibinfo{year}{2023}.
\newblock \bibinfo{title}{{Magnetic frame-dragging correction to the
  electromagnetic solution of a compact neutron star}}.
\newblock \bibinfo{journal}{Monthly Notices of the Royal Astronomical Society}
  \bibinfo{volume}{524}, \bibinfo{pages}{4116--4127}.
\newblock \URLprefix \url{https://doi.org/10.1093/mnras/stad2175},
  \DOIprefix\doi{10.1093/mnras/stad2175},
  \href{http://arxiv.org/abs/https://academic.oup.com/mnras/article-pdf/524/3/4116/50968079/stad2175.pdf}{{\tt
  arXiv:https://academic.oup.com/mnras/article-pdf/524/3/4116/50968079/stad2175.pdf}}.
\bibitem[{{Usov} and {Melrose}(1995)}]{Usov_1995}
\bibinfo{author}{{Usov}, V.V.}, \bibinfo{author}{{Melrose}, D.B.},
  \bibinfo{year}{1995}.
\newblock \bibinfo{title}{{Pulsars with strong magnetic fields : polar gaps,
  bound pair creation and nonthermal luminosities.}}
\newblock \bibinfo{journal}{Australian Journal of Physics}
  \bibinfo{volume}{48}, \bibinfo{pages}{571--612}.
\newblock \DOIprefix\doi{10.1071/PH950571},
  \href{http://arxiv.org/abs/astro-ph/9506021}{{\tt arXiv:astro-ph/9506021}}.
\bibitem[{Vay(2008)}]{Vay2008}
\bibinfo{author}{Vay, J.L.}, \bibinfo{year}{2008}.
\newblock \bibinfo{title}{{Simulation of beams or plasmas crossing at
  relativistic velocitya)}}.
\newblock \bibinfo{journal}{Physics of Plasmas} \bibinfo{volume}{15}.
\newblock \URLprefix \url{https://doi.org/10.1063/1.2837054},
  \DOIprefix\doi{10.1063/1.2837054}. \bibinfo{note}{056701}.
\bibitem[{{Villasenor} and {Buneman}(1992)}]{Villasenor1992}
\bibinfo{author}{{Villasenor}, J.}, \bibinfo{author}{{Buneman}, O.},
  \bibinfo{year}{1992}.
\newblock \bibinfo{title}{{Rigorous charge conservation for local
  electromagnetic field solvers}}.
\newblock \bibinfo{journal}{Computer Physics Communications}
  \bibinfo{volume}{69}, \bibinfo{pages}{306--316}.
\newblock \DOIprefix\doi{10.1016/0010-4655(92)90169-Y}.
\bibitem[{Vivekanand and Joshi(1999)}]{Vivekanand_1999}
\bibinfo{author}{Vivekanand, M.}, \bibinfo{author}{Joshi, B.C.},
  \bibinfo{year}{1999}.
\newblock \bibinfo{title}{Competing drifting radio subpulses in psr
  b0031–07}.
\newblock \bibinfo{journal}{The Astrophysical Journal} \bibinfo{volume}{515},
  \bibinfo{pages}{398}.
\newblock \URLprefix \url{https://dx.doi.org/10.1086/307013},
  \DOIprefix\doi{10.1086/307013}.
\bibitem[{Vranic et~al.(2016)Vranic, Martins, Fonseca and Silva}]{Vranic_2016}
\bibinfo{author}{Vranic, M.}, \bibinfo{author}{Martins, J.},
  \bibinfo{author}{Fonseca, R.}, \bibinfo{author}{Silva, L.},
  \bibinfo{year}{2016}.
\newblock \bibinfo{title}{Classical radiation reaction in particle-in-cell
  simulations}.
\newblock \bibinfo{journal}{Computer Physics Communications}
  \bibinfo{volume}{204}, \bibinfo{pages}{141--151}.
\newblock \URLprefix
  \url{https://www.sciencedirect.com/science/article/pii/S001046551630090X},
  \DOIprefix\doi{https://doi.org/10.1016/j.cpc.2016.04.002}.
\bibitem[{{Xu} et~al.(1999){Xu}, {Qiao} and {Zhang}}]{Xu_1999}
\bibinfo{author}{{Xu}, R.X.}, \bibinfo{author}{{Qiao}, G.J.},
  \bibinfo{author}{{Zhang}, B.}, \bibinfo{year}{1999}.
\newblock \bibinfo{title}{{PSR 0943+10: A Bare Strange Star?}}
\newblock \bibinfo{journal}{\apjl} \bibinfo{volume}{522},
  \bibinfo{pages}{L109--L112}.
\newblock \DOIprefix\doi{10.1086/312226},
  \href{http://arxiv.org/abs/astro-ph/9907132}{{\tt arXiv:astro-ph/9907132}}.
\bibitem[{Yee(1966)}]{Yee}
\bibinfo{author}{Yee, K.}, \bibinfo{year}{1966}.
\newblock \bibinfo{title}{Numerical solution of initial boundary value problems
  involving maxwell's equations in isotropic media}.
\newblock \bibinfo{journal}{IEEE Transactions on Antennas and Propagation}
  \bibinfo{volume}{14}, \bibinfo{pages}{302--307}.
\newblock \DOIprefix\doi{10.1109/TAP.1966.1138693}.
\bibitem[{Yu and Xu(2011)}]{Yu_2011}
\bibinfo{author}{Yu, J.W.}, \bibinfo{author}{Xu, R.X.}, \bibinfo{year}{2011}.
\newblock \bibinfo{title}{{Magnetospheric activity of bare strange quark
  stars}}.
\newblock \bibinfo{journal}{Monthly Notices of the Royal Astronomical Society}
  \bibinfo{volume}{414}, \bibinfo{pages}{489--494}.
\newblock \URLprefix \url{https://doi.org/10.1111/j.1365-2966.2011.18409.x},
  \DOIprefix\doi{10.1111/j.1365-2966.2011.18409.x},
  \href{http://arxiv.org/abs/https://academic.oup.com/mnras/article-pdf/414/1/489/3821220/mnras0414-0489.pdf}{{\tt
  arXiv:https://academic.oup.com/mnras/article-pdf/414/1/489/3821220/mnras0414-0489.pdf}}.

\end{thebibliography}






\end{document}